\documentclass[conference,10pt]{IEEEtran}

\usepackage{amsthm}
\usepackage{algorithm}
\usepackage{algpseudocode}
\usepackage{textcomp}
\usepackage{amsmath}
\usepackage{amssymb}
\usepackage{listings}
\usepackage{stmaryrd}
\usepackage{graphicx}
\usepackage{syntax}
\usepackage{fancyhdr}
\usepackage{textcomp}
\usepackage{semantic}
\usepackage{marvosym}
\usepackage{subfigure}
\usepackage[short,nodayofweek,level,12hr]{datetime}  
\usepackage{yfonts}
\usepackage{rotating}
\usepackage[unicode]{hyperref}

\usepackage{xy}

\newcommand{\myglsentry}[1]{\texttt{\textit{#1}}.\\}
\newcommand{\myglsdesc}[1]{\textit{#1}}

\begin{document}
\xyoption{all}

\title{Using Prolog for Transforming XML Documents}

\author{
  \IEEEauthorblockN{Ren\'{e} Haberland}
  \IEEEauthorblockA{
  Technical University of Dresden, Free State of Saxony, Germany\\
  Saint Petersburg State University, Saint Petersburg, Russia\\
  email: haberland1@mail.ru\\
  (translation into English from 2007)
  }
}

\maketitle
\pagestyle{plain}
\thispagestyle{fancy}
\cfoot{This work is licensed under the Creative Commons Attribution License (CC BY-NC-SA 4.0).}

\newenvironment{these}{
\medskip
 \textbf{Remark:}}{
 \nopagebreak
 \smallskip
}

\newenvironment{myexample}[2]{
\smallskip
\begin{tabular}{l c}
 \textbf{Example #1)} & #2\\
\end{tabular}
\medskip
\newline
}

\newcommand{\switchXML}{\lstset{language=XML,tabsize=5}}
\newcommand{\switchXSLT}{\lstset{language=XML,tabsize=3}}
\newcommand{\switchPascal}{\lstset{language=Pascal,tabsize=2}}
\newcommand{\switchProlog}{\lstset{language=Prolog,tabsize=7}}
\newcommand{\switchJava}{\lstset{language=Java,tabsize=3}}
\newcommand{\switchScheme}{\lstset{language=Lisp,tabsize=1}}
\newcommand{\switchHaskell}{\lstset{language=Haskell,tabsize=4}}

\newcommand{\myrules}[3]{
 #1
 \begin{tabular}{c}{}
  #2\\
  \hline
  #3
 \end{tabular}
}

\newcommand{\myrule}[2]{
 \begin{center}
 	\underline{#1}\\
	#2
 \end{center}
}

\newcommand{\attr}[2]{$\synt{#1}_{#2}$}
\newcommand{\df}{ ::= }
\newcommand{\arr}{ $\rightarrow$ }
\newcommand{\barr}{ $\leftarrow$ }

\begin{abstract}
Proponents of the programming language Prolog share the opinion Prolog is more appropriate for transforming XML documents than other well-established techniques and languages like XSLT.
This work proposes a tuProlog-styled interpreter for parsing XML documents into Prolog-internal lists and vice versa for serialising lists into XML documents.\\

Based on this implementation, a comparison between XSLT and Prolog follows.
First, criteria are researched, such as considered language features of XSLT, usability and expressibility. These criteria are validated.
Second, it is assessed when Prolog distinguishes between input and output parameters towards reversible transformation.
\end{abstract}

\section{Introduction}

\subsection{Preliminaries}

A transformation within XML is a mapping from \textit{XML} onto \textit{XML}.
W.l.o.g. only XML as output is considered in this work.
Unlike programming, no program but document data is being acquired from some sources and outputted later on.
The output is a result of some transformation process.
\textit{Templates} are documents with some parts being \textit{slots}, which are filled  with data from documents as requested. The document obtained is called the \textit{target document}.
Templates are sometimes called \textit{stylesheets}.

Examples of template-based \textit{transformation languages} are Xduce \cite{Hosoya}, Xact \cite{Xact} and XSLT \cite{W3XSLTSpecification}.
Transformation languages are often \textit{markup} languages.
A markup (tag) has a meaning dedicated to a certain domain. For instance, tags can be categorised by command, directive and output information.
Markups encapsulate text sections representing altogether united a corresponding document.
Markups are recursively defined over text. XML consists of markups.

Due to its focus on documents, transformation languages do not have much in common with programming languages.
Despite this circumstance, some concepts are still seamlessly interchangeable:

\begin{itemize}
 \item typing
 \item backtracking
 \item pattern matching
 \item monads
 \item unification
 \item higher-order functions
 \item non-strict functions
 \item modules
\end{itemize}

All of the mentioned and even most of the unmentioned languages have in common that they cannot be integrated at all or with severe restrictions into \textit{hosting languages}, like Java, C++ or Pascal.

In order to resolve this problem, two strategies can be identified as most promising.
First, integrate new features.
The language gets extended.
However, this can only succeed if lingual concepts are universal enough w.r.t. lexemes, idioms.
Second, choose a federated approach.
Depending on the implementation, the hosting language is simulated by introspection.
Solely concepts remain untouched.

The reasons against the first approach are a massive rise in complexity and a notoriously incompatible paradigm.
Hence, the federated approach is chosen to implement the transformation language with Prolog being visible to the user.

\subsection{Motivation}

Prolog has two key features, which make it very powerful. Those two features are \textit{unification} and \textit{backtracking} -- both of which are not present in XSLT.
Unification allows terms to be composed and described easily.
However, the handling may become cumbersome when terms reach a specific size.
Backtracking may trace multiple solutions in a tree-structured search space effectively if applied wisely.
Prolog is also well known for concise programs solving rather complex tasks.
It is expected, unification, backtracking, and additional features may improve expressibility.
However, Prolog is suspected to be inappropriate due to its minimalistic language features on some numeric problems.
The non-distinction between input and output parameters also may indicate a flexible expressibility.
Since there is almost no previous work on this topic available, new characteristics on Prolog as transformation language are expected.

\subsection{Related Work and Foundations}

\textbf{XML-processing with SWI-Prolog}

Seipel \cite{Seipel02} introduced purely experimental transformation language \textit{FNPath} as a subset of SWI-Prolog.
Since Prolog is good at dealing with symbolic terms, it may also be considered by Seipel for transforming XML documents.

XML documents are represented in FNPath as terms.
Queries are composed of navigational operators.
For distinguishing \textit{monotone} from \textit{non-monotone operators}, three classes were introduced: \texttt{FNTree}, \texttt{AssignmentTree} and \texttt{SelectionTree}.
Those classes are sorted in ascending order by abstraction level.
\texttt{FNTree} is the most generalised class.
\texttt{SelectionTree} is the most specialised class.

The FNPath-expression
\texttt{O*[\textasciicircum a:5,\textasciicircum c:6]}
denotes that in a subtree of \texttt{O} attribute a is replaced by 5, and then attribute c is replaced by 6.

Since there are no templates foreseen in FNPath, a direct comparison with \textit{XSLT} is a little concerned.
However, some questions still arise.
For instance, whether all introduced operators are complete w.r.t. a transformation language?
Are there any improvement in usability, and is the representation chosen adequate?

In general, each node in an XML tree is reachable from any other node with FNPath.
However, access may still be very hard due to bloated representations, numerous overloadings and too complex accessor functions.
Another remark on FNPath is both \textit{Parse}/\textit{Serialiser} operators are bound tightly to the SWI-Prolog framework and are by far incompatible otherwise.
All critical operations are written in C and are not part of ISO-Prolog.
Platform independence is violated regardless Prolog programs are interpreted.
These are serious concerns.

Seipel proposes Prolog or another declarative language for transformations due to its high expected abstraction level (\cite{Seipel02}, p.12).
The transformation language should be embedded in a conventional programming language.
Because of the potential non-termination of recursive clauses, a DATALOG-based evaluation manager should be used instead.

\textbf{Scheme-based XSLT-processor}

Kiselyov and Krishnamurthi \cite{Kiselyov} summarise design discrepancies and flaws on XSLT.
The most important of which are:

\begin{itemize}
 \item A few very essential functions require some extraordinary complex templates.
 \item XSLT is not appropriate for invertible transformations because ``\textit{templates are not higher-order}'' (\cite{Kiselyov}, p.1)
 \item XSLT is a closed system, with no extensions possible.
 \item Operators are not complete at all. User-defined operators are hardly available.
\end{itemize}

Apart from the flaws, expressibility and poor readability are also caused by markups.
At this point, the citation from \cite{Kiselyov} from page 4 should be mentioned in \cite{Moertel}: 
 ``\textit{The really bad thing is that the designers of XSLT [...] failed to include fundamental
support for basic functional programming idioms. Without such support, many trivial tasks
become hell.}''.
The third point addresses the same problem as was already mentioned by \cite{Seipel02}.

SXSLT is a new implementation and an extension of XSLT, which is written in Scheme.
In Scheme introspection allows on invoke programs on runtime (so, also templates).

SXSLT offers the following features for free as a result of Scheme as embedded language:

\begin{itemize}
  \item higher-order functions are handled as so-called S-expressions.
It allows calling an associated function by name during runtime.
\item local templates
\item flexible iteration ordering
\item access to a resulting tree
\end{itemize}

Although the authors criticise both the syntactic discrepancy and operations (\cite{Kiselyov}, p.3) between \textit{XPath} and XSLT in matching expressions, they still introduce the event-based navigation language SSAX.

In fig.\ref{fig:pre-post-order}, the SXSLT-function is shown that traverses an XML tree.
The result of the function \texttt{pre-post-order} is an event tree generated from \texttt{bindings} by successive application of templates.
When the function is called with an element node and a traversal function as arguments, the latter is tried.
If traversal fails, then the current node is traversed in pre-order.
Child nodes are handled similarly.

\begin{figure}[h]
\begin{verbatim}
(define (pre-post-order tree bindings)
 (cond
   ((nodeset? tree)
    (map (lambda (a)
           (pre-post-order a bindings)) 
         tree))

   ((not (pair? tree))
    (let (trigger '*text*)	 
     (cond 
      ((or (assq trigger bindings) 
           (assq '*default* bindings))
       (lambda
         (b) 
         (((procedure? (cdr b))
           (cdr b) (cddr b)) 
           trigger tree)))

       (error "Unknown binding"))))

   (error handle-children-nodes...)))
\end{verbatim}
\caption{Scheme-function pre-post-order (after \cite{Kiselyov})}
\label{fig:pre-post-order}
\end{figure}

In the next step both, Kiselyov and Krishnamurthi want to integrate additional features into SXSLT like context propagation, additional traversal strategies and a type system.

\textbf{Hypothetical XML-transformation processor in Haskell}

Meijer and Shields \cite{Meijer} proposed XM$\lambda$ for typing transformation languages.

Typing was considered too often dropped in favour of a shorter and easier notation.
That is why both designed the language XM$\lambda$.
XM$\lambda$ is based on Haskell, so it is a statically typed transformation language and provides higher-order functions, type polymorphism, pattern matching, type constructors and \textit{monads} (\cite{Meijer}, p.6).
Transformation directives are modelled as tags, which are evaluated in Haskell.
So the transformation script uniquely consists of tags encapsulating element constructors and Haskell expressions internally.

Fig.\ref{fig:getParaHaskell} shows typing and definition of the example function \texttt{getPara}.
The tag in paragraph \texttt{<P>} contains the bound variable \texttt{p}, which occurs on the lambda-term right-hand side.
The call \texttt{getPara <P>Hello World!</P>} returns \texttt{Hello World!}.

\begin{figure}[h]
\begin{verbatim}
getPara::P->String
getPara = \<P><%= p%></P> -> p
\end{verbatim}
\caption{The XM$\lambda$-function \texttt{getPara}}
\label{fig:getParaHaskell}
\end{figure}

\textbf{Higher-order functions in XSLT}

Many scientists are convinced about XSLT is declarative (\cite{Novatchev}, p.1).
It is strictly functional, s.t. in practice, this may even become a notable hinder.

Example by example, Novatchev explains in detail how \textit{functionals} are defined and used in XSLT.
XSLT is not changed herewith.
New functions are defined in new namespaces.

\begin{figure}[h]
\begin{center}
 \includegraphics[width=8cm]{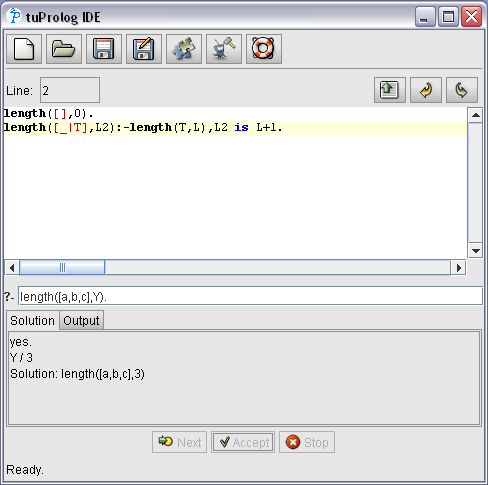}\\
\end{center}
\caption{tuProlog IDE window}
\label{fig:tuPrologIDE}
\end{figure}

\textbf{tuProlog}

tuProlog has actively been developed at the University of Bologna, Italy.
It implements a Proxy-pattern in Java \cite{Denti01}, \cite{Denti05} and allows it to define its \textit{functors} and predicates (see fig.\ref{fig:tuPrologIDE}).
Once defined, these can be used in Java and a Prolog context.
Even a combination of both is possible.

\subsection{Use Cases}

Due to its semi-structuredness, XML is popular for cross-platform communications.
XSLT is often used on the server-side to transform document, which is exchanged during communication between server and client.

For example, let a typical client/server warehouse architecture be given where the communication is based upon XML in the application layer.
Let client requests be stored before an invoice is issued.
Multiple data needs to be gathered from different sources, like customers ID, order ID, to create the invoice.
The generation would be implemented as XSLT.

\textbf{Case Study No.1) Business rules policy in contractual agreements}

In \cite{Grosof}, Grosof, Labrou and Chang present Prolog to process descriptions and strategy rules in an E-commerce background with business rules.
Business rules in Prolog seem to have a significant advantage over imperative approaches or even SQL views, namely a semantically adequate representation.
That can easily be seen by the appropriate and still flexible description in comparison to other approaches.

For example, a particular business rule may be: ``\textit{If buyer returns the purchased good because it 
is defective, within one year, then the full purchase amount will be refunded}''
(\cite{Grosof}, p.69).
Business rules are in a knowledge base, which may be adapted by need \cite{Baumeister}. 

The authors recommend -- even if that would technically be possible, still to minimise integrational risks and leave existing routine work with the existing software infrastructure for stability purposes, such as triggering orders in case of running out of stock, several event-based database triggers.\\

\begin{figure}
\begin{center}
 \includegraphics[width=8cm]{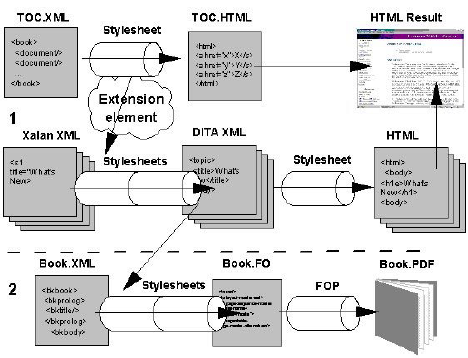}\\
\end{center}
\caption{Multichannel publishing with ``\textit{DITA}''}
\label{fig:DITA}
\end{figure}

\textbf{Case Study No.2) Multichannel Publishing}

In \cite{Leslie}, Leslie proposes using three resulting documents for an incoming XML document.
For an XML document, as input, an HTML list is generated using an XSLT-stylesheet, representing a table of contents.
Second, an HTML document is generated, so it is human-readable.
Third, PDF is generated for the same HTML document (see fig.\ref{fig:DITA}).

\subsection{Objectives}

The objectives can further be set as follows:

\begin{itemize}
 \item Analysis and Design of how to represent XML documents best in Prolog.
 \item Implementation of XML-parser and serialiser.\\
 How can XML-data be read into Prolog and be written back into an XML-file?
 \item Implement typical transformations within Prolog.\\
 In analogy to XSLT, a relatively complete set of examples should be implemented in Prolog and compared with XSLT.
 Additionally, appropriate operators shall be defined over Prolog terms -- which eventually make up Prolog as transformation language. This shall be investigated w.r.t. completeness and usability.
 Numerous tests shall assure the quality of all designed operators.

 What are typical examples of transformations?
 How to divide essential from additional operators which could improve usability and appear plausible?
 Are there any restrictions, new possibilities or exceptional cases due to Prolog's logical nature?

 \item Determining comparison criteria.

Which \textit{software metrics} as known from procedural programming languages may be adapted to XSLT and Prolog?
Is there anything to take into consideration herewith?
What other measures apart from metrics shall be considered?
How to measure qualitative features?
Which kind of transformations allows to flip original and target documents (\textit{invertibility})?

 \item Comparison of Prolog and XSLT by criteria.

In which cases does XSLT better, and in which does Prolog better?

How significant are these advantages?
\end{itemize}

\subsection{Structure of this Work}

Sect.\ref{sect:Foundations} introduces basic terms needed.
XSLT as XML-transformation language and Prolog as a logical programming language are briefly introduced together with its most essential concepts.

Sect.\ref{sect:Design} deals with the processing of XML documents in Prolog.
The mapping from an XML tree into a Prolog-term and vice versa are discussed and implemented.
Afterwards, transformation operators are introduced.
Specialities of Prolog towards transformations are presented.

Sect.\ref{sect:Implementation} gives a short overview of the implementation of the Prolog components.
It briefly shows the user interface, the overall architecture and the most critical components of the system.

Sect.\ref{sect:Comparison} defines the essential comparison criteria, and comparison is mentioned by selected examples and assessed.
Overall results and tendencies are discussed.
Finally, the invertibility is probed for document transformation in general.

Sect.\ref{sect:Summary} summarises all previous sections and provides an outlook on the future development of the logic-oriented approach presented in this work.

\section{Foundations}
\label{sect:Foundations}

\subsection{XML and XSLT}

\subsubsection{Common}

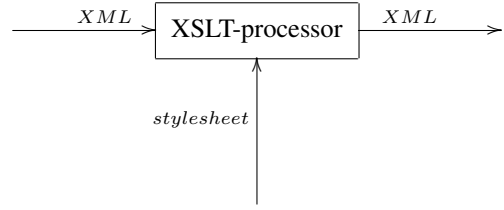
\begin{figure}
\begin{center}
\begin{tabular}{l}
\xymatrix{
   \ar[rr]^{XML \qquad } && *++[F]\txt{XSLT-processor} \ar[rr]^{\qquad XML} && \\\\
          &&  \ar[uu]^{stylesheet}       && 
}
\end{tabular}
\end{center}
\caption{XSLT processor}
\label{fig:xsltProc}
\end{figure}

XSLT is a specification language in XML for document transformations.
A XML document is transformed into another XML document (see fig.\ref{fig:xsltProc}).
Even if the result does not necessarily be XML in the case of XSLT, it is still agreed upon XML for the sake of simplicity.
W.l.o.g., in the flat text, it is agreed upon that a surrounding XML tag always embraces the utmost text.
XML is a semi-structured markup language.
Tags capture the data in an XML document.
Tags may contain numerous other tags.
Hence, tags have a tree structure.
A brief review on actual problems with XML and formal semantics can be found in \cite{Renar}.

Many states of the art transformation languages are descriptive.
XSLT does not have side effects nor explicit variable assignments.
It is functional.
XSLT integrates programming features originating from imperative and object-oriented transformation languages, as in XACT \cite{Xact}, \cite{Kirkegaard}, \cite{Christensen04}.

\subsubsection{Backus-Naur-Form of XML}

The exact syntax of an XML document is defined in Backus-Naur-form as

\begin{center}
\label{intext:bnfxmldefinition}
\begin{grammar}
<XMLNode> ::= <Element> \alt <PI>
	\alt <Comment>
	\alt <Text>

<Element> ::= \textless <Id> <Atts> /\textgreater
	\alt \textless <Id> <Atts> \textgreater <XMLNode> <XMLNode2> \textless / <Id> \textgreater
	
<XMLNode2> ::= $\varepsilon$
	\alt <XMLNode> <XMLNode2>

<PI> ::= \textless ? <Text> \textgreater

<Comment> ::= \textless ! - -  <Text> - - \textgreater

<Atts> ::= $\varepsilon$
	\alt <Id> = $''$ <Text> $''$ <Atts>
\end{grammar}
\end{center}

\synt{Id} denotes some identifier.
Identifiers start with a letter and are followed by arbitrary many alpha-numeric characters.
The alternative of \synt{Element} implies both \synt{Id}s are identical.

\synt{Text} denotes an arbitrary XML-conform string.
Arbitrary text may occur, except brackets, for example, $<$ is escaped as \&lt; and $>$ as \&gt;.

\synt{Comment} denotes a common SGML-conform comment.

\synt{PI}s denote so-called ``\textit{processing instructions}'' and are used by some dedicated applications only.
Data that does not belong to the XML document may be encoded into \textit{PI}s.

\subsubsection{XPath}
\label{sect:xpath}

XPath is a navigation language for XML \cite{XPathSpec}, and its distinction is high expressibility and extensibility.
XPath is a ``\textit{sub}''-language of XSLT and XQuery.
``\textit{Sub}'' refers here not to set inclusion w.r.t. formal languages, instead it denotes here a mechanism, s.t. expressions in XPath may be embedded into XML.
Although XSLT is XML, XPath is not.
That is one reason why all three languages require different interpretation and tools for each language.
XPath expressions are embedded into XSLT by \texttt{select} and \texttt{match} attributes.
XPath allows locating nodes and attributes within a tree representing an XML document.
XPath offers aggregate functions over numbers and strings also.

The operator \lit{/} is used in order to address the top-level node in a tree model.
In general, however, the top-level element node may have a sibling node right before.
Beginners often commit this mistake.
A path expression needs to be built up using the operators \lit{/} and \lit{//} to locate a node or attribute.
\lit{/} searches from the current node the node immediately one level below.
\lit{//} searches for any level below (implicitly also assuming the current level).
So, \texttt{//person/address/city} looks from the current node downwards until it finds an element node called \texttt{person}.
There must be another element node called \texttt{address} for each node found underneath, which is directly above the \texttt{person}.
If found, then one level below there must be just another element node called \texttt{city}.
If now all conditions match, then the entire content from \texttt{city} is returned.
If a specification requires different conditions to be applied to found \texttt{city} nodes, then \lit{[]} needs to be suffixed with a meaningful condition.
The expression within the square brackets is evaluated for every matching node.
Only if the predicate is satisfied, the found \texttt{city}-node is added to the resulting multi-set.
For example, \texttt{\$X[@id]} filters all nodes \texttt{\$X}, satisfying the predicate \texttt{@id}.
So, only if an element node has an attribute \texttt{id}, then this node qualifies as a result.

Once XPath gathers all qualified results, it passes them to the host language processor, which here is XSLT.
Qualifying results may contain nodes, node sets, boolean values, strings and numbers.
A node is returned whenever a path expression locates at least one matching node from the XML model.
Otherwise, a node-set is returned.
Boolean values, strings, integers and real numbers are implicitly turned into strings by matching aggregations for strings and core arithmetics.

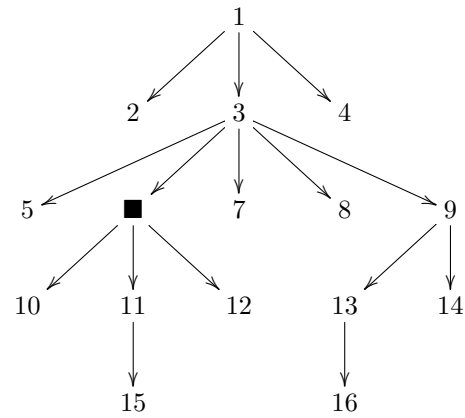
\begin{figure}[h]
\begin{center}
\begin{tabular}{l}
\xymatrix{
  && 1 \ar[dl] \ar[d] \ar[dr] &&\\ 
 & 2 & 3 \ar[lld] \ar[ld] \ar[d] \ar[rd] \ar[rrd] & 4 &\\
 5 & \blacksquare \ar[dl] \ar[d] \ar[dr] & 7 & 8 & 9 \ar[dl] \ar[d] \\ 
 10 & 11 \ar[d] & 12 & 13 \ar[d] & 14\\ 
  & 15 &  & 16 & 
}
\end{tabular}
\end{center}
\caption{XPath-Axes}
\label{fig:xpathAxes}
\end{figure}

Once desired nodes have been found, XPath easily locates its neighbours (see fig.\ref{fig:xpathAxes}).
In the following, the black node \texttt{\$X} is assumed to be the starting node.
XPath-axes stand before the operator \lit{::}, the navigation expression follows.
The \texttt{ancestor}-axis seeks for a predecessor.
If \texttt{\$X} is applied to an \texttt{ancestor}, then \{1,3\} is returned.
The path expression is \texttt{\$X::ancestor}.
The only parent node three is obtained using \texttt{parent}.
Node 6 is obtained by \texttt{\$X::self}.
\texttt{\$X::descendant} returns \{10,11,12,15\}.
\texttt{following} returns \{7,8,9\}.
\texttt{preceding} returns \{5\}.
Besides, \texttt{ancestor-or-self} and \texttt{descendant-or-self} unite two axes.
\texttt{ancestor-or-self} returns \{1,3,6\}.
\texttt{descendant-or-self} returns \{6,10,11,12,15\}.
Besides the presented axes, there are shortened operator synonyms.
For instance, \texttt{::attribute} stands for \lit{@}, where \lit{//} stands for \texttt{::descendant-or-self}.
\lit{..} stands for \texttt{::ancestor}.

The \lit{@}-operator selects attributes.
Left of \lit{@} is the selected node.
Right of it is the attribute name to be selected.
If the attribute does not exist, an empty node-set is returned.
If \lit{@} occurs inside the \lit{[]}-predicate, then the XPath-expression is only a check for the occurrence of an attribute specified.
\texttt{\$X@id} returns attribute \texttt{id}'s value, where \texttt{\$X[@id]} checks if the actual node \texttt{\$X} has an attribute with name \texttt{id}.

\subsubsection{Rule-based Transformation}
\label{sect:rulebasedTransformApproach}

As mentioned in the introduction, an XML transformation takes one or more XML documents and turns them into some target XML document.
A given set of rules can do this, hence the rule-based approach.
The rules appear unprioritised, covering parts of the transformation.
Transformation rules may be modelled as $x \mapsto y$.
Here, \textit{x} provides the pattern an element node has to match with, and \textit{y} denotes the resulting node-set.
They are coming up with transformation rules in XSLT.
Transformation rules are templates (see sect.\ref{sect:templates}) of the kind:

\begin{center}
\texttt{<xsl:template match=X>Y</xsl:template>}
\end{center}

\textit{X} is the source document.
\textit{Y} is the target document.
A valid binding would, for example, be \texttt{X="/"},  \texttt{y="<a>hello world!</a>"}.

An XML-tree is always \textit{traversed} in pre-order except specified differently explicit.
Every element node is visited by default exactly once.
The XSLT processor attempts to apply a matching template.
If the attempt fails, then the traversal is continued in pre-order.
Otherwise, a template is selected from a potential set of templates and applied.
The result set is returned to the caller.

The user may alter the implicit traversal with \texttt{apply-template}.
\texttt{call-template}, together with the attribute \texttt{mode}, may deactivate the implicit traversal.
Instead, a template is explicitly called.
The attribute \texttt{priority} can prioritise among matching rules so that a particular result set may be favoured.

The introduced tags allow writing user-defined traversal functions.
The most crucial functions (in-order, post-order) shall be standardised.
For instance, an inverse polish intermediate representation of terms encoded as tags are in MathML.

\begin{these}
By pre-defined traverse functions, a transformation language's expressibility is increased (cf. \cite{Kiselyov}).
\end{these}

Naturally, parametrisation comes for an additional tax.
A switch for controlling traversal would be highly recommended.
Every user should pass traversal ordering as an argument to the template or the actual traversal function (cf. \cite{Kiselyov}) to assure \textit{referential transparency}.
That would avoid global variables and I/O-operations.
Both, left-hand and right-hand sides of a transformation rule may be extended by ``\textit{syntactic sugar}'', s.t. tests for membership could be simplified as well as more sophisticated \textit{pattern-matching} (\cite{Vion-Dury03}, pp.21).
Unfortunately, improvements are made only from XSLT 2.0 onwards.
These improvements include loop extensions over arbitrary types, regular expressions in strings and a more parametrised matching on selection templates.
However, XSLT is closed by its design, and no variation nor extension are easily doable.

\subsubsection{XSLT}

As already mentioned, XPath is part of XSLT.
Each XSLT-stylesheet is in a separate XML-file.
Numerous tools for XML contributed also to the popularisation of XSLT, for instance \cite{XalanJ}, \cite{XMLSpy}, etc.\\

Fig.\ref{fig:xsltTags} summarises the essential XSLT tags.
\texttt{stylesheet} is the root element of every XSLT transformation.
It unites many arbitrary \texttt{template} \cite{W3XSLTSpecification}.
The attribute \texttt{match} contains an XPath expression associated with a particular node.
The attribute \texttt{name} provides named templates.
Template calls can either implicitly use \texttt{apply-templates} or explicit by calling \texttt{call-template} with a previously defined template name.
\texttt{nodesetexpression} denotes on an implicit traversal the node-set to be traversed.
The nodes matching for a given XPath expression are applied to the corresponding template.

Expressions may be evaluated via \texttt{value-of}.
Expressions are strings, numbers, variables and trees.
\texttt{copy-of} returns an exact copy of a tree node.

Controlling a template includes conditions with/without an alternative, case selection and loops.
Conditions are specified with "\texttt{if}" if there is no alternative and with \texttt{choose} if there is at least one alternative.
"\texttt{choose}" \texttt{otherwise} denotes all the cases united that are not previously covered by any of the specified cases.
\texttt{sort} sorts a node-set by a specified attribute.
Sorting by multiple attributes is available only since version 2.0.
Sorting may be ascending or descending.

Variables and parameters in XSLT can be declared by \texttt{variable} and \texttt{param}.
Except for the syntax, they seem to be identical.
For instance, identifiers from either one are preceded by a dollar sign within an XPath expression, e.g. for a  \texttt{select} or \texttt{match} attribute.
Parameters and variables must be declared in a template.
Calls to named templates require the parameter \texttt{with-param} is passed for recognition purposes.

\switchXML
\lstset{emph={nodesetexpression,attributename,templatename, stringexpression,condition,qname,pattern,qnumber,expression},emphstyle=\underbar}

\begin{figure}[h]
\begin{center}
\begin{tabular}{c}
Apply further templates\\
\begin{lstlisting}
<xsl:apply-templates 
  select=nodesetexpression
  node=... />
\end{lstlisting}\\[0.7cm]

Explicit template-call\\
\begin{lstlisting} 
<xsl:call-template
  name=templatename />
\end{lstlisting}\\[0.7cm]

1:1 deep copy of an arbitrary expression\\
\begin{lstlisting}
<xsl:copy-of
  select=nodesetexpression />
\end{lstlisting}\\[0.7cm]

Bound loop (no early break) over XPath-expression\\
\begin{lstlisting}
<xsl:for-each
  select=nodesetexpression />
\end{lstlisting}\\[0.7cm]

Parameter definition\\
\begin{lstlisting}
<xsl:param name=qname
   select=nodesetexpression />
\end{lstlisting}\\[0.7cm]

Soring in \textit{for-each}-blocks\\
select .. attribute to be sorted by\\
order .. \textit{ascending}/\textit{descending}\\
\begin{lstlisting}
<xsl:sort select=stringexpression
  order="ascending"|"descending"
  ... />
\end{lstlisting}\\[0.7cm]

XSLT-document definition\\
\begin{lstlisting}	 
<xsl:stylesheet
   version=number />
\end{lstlisting}\\[0.7cm]

Template-declaration\\
\begin{lstlisting}
<xsl:template match=pattern
  name=qname ... />
\end{lstlisting}\\[0.7cm]

Evaluated expression\\
disable-output-escaping .. \&gt; as $>$, etc.\\
\begin{lstlisting}
<xsl:value-of
  select=stringexpression 
  disable-output-escaping="yes"|"no" />
\end{lstlisting}\\[0.7cm]

Variable declaration\\
\begin{lstlisting}
<xsl:variable
  name=qname
  select=expression />
\end{lstlisting}\\

\end{tabular}
\end{center}
\caption{The essential XSLT-tags}
\label{fig:xsltTags}
\end{figure}

No side-effects are possible by global variables.
The execution ordering in a stylesheet is always sequential.
If an error occurs, neither is a step rejected nor is an alternative tried.

\begin{these}
By integration of a DATALOG-oriented clause scheduler, existing declarative transformation languages are enabled with multi-solution recognition.
If there is a finite solution, the scheduler can get it before running into a non-terminating loop.
Counter-measures may effectively be taken so against non-terminating loops.
\end{these}

If more than one template matches in XSLT, then a warning is issued, and transformation continues with only the first of all alternatives.
Alternatively, the transformation process may be cancelled.

In Prolog-based systems, multiple solutions can be evaluated.
A DATALOG-scheduler allows tracing all paths equally in breadth and width due to modified backtracking.\\

Variables in XSLT can only be assigned once -- this is the semantic equivalent to a symbol.
That is different to imperative programming languages, where redefinitions are allowed.
Once a symbol is bound, it remains forever within a particular template.
Multiple assignments within a template are not valid.
In the following example, the template returns \texttt{1}.
The second binding is suppressed, and a warning is issued, depending on the concrete XSTL-processor.

\begin{verbatim}
<xsl:template match="/">
 <xsl:variable name="a">1</xsl:variable>
 <xsl:variable name="a">3</xsl:variable>
 <xsl:value-of select="$a"/>
</xsl:template>
\end{verbatim}

A comprehensive semantic is given in \cite{Wadler}.
Ongoing expressibility research can be found, e.g. in \cite{Janssen}.

Parameters are used as usual, and there is no semantic discrepancy on variables from imperative languages (cf. fig.\ref{fig:gcdXslt}).
The loop introduced by \texttt{for-each} with a finite number of iterations defined before entering the loop body for the first time may be escaped earlier by an additional condition passed to \lit{[]}.
For example, the number of iterations may be limited to the first 10 in case there is such a consecutive sequence first:

\begin{verbatim}
<xsl:for-each select="//a[position()<10]">
\end{verbatim}

In this work Xalan/J is used as XML-processor \cite{XalanJ}.

\subsection{Prolog}

\subsubsection{Common}

Prolog is a general-purpose programming language with a high abstraction level.
This statement can easily be checked when comparing typical Prolog applications with those in C or Pascal, for instance.
For a profound look into numbers, numerous publications and reports on software metrics, like the MacCabe complexity, should be taken into closer consideration.

In contrast to imperative programming languages, Prolog describes a problem rather than providing an instruction sequence for solving it.
Hence, Prolog is descriptive and allows the user to focus on the problem description more.\\

A Prolog-program consists of facts, rules and queries (\cite{Sterling}, p.11).
The actual calculation is a constructive proof or a query refutation (\cite{Sterling}, p.4).
If a goal can be derived from a given knowledge base's rules, then the calculation succeeds, and the result is returned whatever is specified to the query as a result.
Otherwise, a query is refuted, or it does not terminate.
So, the result has three options: \texttt{Yes}, \texttt{No} or ``\textit{No result decidable}''.

Facts have the form  \texttt{$A_{0}$.}, where $A_{0}$ is a placeholder for any predicate without any premise.
$A_{0}$ can be interpreted as relation \texttt{A0($r_{0},...,r_{n}$)}.
$\forall i.r_{i}$ elements of the relation $A_{0}$ can be atoms, symbols, lists or any of its composition.
Atoms are numbers and identifiers.
Identifiers are strings with a first lowercase character, and symbols are identifiers with a first uppercase character or \lit{_}.
A \textit{list} is a comma-separated enumeration, guarded by square brackets.
Terms are constructs, consisting of a functor and an arguments list, for instance, \texttt{element(top,[],[])}.
Let the following two facts be given:

\begin{verbatim}
mortal(socrates).
immortal(zeus).
\end{verbatim}

Sentences of the form "$A_{0}\leftarrow A_{1},A_{2},...,A_{n}.$" denote rule.
"$A_{0}$" is the head of a rule, where "$A_{1},..,A_{n}$" are subgoals (\cite{Sterling}, p.18).
The comma in between subgoals is logical conjunction.
Only if all subgoals $A_{i}$ are satisfied, the entire rule also satisfies.
In Prolog $\leftarrow$ is replaced by \texttt{:-}.

Now another fact and a new rule \texttt{mortal} are introduced.
It reads as: ``\textit{If X is a human, then X is mortal}''.

\begin{verbatim}
human(socrates).
mortal(X):-human(X).
\end{verbatim}

Queries have the form "?-$A_{0},A_{1},...,A_{n}$.".
$A_{0}$ till $A_{n}$ are subgoals. They are run successively.
If one subgoal succeeds, then the subgoal with the following index is tried.
If one subgoal fails, the remaining are not run, and the entire rule fails.

The evaluation ordering can be different if a subgoal fails.
The rule scheduler may switch to the next subgoal and skip one to return back later.
If all subgoals succeed, \texttt{YES} is returned, \texttt{NO} otherwise.
The query \texttt{Y=socrates,mortal(Y)} binds the atom \texttt{socrates} to the symbol \texttt{Y} and checks then if \texttt{mortal(socrates)} is true.
According to the given rules, \texttt{YES} is derived.\\

The computation model is reduction (\cite{Sterling}, p.23).
Queries can be considered procedural statements.
Rules are procedures, where a rule-head corresponds to a procedure head, and conjunction of a sequence of subgoals corresponds to a procedure-body.
Facts can be modelled as constants.
In a reduction step, a procedure call is performed.
The reduction is made \textit{lazily} in Prolog (\cite{SterlingPractice}, p.178).
So, a subgoal is only evaluated if needed for the overall calculation.
Fig.\ref{fig:evalPrologQuery} shows the complete derivation for query \texttt{a2(X,1,3)}.
\texttt{a2} implements the Ackermann function as following:

\begin{verbatim}
a2(0,M,Res):-Res is M+1.
a2(N,0,Res):-N1 is N-1, a2(N1,1,Res).
a2(N,M,Res):-N1 is N-1, M1 is M-1,
             a2(N,M1,Res2), 
             a2(N1,Res2,Res).
\end{verbatim}

\begin{figure}
\begin{center}
\includegraphics[width=9cm]{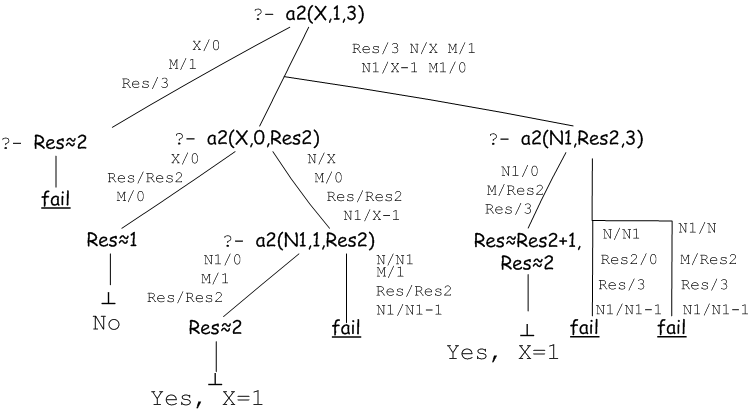}
\end{center}
\caption{Evaluation of a Prolog-query}
\label{fig:evalPrologQuery}
\end{figure}

The most crucial construct in Prolog is the logical term expression.
It is defined as following (according to \cite{Sterling}, p.27):

\begin{grammar}
<term> ::= <constant> \alt <variable> \alt <compound term>
\end{grammar}

\synt{constant} denotes a constant like \texttt{a, tmp} or \texttt{a13b}.
\synt{variable} denotes a Prolog "\texttt{variables}".
Strictly speaking, Prolog does not know of real variables but symbols, although they are called variables in Prolog.
Except said differently, we refer to Prolog variables (symbols) when talking about variables in Prolog.
Symbolic evaluation is another strength of Prolog.
Symbols are not typed and can be restricted.
The value of a symbol is a term, which may contain symbols again.
A symbol's denotation may be determined on runtime.
Therefore it is dynamic depending on the execution and bound to a scope.
If sub-expressions of a symbol value change, then the meaning of the symbol instantaneously changes too.
Unfortunately, symbols also may have disadvantages.
One severe is speed.
However, making generalisations is neither precise nor appropriate because a symbol may not make hard estimates on performance whatsoever.
It may depend on other parameters like algorithm, execution model.
\synt{compound term} denotes composed terms like \texttt{f(0),g(f(0),f(1))}, where \texttt{f} and \texttt{g} are functors.

Sterling made on page 87 in \cite{Sterling} a true comment: ``\textit{unification is the hard of the computational model of logical programs}''.
In logical programming it is often required to check if two terms are the same or if those may be transformed into each other.
It is called unification.
For example, the two terms \texttt{g(X,f(f(0)))} and \texttt{g(f(0),f(X))} are unifiable, if \texttt{X} is bound to \texttt{f(0)}.
However, \texttt{g(1)} is not unifiable with \texttt{g(f(X))}, since \texttt{1} is not unifiable with \texttt{f(X)}.
In contrast to classic equality checks, the original symbols state within a term may after unification not be the same as before -- among the current list of subgoals at least, because unification "overwrites" (in fact, it only binds once) a symbol as soon as a unification attempt succeeds.\\

A trivial result set is a solution for a request  which is derived directly from a fact.
In Prolog, \textit{trivial solutions} can often easily be derived.
All other solutions need to be derived from facts and rules and are called non-trivial solutions.
When looking for non-trivial solutions, the evaluation manager may get stuck in non-terminating cycles in general.
Rules that do not assure this is avoided and called \textit{insecure} rules (\cite{SterlingPractice}, p.147).
Insecure rules need to be avoided.\\

The semantics for a given Prolog-program $P$ can be described by a Herbrand-universe, the Herbrand base and an interpretation and the model.
For further consideration, let $P$ be the following program:

\begin{verbatim}
 natural_number(0).
 natural_number(s(X)):-natural_number(X).
\end{verbatim}

The Herbrand-universe \textgoth{U}(P) is all \textit{ground terms}, which are composed of function symbols and constants (\cite{Sterling}, p.102).
For example, \textgoth{U}(P)=\{0, s(0), s(s(0)), ...\} holds.
The Herbrand-base \textgoth{B}(P) denotes the set of all ground terms that can be composed of all predicates over $P$ and terms from \textgoth{U}(P).
So, whenever \textgoth{U}(P) is infinite, \textgoth{B}(P) is also infinite.
For the given example \textgoth{B}(P) = \{\texttt{natural\_number(0)}, \texttt{natural\_number(s(0))}, ... \} holds.

An interpretation calculates the subset of the Herbrand-base \textgoth{B}(P).
It consists of mapping rules for constants, functors and relations.
A model is the set of all possible interpretations \cite{Sterling}.\\

The $\lambda$-calculus is the underpinning apparatus used later (\cite{Sterling}, p.119) --- the ordering of the rules matters.
A rule coming first has the highest \textit{priority}, and a rule is coming last the lowest.
That is why a rule with left-recursion may not come before a base case.
Otherwise, the interpretation of a corresponding predicate may not be determined.
The ordering of subgoals also matters.
If the Prolog interpreter's calling convention is violated, a fail or a massive overload may be the consequence.
That is due to an exploding search space that may not have sufficient constraints until the interpreter runs out of memory or time. 
For instance, \texttt{fact(2,X)} $\vdash$ \texttt{fact(1,_)} ... $\vdash$ \texttt{fact(-2,_)} $\cdots$ is critical for this program:

\begin{verbatim}
fact(N,Res):-N1 is N-1,
  fact(N1,Res2),
  Res is N*Res2.
fact(0,1).
\end{verbatim}

\subsubsection{Advanced Concepts}
\label{sect:baseConcepts}

Although Prolog does not know of explicit type casts, it is sometimes vital to know which category symbols belong.
Thus, \textit{meta-logical type predicates} were introduced into ISO Prolog (\cite{Sterling}, p.176).
It allows coarse type checks on term expressions.
Such predicates include \texttt{atom/1, var/1, list/1, compound/1, atomic/1}
and \texttt{ground/1}.

A remarkable feature in Prolog is a \textit{cut}, which is introduced as the built-in subgoal \lit{!}.
Cuts allow for cutting off multiple solutions during runtime.
Since there is always the risk to cut off the right solutions, the proper usage of a cut shall always be done carefully.
Depending on the problem to be addressed, cutting off wanted results is a so-called ``\textit{forbidden}'' (or RED) cut, where a cut only drops redundant results without losing relevant information is so-called permissive (GREEN).

Often GREEN cuts are used in order to find the first solution only, not necessarily the optimal solution.

Let us consider the programs $P1$ and $P2$ in fig.\ref{fig:factorial} in order to demonstrate cuts. 

\begin{figure}
\begin{center}
\begin{tabular}{lr}
\begin{tabular}{l}
\begin{minipage}{3.3cm}
Program P1:
\begin{verbatim}
fact(N,Res):-
  N>0,N1 is N-1,
  fact(N1,Res2),
  Res is N*Res2.
fact(0,1):-!.
\end{verbatim}
\end{minipage}
\end{tabular} &
\begin{tabular}{l}
\begin{minipage}{3.3cm}
Program P2: 
\begin{verbatim}
fact(N,Res):-!,
  N>0,N1 is N-1, 
  fact(N1,Res2),
  Res is N*Res2.
fact(0,1).
\end{verbatim}
\end{minipage}
\end{tabular}
\end{tabular}
\end{center}

\caption{Cut-variants of factorial}
\label{fig:factorial}
\end{figure}

Program $P1$ demonstrates a GREEN cut.
When \texttt{fact} is called, the input value is checked if greater than zero.
It covers the recursive part of the definition.
The case "$N=0$" is covered by the second rule.
By the preceding comparison in the first rule, non-termination is avoided.
Without the comparison, the first rule would always match, and so the program would 
not terminate.
In the second rule, the cut is in the body of the rule.
At this position, it cuts off all following alternatives, s.t. interpretation resumes from the caller's position.
$P1$ calculates the factorial function for a given natural number.

Program $P2$ demonstrates a red cut.
The call \texttt{fact} returns the incorrect result \texttt{NO} since the first rule always matches.
The second rule is never considered as an alternative because the cut occurs before the first rule.

In consequence, the first rule calls itself until the condition $N>0$ does no more hold.
As soon as the second subgoal does not match, no other alternative is sought.
So, the query fails finally.
The program $P2$ calculates the factorial function.\\

Since there is no exact definition of a boolean negation in the same way as in Pascal or Java, negation needs to be defined on the solution sets w.r.t. a predicate or interrupt the calculation of negation at an appropriate position only.
To negate a predicate means actually \texttt{YES} if correct reasoning turns into a \texttt{NO}, and \texttt{NO} turns into \texttt{YES}, but only under the condition the result is determined and terminates before exiting with success or fail in both directions.
Although it does not always hold, the predicate \texttt{not/1} can achieve it.
Remarkably, all following subgoals are eventually not triggered in case of a fail.

If a cut is used as negation, then this means that all alternatives are excluded on fail, which needs explicitly to be triggered by calling the built-in predicate \texttt{fail/0}.

Non-deterministic programming is also characteristic of Prolog (\cite{Sterling}, p.250-281).
Here, a systematic check for multiple solutions is meant.
The solution set is gathered by \textit{sequential search} and backtracking.
This strategy is better known as "\textit{generating and test}" and is meant metaphysically as in fig.\ref{fig:generateNTest}:

\begin{figure}[h]
\begin{verbatim}
 find(X):-generate(X),test(X).
\end{verbatim}
\label{fig:generateNTest}
\caption{generate-and-test meta-rule}
\end{figure}

So, first, some solution \texttt{X} is generated, then it is tested if this solution fulfils all required constraints.
Mixing up both subgoals does not lead to a universal solution because, in general, \texttt{X} may not be unified, s.t. a concrete singleton result matches a generated particular result.
The predicate \texttt{find} returns many arbitrary results because, during evaluation, all \texttt{X} are sought to match with \texttt{generating}.
The test should be as close as possible to the generational subgoal to quickly reduce search space to reduce evaluation overhead.
The non-determinism in Prolog has two instances according to \cite{Sterling} (p.250-281):

\begin{itemize}
 \item "\textit{do not care}"-non-determinism:
   Among multiple rules, any arbitrary rule is selected.
   In this case, it is initially clear by the given rules set that an arbitrary application ordering of the rules will always bring the same result.
   So, the ruleset is confluent.

 \item "\textit{do not know}"-non-determinism:
   Here a rule is chosen, and it is tried to prove a subgoal.
   If it fails, backtracking will bring control back to the last still valid state and search from there onwards for further alternatives.
   As soon as a subgoal finds a successful path till the end, the algorithm succeeds.
It is assumed for a correct solution exists for that algorithm.
Otherwise, an exhaustive search for alternatives may be the result.
 If no alternative exists, then the subgoal fails.
   The difficulty lies in deciding which rule is the correct one.
\end{itemize}
\label{intext:defInvertibility}

Invertibility of relations is closely related to non-determinism, and it implies a relation has more than one valid input.
$P1$ from above is not invertible since only the first argument \texttt{N} qualifies as input.
If \texttt{fact} is bound to input with a symbol as the first argument and any other argument as second, it may not be inferred.
Hence, \texttt{N} is unknown, so is the subgoal \texttt{N1 is N-1}, which cannot be evaluated arithmetically.
Invertibility allows in a practical sense to determine an inverse mapping for n incoming arguments.
If for a function, all inverse functions are determined, then this function is called \textit{fully invertible}.

\textit{Invertible transformations} are mappings, which uniquely generate for a valid document a resulting document.
Such mappings are bijective since documents are injective and are mapped, preserving the structural information.
So, the invertible transformation is an isomorphism here.\\

It is not evident that employing the logical paradigm general-purpose Prolog still has restrictions in the design, that is why it (\cite{SterlingPractice}, p.146).
It has, for instance, the predicates \texttt{read/1, write/1} for input and output, which every general-purpose programming language should provide.
The predicates \texttt{findall/3} and \texttt{!/0} are due to an effective control with solution sets.
\texttt{assert}-commands can simulate variable assignments.

\section{Design}
\label{sect:Design}

\subsection{Prolog Data Structure}

\label{sect:modellingPrologDatastructure}

\subsubsection{Common}

In this section, the translation from XML into a Prolog representation is designed.
The sections on parser and serialiser formalisation attributed grammars are introduced for reading and writing files to the Prolog system.\\

Element, text, PI and comment nodes shall be processed.
Node constructors are to distinguish node types and are called functors in Prolog and are denoted by an atom.
The concrete \textit{kind} of a node constructor is specified with brackets containing a list of parameters.
Subnodes are defined recursively.
XML documents can be modelled as Prolog-terms.

\subsubsection{Element Nodes}
\label{sect:elementNodes}

An XML-element node is uniquely identified by its name, an attribute list and a sorted node list.
The name is modelled as an atom.
Both lists are modelled as lists.
A list is always ordered in Prolog.
So, the attribute list has always an ordering.
For instance, if an ordered list is needed when checking for equality, then canonisation is needed on the corresponding element node (see fig.\ref{fig:canonisationPascal}).

In \cite{PrologWiki}, in subsection ``\textit{Processing hierarchical structures}'', a list rather than a constructor is recommended.
Although lists and tuples do not differ much when it comes to an appropriate representation, the data structure has a finite number of elements.
Hence, tuples seem to be the right decision.

Lists associate with an ordered collection of a homogeneous element.
The homogeneity may look far much more extensible at first glance.
However, processing rules still need to be aware of position and data.
A violation of this convention would result in malfunctioning.
Apart from that, the requirement to have separate attributes from child nodes could be weakened, so arbitrary interleavings could be allowed.
That means attributes and child nodes could be merged into one list.
However, such weakening would invalidate a unique representation since there is no more correspondence than left between attributes and child nodes.
Another difficulty would be the undefined \textit{arity} and list length.
That would make processing complicate.

Seipel suggests in \cite{Seipel02}, \cite{Seipel05}, \cite{Heumesser} to use the triple \texttt{Tag-Name:AttributeList:Content} for element nodes.
\lit{:} stands for the list constructor.
\texttt{a:b:c} stands for list \texttt{[a|[b|[c|[]]]]} or \texttt{[a,b,c]}.
The third tuple, component \texttt{Content} denotes the children list.
This representation is complete and is taken, except for these three modifications:

\begin{enumerate}
 \item The triple mentioned is written in brackets and is prefixed with an \texttt{element}.

\item The list functor \lit{:} is substituted by a comma.

\item \texttt{Content} is renamed into \texttt{ChildrenNodeList}.
\end{enumerate}

So, for example, the element node \texttt{<a>hallo</a>} becomes \texttt{element(a,[],[text(hallo)])}.

\subsubsection{Attribute List}
\label{sect:attsList}
 
In XML, an attribute entry has the semantics mapping:

$$\textit{Attribute-Identifier} \mapsto \textit{Text}$$

Since an element node has an arbitrary number of attributes, a list is the data structure of choice.
An attribute entry has two possible representations.\\

\textbf{Variant 1: Tuple notation}

\begin{verbatim}
[(id,"value"),(id2,"value2"),...]
\end{verbatim}

This representation is minimal.
Brackets and quotes are required but could lead to heavily nested expressions.\\

\textbf{Variant 2: with equality sign}

\begin{verbatim}
['id="value"','id="value2"',...]
\end{verbatim}

This variant is closer to XML.
Equality and quote signs are separators.
Besides, the expression requires guarding single quotes, so the entry becomes an atom.
Anything different from an atom leads to a problem.
Transformation rules must split this atom in order to extract attribute identifier and associated content.

Due to less probability of errors, the second variant is preferred.
Even so, this means additional overhead.
The overhead of accessing attribute names is linear in complexity.
It is accepted in order to get better usability.
So, the element node \label{intext:example2atts}
\texttt{<a id="1" name="i">...</a>}\\
turns into the Prolog element node:\\
\texttt{element(a,['id="1"', 'name="i"'],[...])} .

\subsubsection{Text Nodes}
\label{sect:textNodes}

Text in between element nodes can be considered as a text node.
No additional characters are required.
Such an approach is easily implemented in Prolog.
However, this is in contrast to the just agreed convention, that its constructor distinguishes each node type.
In order to unify text, comment and PI nodes, the constructor \texttt{text} is introduced.
Text in between two element nodes, for instance, "\texttt{hello world}", is transformed into \texttt{text('hello world')}.

\subsubsection{Child Nodes List}

The sequence of child nodes that appear in an XML document is essential.
Changing the order of child nodes appear result in a fundamentally new XML document.
Child nodes are effectively implemented using lists.
Concrete child nodes can be different herewith.
For instance, when two element nodes are followed by two comment nodes and a text node:

\begin{verbatim}
 [ element(a,[],[]),
   element(b,['name="i"'],[]),
   comment('hallo'),
   comment('welt'),
   text('A') ]
\end{verbatim}

A missing typing can cause the following syntax errors:

\begin{enumerate}
 \item List is read instead of nodes.
 \item Child nodes list is read instead of non-element nodes.
\end{enumerate}

These errors can be excluded by meta-logical type predicates (see sect.\ref{sect:baseConcepts}).
Otherwise, the error may occur only during serialisation.

\subsubsection{Comment Nodes}
\label{sect:commentNodes}

Comment nodes are represented as following in XML:

\begin{verbatim}
 <!-- this is comment -->
\end{verbatim}

In Prolog the exact text is guarded by single quotes and is passed as an argument to a \texttt{comment}-constructor, such as \texttt{comment('this is comment')}.

It is worth noting that arbitrary text is guarded by single quotes, regardless of whether it is only a single word or text over several pages.
The first letter must be in lower-case.
It is in analogy to text and PI nodes.
In Prolog, it is not mandatory to encode \texttt{a} as \texttt{'a'}, since Prolog always treats atoms without single quotes to its knowledge base.
Hence, guarding should always be present.
Quotes within the text do not change.
Triple single quotes must escape single quotes.
As before, potential errors may only be recognised during serialisation.

\subsubsection{Processing Instruction Nodes}
\label{sect:piNodes}

A PI-node \texttt{$<?$ something $>$} in XML is represented in analogy to sect.\ref{sect:textNodes}, sect.\ref{sect:commentNodes} as \texttt{pi}-constructor: \texttt{pi('something')}.
PI-nodes are only passed but never processed.
That is why they are never considered separately in this work.

\subsubsection{Parser}
\label{sect:parserFormal}

A Backus-Naur form defines the syntax of an XML tree in sect.\ref{intext:bnfxmldefinition}.
After each XML node is mapped onto Prolog-nodes, the corresponding attributed grammar $G_{X2P}$ sums up the section (see fig.\ref{fig:parserAttributedGrammar}).
The set $N$ contains all non-terminal symbols, $T$ denotes the set of terminals, and $P$ contains productions.
\synt{XMLNode} denotes the starting non-terminal symbol, which is $X$ in the short form here.
The function \texttt{cat} concatenates all parameters.

\begin{figure}

$G_{X2P}$=($N,T,X,P$)\\

N=\{$X,X^2,Atts$\}\\
T=\{\lit{\textless},\lit{\textgreater},\lit{/},\lit{?},\lit{!--},\lit{--},Id,Text\}\\

P=

\begin{tabbing}
\linespread{0}
$X_{Text}$\arr \lit{<} $Id_{Text1}$ $Atts_{Text2}$ \lit{/>}\\
\footnotesize \qquad Text\barr  cat( \lit{element(}, Text1, \lit{,}, \lit{[}, Text2, \lit{]}, \lit{[])} )
\normalsize	
\linespread{1}
\end{tabbing}

\begin{tabbing}
\linespread{0}
$X_{Text}$\arr \parbox[t]{7cm}{\lit{<} $Id_{Text1}$ $Atts_{Text2}$ \lit{>} $X_{Text3}$ $X^2_{Text4}$ \lit{</} $Id_{Text5}$ \lit{>}  }\\
\footnotesize \qquad Text5\barr  Text1\\
\footnotesize \qquad Text\barr  \parbox[t]{7cm}{cat( \lit{element(}, Text1, \lit{,}, \lit{[}, Text2, \lit{]}, \lit{[}, Text3, \lit{,}, Text4, \lit{])} ) }
\normalsize	
\linespread{1}
\end{tabbing}

\begin{tabbing}
\linespread{0}
$X_{Text}$\arr \lit{<} \lit{?} $Text_{String1}$ \lit{>}\\
\footnotesize \qquad Text\barr  String1
\normalsize	
\linespread{1}
\end{tabbing}

\begin{tabbing}
\linespread{0}
$X_{Text}$\arr \lit{<} \lit{!--} $Text_{String1}$ \lit{--} \lit{>}\\
\footnotesize \qquad Text\barr  String1
\normalsize	
\linespread{1}
\end{tabbing}

\begin{tabbing}
\linespread{0}
$X_{Text}$\arr $Text_{String1}$\\
\footnotesize \qquad Text\barr  String1
\normalsize	
\linespread{1}
\end{tabbing}

\begin{tabbing}
\linespread{0}
$X^2_{Text}$\arr $\varepsilon_{Text1}$\\
\footnotesize \qquad Text1\barr  \empty\\
\footnotesize \qquad Text\barr  Text1
\normalsize	
\linespread{1}
\end{tabbing}

\begin{tabbing}
\linespread{0}
$X^2_{Text}$\arr $X_{Text1}$ $X^2_{Text2}$\\
\footnotesize \qquad Text\barr  cat( \lit{,} Text1 \lit{,} Text2 )
\normalsize	
\linespread{1}
\end{tabbing}

\begin{tabbing}
\linespread{0}
$Atts_{Text}$\arr $\varepsilon_{Text1}$\\
\footnotesize \qquad Text1\barr \\
\footnotesize \qquad Text\barr  Text1
\normalsize	
\linespread{1}
\end{tabbing}

\begin{tabbing}
\linespread{0}
$Atts_{Text}$\arr $Id_{Text1}$ \lit{=} \lit{$''$} $Text_{Text2}$ $Atts_{Text3}$ \lit{$''$}\\
\footnotesize \qquad Text4\barr \textit{if Text3 is empty:} '', \textit{otherwise:} ',' Text3\\
\footnotesize \qquad Text\barr  cat( \lit{'}, Text1, \lit{=}, Text2, \lit{'}, Text4)
\normalsize	
\linespread{1}
\end{tabbing}
\caption{Attributed Grammar $G_{X2P}$}
\label{fig:parserAttributedGrammar}
\end{figure}

\subsubsection{Serialiser}

Serialisation is the inverse operation of parsing.
So, a Prolog data structure is turned back into an XML document in analogy to sect.\ref{sect:parserFormal}.
The Backus-Naur form is in fig.\ref{fig:BNFPrologTerm}, and the attributed grammar is in fig.\ref{fig:bnfPrologNode}.

\begin{figure}
\label{fig:bnfprologdefinition}
\begin{grammar}
<PrologNode> ::= element( <Id> , [ <Atts> ] , [ <PrologNode> <Nodes> ] )
	\alt text( <Text> )
	\alt comment( <Text> )
	\alt pi( <Text> )
	\alt $\varepsilon$

<Atts> ::= $\varepsilon$
	\alt ' <Id> =  $''$ <Text> $''$ ' <Atts2>
	
<Atts2> ::= $\varepsilon$
	\alt ,  ' <Id> =  $''$ <Text> $''$ ' <Atts2>

<Nodes> ::= $\varepsilon$
	\alt , <PrologNode> <Nodes>
\end{grammar}
\caption{Backus-Naur-Form of a Prolog-term}
\label{fig:BNFPrologTerm}
\end{figure}

\begin{figure}
$G_{P2X}$=($N,T,PN,P$)\\

N=\{$PN$,$Atts$,$Atts2$,$Nodes$\}\\
T=\{  \lit{(}, \lit{)}, \lit{,}, \lit{[}, \lit{]}, \lit{'}, \lit{=},
\lit{"'}, \lit{element}, \lit{comment}, \lit{pi}, \lit{text} \}\\

P=

\begin{tabbing}
\linespread{0}
$PN_{Text}$\arr \parbox[t]{6.7cm}{\lit{element} \lit{(} $Id_{Text1}$ \lit{,} \lit{[} $Atts_{Text2}$ \lit{]} \lit{,} \lit{[]} \lit{)} }\\
\footnotesize \qquad Text\barr \lit{<} Text1\ Text2\ \lit{/>}
\normalsize	
\linespread{1}
\end{tabbing}

\begin{tabbing}
\linespread{0}
$PN_{Text}$\arr  \parbox[t]{6.7cm}{\lit{element} \lit{(} $Id_{Text1}$ \lit{,} \lit{[} $Atts_{Text2}$ \lit{]} \lit{,} \lit{[} $PN_{Text3}$ $Nodes_{Text4}$ \lit{]} \lit{)} }\\
\footnotesize \qquad Text\barr \lit{<} Text1\ Text2\ \lit{>} Text3 Text4 \lit{</} Text1 \lit{>}
\normalsize	
\linespread{1}
\end{tabbing}

\begin{tabbing}
\linespread{0}
$PN_{Text}$\arr \lit{text} \lit{(} $Text_{Text1}$ \lit{)}\\
\footnotesize \qquad Text\barr Text1
\normalsize	
\linespread{1}
\end{tabbing}

\begin{tabbing}
\linespread{0}
$PN_{Text}$\arr \lit{comment} \lit{(} $Text_{Text1}$ \lit{)}\\
\footnotesize \qquad Text\barr \lit{<!--} Text1 \lit{-->}
\normalsize	
\linespread{1}
\end{tabbing}

\begin{tabbing}
\linespread{0}
$PN_{Text}$\arr \lit{pi} \lit{(} $Text_{Text1}$ \lit{)}\\
\footnotesize \qquad Text\barr \lit{<?} Text1 \lit{/>}
\normalsize	
\linespread{1}
\end{tabbing}

\begin{tabbing}
\linespread{0}
$Atts_{Text}$\arr $\varepsilon_{Text1}$\\
\footnotesize \qquad Text1\barr \empty\\
\footnotesize \qquad Text\barr Text1
\normalsize	
\linespread{1}
\end{tabbing}

\begin{tabbing}
\linespread{0}
$Atts_{Text}$\arr   \parbox[t]{6.5cm}{\lit{'} $Id_{Text1}$ \lit{=} \lit{``} $Text_{Text2}$ \lit{``} \lit{'} $Atts2_{Text3}$ }\\
\footnotesize \qquad Text\barr Text1 = \lit{``} Text2 \lit{``} Text3
\normalsize	
\linespread{1}
\end{tabbing}

\begin{tabbing}
\linespread{0}
$Atts2_{Text}$\arr $\varepsilon_{Text1}$\\
\footnotesize \qquad Text1\barr \empty\\
\footnotesize \qquad Text\barr Text1
\normalsize	
\linespread{1}
\end{tabbing}

\begin{tabbing}
\linespread{0}
$Atts2_{Text}$\arr  \parbox[t]{6.3cm}{\lit{,} \lit{'} $Id_{Text1}$ \lit{=} \lit{``} $Text_{Text2}$ \lit{``} \lit{'} $Atts2_{Text3}$ } \\
\footnotesize \qquad Text\barr Text1 = \lit{``} Text2 \lit{``} Text3
\normalsize	
\linespread{1}
\end{tabbing}

\begin{tabbing}
\linespread{0}
$Nodes_{Text}$\arr $\varepsilon_{Text1}$\\
\footnotesize \qquad Text1\barr \empty\\
\footnotesize \qquad Text\barr Text1
\normalsize	
\linespread{1}
\end{tabbing}

\begin{tabbing}
\linespread{0}
$Nodes_{Text}$\arr $PN_{Text1}$ $Nodes_{Text2}$\\
\footnotesize \qquad Text\barr Text1 Text2
\normalsize	
\linespread{1}
\end{tabbing}
\caption{Attributed grammar $G_{P2X}$}
\label{fig:bnfPrologNode}
\end{figure}

\subsection{Transformation Rules}

\subsubsection{Discussions}

As already mentioned in sect.\ref{sect:rulebasedTransformApproach}, the rule-based transformation approach is widespread.
Rule application is pre-order and top-down.
A rule is selected from all matching rules, e.g. by priority, and called afterwards.
As soon as no other template is available, the resulting node-set is returned to the caller.

Several transformations process a sub-tree.
The result of a transformation step is a tree sequence.
The sub-transformations have specific properties in common (see sect.\ref{sect:modellingTransformations}).\\

Bruno, Le Maitre and Murisasco \cite{Bruno} consider non-monotone transformation operations in XQuery.
They extend the query language XQuery by the operations: insertion, moving, renaming.
These operations are needed for short transformation specifications to avoid specifying all non-modified elements of the incoming document.

Christensen et al. \cite{Christensen04} implement templates in their Java-based framework Xact.
Xact makes use of path expressions that are similar to XPath.
A path expression is modelled as composite.
It is very close to an XML document.
By doing so, path expressions, XML documents, and other framework classes can easily be handled.

Kiselyov and Krishnamurthi \cite{Kiselyov} try to resolve XSLT's restrictions on functionality and expressibility.
However, they stick to template-centric transformations.
It can be stated SXSLT matches the actual node with the left side of the rule during the iteration and evaluates afterwards.
In case the evaluation succeeds, this is identical to the \texttt{generate-and-test} approach (cf. fig.\ref{fig:generateNTest}).
First, a result is generated, then this result is passed to the calling instance and checked.\\
The consequences are:

\begin{enumerate}
 \item Templates must be traversed in pre-order by default in order to be comparable with XSLT.
 \item Invertibility assumes invertible sub-transformations.
 \item If a template matches and generates a result, then the template takes over control of the further transformation.
\end{enumerate}

\subsubsection{Templates}
\label{sect:templates}

In order to compare XSLT with Prolog, templates are needed in Prolog.
The following is agreed upon:

\begin{itemize}
 \item \textbf{Traversal order:}\\
The document is traversed without any further notice in pre-order, but the user can alter it.
Each node is matched against a template.
The specification of a node is unified with the left side of the transformation rule herewith.
In case unification succeeds, the right side of the transformation rule is returned.
Otherwise, the traversal proceeds with the following child nodes.

\item \textbf{Traversal continuation:}\\
The template must have a possibility to control the traversal of child nodes.
After the actual child nodes are visited, the next sibling element node is visited.
The application of the template to arbitrary nodes differs from the application to a list of nodes.
In the case of a nodes-list, results are successively put into a list of all results.
Each result is also a list of resulting trees.
If a rule does not match, then this rule is ignored, and an alternative is sought.
If no other rules matches, then an empty list is returned as a result.

\item \textbf{Calls:}\\
\texttt{xsl:call\--template} explicitly calls templates in XSLT.
\texttt{xsl:apply-templates} allows calling templates implicitly.
This call is appropriate whenever the structure of the child nodes is not clear.
There are no restrictions on recursion.
Callable templates are reachable, and non-recursive templates can be removed.

In Prolog, explicit calls are enforced by predicate calls as a subgoal.
The predicates \texttt{traverseElements} and \texttt{traverse} trigger implicit calls.
There is a \texttt{template} apart from the two just mentioned which controls the traversal of an XML tree.

\texttt{traverse(@in,-out)/2} traverses the input tree as just described and returns a result list.
The list is needed for output since, in general, there is no single target node.

\texttt{traverseElements/2} is in analogy to \texttt{traverse/2}.
The input is a nodes list, which is traversed successively.
Gained results are unified and then stored in the list.
\texttt{traverseElements} is needed when child nodes are explicitly processed.

\texttt{template(+node,-node)/2} is a template to be defined by the user.
\end{itemize}

\subsubsection{Examples}
\label{sect:examples}

Effectively, the user writes templates and helper predicates.\\

\begin{myexample}{1}{Simple Example}
Match an arbitrary node whose two child nodes are \textit{a}.

\begin{verbatim}
template(element(top,_,[A,A]),
         [text('a')]):-
  A=element(a,_,_).
\end{verbatim}

is equivalent to the XSLT-variant.

\begin{verbatim}
<xsl:template match="top[count(child::*)=2]
 and a[1] and a[2]">
  <xsl:text>a</xsl:text>
</xsl:template>
\end{verbatim}
\end{myexample}

\begin{myexample}{2}{Example using \texttt{append}}
Match arbitrary nodes whose attribute \texttt{id} equals 1234.

\begin{verbatim}
template(element(_,A,_),[text('.')]):-
  append(_,['id="1234"'|_],A).
\end{verbatim}

or by using the \texttt{transform}-predicate from sect.\ref{sect:modellingTransformations}.

\begin{verbatim}
template(E,[text('.')]):-
  transform(E@id,'1234').
\end{verbatim}

is equivalent to it XSLT-variant.

\begin{verbatim}
<xsl:template match="/[@id="1234"]">
  <xsl:text>.</xsl:text>
</xsl:template>
\end{verbatim}
\end{myexample}

\subsubsection{Open Questions}

May \cite{May} asks on page 21 for a more straightforward rule representation.
The representation suggests no \texttt{template}-predicate is used as introduced.
Instead, the mapping $\tilde x\mapsto \tilde y$ shall be implemented in Prolog as easy as possible.
The first example from sect.\ref{sect:examples} would look like the following in Prolog:

\switchProlog
\begin{verbatim}
 element(top,_,[A,A]):  [text('a')]
\end{verbatim}
\vspace{-0.3cm}
\begin{center}\small{$\tilde x \mapsto \tilde y$}\end{center}
\vspace{-0.5cm}
\begin{center}\small{x $\leftarrow$ y}\end{center}

Here, $x$ could be taken as head and $y$ as the Horn-clause body.
Semantically, this is inverse to the transformation mapping since $x$ is an implication of $y$.
$x$ would still match in a traversing algorithm and $y$ results.
However, $y$ may have at most one result.
Needed intermediate results cannot be contained in $y$ -- even if $y$ is a result tree list.
Prolog is relational, so technically results of a minor calculation could be arguments to a relation.

In order to swap $x$ with $y$, helper predicates may be used.
A traversal would be useless because the node to be matched is in the ruling body.
May's proposition would not be appropriate for the goal of this work.
This work shows how navigation expressions without conditions can be developed towards the transformation system.
A practical use would therefore not be considered due to high implementation efforts.\\

A constructor \texttt{template} is needed, and the term ``constructor'' entirely remains its original meaning.\\

Seipel \cite{Seipel05}, \cite{Heumesser}, Meijer and Shields \cite{Meijer} as well as \cite{PrologWiki} criticise the template approach in real applications.
In \cite{Heumesser} and \cite{Seipel05}, helper predicates are used as a replacement for functions. Templates are avoided.

Meijer and Shields choose a functional approach.
Queries are processed on the input document, and the calculated results are written to the output document.
This approach has high functionality, encapsulation and reuses in comparison with stylesheets.
Polymorphism and a type system \cite{Meijer} support the reached high level.
However, this approach still has the disadvantage to be too bloated the more the generated output document differs from the input document.
Therefore, it shall be clarified how the separation of transformation tags from Haskell functions in XM$\lambda$ can positively impact usability.

\cite{PrologWiki} suggests, similar to Meijer and Shields, only to a few base operations like (deletion, insertion, chaining) and pass the control and result management entirely over to the user.
This contribution can, due to its complexity, only be considered as motivation in this work. However, it shall be investigated further for certainty.\\

All examples mentioned in this section avoid templates.
It is implicitly assumed, control over the input documents stays with the user.
This work investigates whether or not template-based or template-free approaches are more accurate and appropriate for transformations.

\subsection{Transformation Operators}

\subsubsection{Relations}
\label{sect:modellingTransformations}

Before XSLT-equivalent operators for Prolog are introduced, it is essential to reason about operators for transformations.
In Prolog, the \textit{Relational Algebra} operations can be introduced (cf. \cite{Sterling}, p.42-44) as follows:

\begin{enumerate}
	\item \textbf{Relational Union} T=R$\cup$S:\\
$t(x_{1},...,x_{m}):-r(x_{1},...,x_{m}).$\\
$t(y_{1},...,y_{n}):-s(y_{1},...,y_{n}).$

	\item \textbf{Relational Difference}\footnote{\textit{not} is a logical \textit{not} here for interpretation} T=R/S:\\
$t(x_{1},...,x_{n}):-r(x_{1},...,x_{n}),not(s(x_{1},...,x_{n})).$

	\item \textbf{Cartesian Product} T=R$\times$S:\\
$t(x_{1},...,x_{m},y_{1},...,y_{n}):-r(x_{1},...,x_{m}),s(y_{1},...,y_{n}).$

	\item \textbf{Relational Projection} T=$\Pi_{S}(R)$:\\
$
t(s_{1},...,s_{n}):-r(x_{1},...,x_{m}). \qquad s_{i}\in \{x_{1},...,x_{m}\}
$
	\item \textbf{Relational Selection} T=$\sigma_{S}(R)$:\\
$t(x_{1},...,x_{n}):-r(x_{1},...,x_{n}),s(x_{1},...,x_{n}).$\\
where $s(x_{1},...,x_{n})$ makes a selection \footnote{\textit{s} is arbitrarily composed}.

	\item \textbf{Relational Renaming} T=$\rho_{S}(R)$:\\
$t(x_{1},...,x_{n}):-r(x_{1},...,x_{n}).$
\end{enumerate}

Here, $r$, $s$ and $t$ denote relations with arity $n$ or $m$.
$not$ is a syntactical convention here for the negation of a given relation.
It could be rewritten  as $no-s$ inverting the boolean return value of the relation interpretation $s$.

So, Prolog is at least as powerful as Relational Algebra.

Unfortunately, the operators from the Relational Algebra are not sufficient for document transformations in general.
For instance, operators are missing, which allow browsing and manipulating.
Hence those are going to be introduced next.

\subsubsection{Base Operations}
\label{sect:baseOperations}

Operators process terms.
They both uniquely appear together in the predicate \texttt{transform}.
It allows all transformations can be handled uniquely.
\texttt{transform} has arity two since a transformation defines itself as the state before and after a transformation.

A document transformation's fundamental duties can be derived from observation as navigation, construction and manipulation.
Navigation accesses the document's fragments, and construction means the insertion of new nodes and attributes into the target document.
Manipulation means the alteration or deletion of document fragments.
Although manipulation may be replaced by navigation and construction, those may still be crucial in exceptional cases.
For example, if a single element node needs to be removed, then in general, the need for a total reconstruction of all not-related bits is doubtful.
It would be an extraordinarily complicated way to describe a simple modification.

\textbf{Navigational Operators}
\label{intext:navigationalOperators}

In XSLT, XPath is used for navigation.
XPath uses location paths and axes \cite{XPathSpec}.

\begin{figure}
\begin{grammar}
<Element> ::= <LocationPath> \lit{/} \synt{Name}
	\alt <LocationPath> \lit{//} \synt{Name}

<Comment> ::= <LocationPath> \lit{\#} \synt{Integer}

<PI> ::= <LocationPath> \lit{?} \synt{Integer}
	
<AttributeValue> ::= <LocationPath> \lit{@} \synt{Name}

<AttributeName> ::= <LocationPath> \lit{id} \synt{Name}
\end{grammar}
\caption{Backus-Naur-Form for navigational operators}
\label{fig:bnfNavigationOperators}
\end{figure}

The location path in XPath uses the operators \lit{/}, \lit{//}, and \lit{@} (cf. fig.\ref{fig:bnfNavigationOperators}).
Two kinds of paths exist (i) relative paths and (ii) absolute paths.
In the case of (i), the path starts with a node name.
In the case of (ii), the path starts with either \lit{/} or \lit{//}.
Even if Prolog characterises any operator as relative, it is still important to note that the concrete context must uniquely identify a node.
Hence, nearly all path expressions in practice are indeed relative.

\lit{@} and \lit{id} is used as attribute accessors.
\lit{@} checks, if an element node contains an attribute or not.
\lit{id} is a core function of XPath \cite{XPathSpec}.
The difference to \lit{@} is that \lit{id} seeks for a matching attribute name.
\lit{id} is introduced as a navigational operator to Prolog.
Since the dual operator to \lit{@} \lit{id} is already a navigational operator, an additional distinction between base operators and helper functions is not meaningful.
In a location path in Prolog both \lit{id} and \lit{@} stand utmost-right.
Because it does not accumulate further nodes as to the result, the inductive generation for navigation paths stops.

\lit{\#} and \lit{?} access comment and pi-nodes.
They are related to the XPath-correspondences \texttt{comment()} and \texttt{processing-instruction()}.
However, the operators differ in their semantics.
In XPath, operators are introduced as functions to the node-test \lit{[]}.
They realise a filter.
In contrast to that, in Prolog, they perform a transformation and return a node as a result.
Since, in general, the content and the corresponding position are desired, the operators \lit{\#} and \lit{?} were introduced.
\lit{\#} and \lit{?} may be considered as projection in terms of Relational Algebra (see fig.\ref{fig:rulesNavigationOperators}).\\

In sect.\ref{sect:xpath}, XPath-axes were added.
The axis \texttt{namespace} \cite{XPathSpec} was dropped since no namespaces shall be considered in this work.
\texttt{attribute} was not newly defined.
This axis can be replaced by using the operator \lit{@}.
In order to implement the axis \texttt{parent}, the node above must be given.
However, this is not tractable unless either an additional data structure stores all relevant nodes (see sect.\ref{sect:contextEnv}).
A worsened usability is due to a direct consequence of bloated transformation rules and increased complexity.
It speaks against the introduction of axis \texttt{parent}.

The axes \texttt{ancestor} and \texttt{ancestor-or-self} can also not be introduced due to their direct dependency on \texttt{parent}.
The axes \texttt{following} and \texttt{preceding} access parent nodes according to their definitions.
That is why these, and the operators \texttt{following-sibling} and \texttt{preceding-sibling} can also not be introduced.
The same does not count for \texttt{child}, \texttt{descendant}, \texttt{self} and \texttt{descendant\--or-self}.\\

Except for joins, \lit{\#} and \lit{?}, all operators mentioned above are replaceable by one or more relational operations applications (see fig.\ref{fig:rulesNavigationOperators}).
\lit{//} is described in Prolog as \lit{\^} because binary operators are not permitted as operator symbol in tuProlog.
The operator \lit{/} seeks a given element node $E$ and a corresponding node name $N$ the corresponding child node.

In fig.\ref{fig:rulesNavigationOperators}, for each navigational operator, the corresponding rule is defined.
The operator is linked to the operands from the left to right with ascending index.
The rule determines the specification of the operands.
It is worth noting, the rule for \lit{//} is non-deterministic because the third and fourth rules have the same premise.
Consequently, a derivation tree may follow in general with an unknown amount of solutions.
If a \texttt{fail} is reached, Prolog proceeds with the last successful alternative.
Since the \texttt{descendant} bases on \lit{//}, it also is non-deterministic.
The axis \texttt{descendant-or-self} is not separately listed because it is composed of \texttt{descendant} and \texttt{self}.

\begin{figure}
\begin{tabular}{r  c}
$/_{E},_{N}$: & \parbox{5.5cm}{\myrule{E=element(_,_,[...,element(N,A,C),...])}{element(N,A,C)}}\\
$//_{E},_{N}$: & \parbox{3cm}{\myrule{E=element(N,_,_)}{E}}\\
$//_{E},_{N \not= X}$: & \parbox{3cm}{\myrule{E=element(X,_,[])}{\textit{fail}}}\\
$//_{E},_{N \not= X}$: & \parbox{3.3cm}{\myrule{E=element(X,_,[H$|$T])}{H$ // $N}}\\
$//_{E},_{N \not= X}$: & \parbox{3.3cm}{\myrule{E=element(X,_,[H$|$T])}{element(X,_,T)$ // $N}}\\
$@_{E},_{Att}$: & \parbox{2.8cm}{\myrule{E=element(_,[],_)}{\textit{fail}}}\\
 & \parbox{5.3cm}{\myrule{E=element(_,[...,'Att="'Val"'',...],_)}{'Val'}}\\
$id_{E},_{Val}$: & \parbox{3cm}{\myrule{E=element(_,[],_)}{\textit{fail}}}\\
 & \parbox{5.3cm}{\myrule{E=element(_,[...,'Att="'Val"'',...],_)}{'Att'}}\\
$\#_{E}$: & \parbox{3cm}{\myrule{E=comment("'X"')}{'X'}}\\
$?_{E}$: & \parbox{2cm}{\myrule{E=pi("'X"')}{'X'}}\\
$child_{E},_{C}$: & \parbox{4.3cm}{\myrule{E=element(_,_,[ ... , C, ... ])}{C}}\\
$descendant_{E}$: & \parbox{3cm}{\myrule{E=element(_,_,[]), }{\textit{fail}}}\\
 & \parbox{4cm}{\myrule{E=element(_,_,_),E2=E/_}{E2$ // $_}}\\
$self_{E}$: & \parbox{3cm}{\myrule{E=element(_,_,_)}{E}}
\end{tabular}
\caption{Navigation rules}
\label{fig:rulesNavigationOperators}
\end{figure}

\textbf{Constructors}

In XSLT \texttt{<xsl:element name="a"/>} constructs the element node \texttt{<a/>} (cf. sect.\ref{intext:example2atts}).
In analogy to that, constructors exist for attributes, comments and processing instructions.
The \texttt{element}-constructor may be bypassed, and the element node may be provided directly instead, for example, by \texttt{$<$a$/>$}.

In Prolog, the only possibility is to provide the element node.
As soon as a node type is specified, then the constructor is fully determined.
Other nodes are built up analogously.
Only attributes may be placed inside an element node as described in sect.\ref{sect:attsList}.

\textbf{Non-monotone Operators}

Monotony is a property of transformation operators.
An operator is non-monotone w.r.t. a term representing a given document if a small fragment is either deleted or altered.
Otherwise, the operator is monotone.
That is in analogy to building a sculpture: either by adding or by removal material.
For example, the operator \texttt{copy} is monotone since the document remains untouched.

\begin{center}
\myrules{$copy_{X}$:}{X=element(_,_,_)}{X}
\end{center}

In contrast to that, \texttt{copy_of} is non-monotone since the child nodes are cut off.

\begin{center}
\myrules{$copy\_of_{X}$:}{X=element(N,A,C)}{element(N,A,[])}.
\end{center}

Now operators for deletion are introduced informally, relying on Prolog's semantics on predicates:

\begin{center}
\myrules{$removeElement_{E},_{N}$:}{\\E=element(Name,A,C)\\\texttt{append(}Pre,[element(N,_,_)$|$Post],C\texttt{)}\\\texttt{append(}Pre,Post,C2\texttt{)}}{element(Name,A,C2)}
\end{center}

\begin{center}
\myrules{$remove_{E},_{N}$:}{E=element(Name,A,C)\\\texttt{append(}Pre,[N$|$Post],C\texttt{)}\\\texttt{append(}Pre,Post,C2\texttt{)}}{element(Name,A,C2)}
\end{center}

\begin{center}
\myrules{$removeAttribute_{E},_{Att}$:}{E=element(N,A,C)\\\texttt{append(}Pre,[' Att = " ' _ " ''$|$Post],A\texttt{)}\\\texttt{append(}Pre,Post,A2\texttt{)}}{element(N,A,A2)}
\end{center}

Operators for insertion and manipulation at a certain position are not introduced because those may be sufficiently specified by existing attribute and child nodes.

\textbf{Canonisation}

During transformation, it may become useful to check two documents or parts of it on equality.
Here single measures may differ in their orderings, but not in values, such as an invoice representation.
For this example, measures shall be placed into attributes.

Syntactically terms represent documents differently if only the attribute ordering differs and all values and contents are identical.
If attributes appear ordered by their identifiers, then attributes are ``canonised''.
Prolog does not provide canonisation apriori.
In order to apply canonisation, the helper \texttt{canon} shall be used.
Fig.\ref{fig:canonisationPascal} sketches the functionality of the predicate \texttt{canon} denoted as a Pascal-like function.
It performs a lexicographical sorting ascending by attribute identifier.
\texttt{curry} is as \texttt{curry2} is an array with a two-element record as the base type.
The first part \texttt{first} denotes the attribute identifier.
The second part \texttt{second} denotes the attribute value.
\texttt{curry2} is the result of sorting the array \texttt{curry} after the first record is edited.
Now the canonised equality of nodes can be defined (see fig.\ref{fig:canonizedEquality}).
The definition is done in Prolog because of a concise application of pattern-matching to the input data and a compact function definition.
The following fragment illustrates predicate \texttt{canon}:

\begin{center}
\begin{tabular}{l}
\begin{tabular}{rlcll}
  \texttt{L1} & $=$ & \texttt{[width="100", border="black"],} & (1)\\
  \texttt{L2} & $=$ & \texttt{[border="black", width="100"],} & (2)\\
  && \texttt{canon(L1,CL1),} & (3)\\
  && \texttt{canon(L2,CL2).} & (4)
\end{tabular}
\end{tabular}
\end{center}

The list \texttt{L1} is not canonised.
The results of the canonisation are \texttt{CL1} and \texttt{CL2}, which are directly unifiable.
\texttt{CL2} is also unifiable with \texttt{L2}.
\texttt{CL1} is not unifiable with \texttt{L1} because, in \texttt{CL1}, the first attribute identifier is not \texttt{width}.

\switchPascal
\begin{figure}[h]
\begin{verbatim}
function canon(L)
begin
  for i:=1 to length(L) do begin
    (* L[i]='id="value"' *)
    curry[i]:=(Identifier(L[i]),
               Value(L[i]));
  end;
  curry2:=sortByElement(curry,first);
  for i:=1 to length(curry2) do
    L2[i]:=''''+curry2[i].first+'="'
      +curry2[i].secound+'"';
  canon:=L2;
end;
\end{verbatim}
\caption{Canonisation in a Pascal-like listing}
\label{fig:canonisationPascal}
\end{figure}

\begin{figure}[ht]
\begin{verbatim}
% equals::Node->Node->Boolean
equals(element(N,A1,[]), 
       element(N,A2,[])):-
  canon(A1,CA1),
  canon(A2,CA2),
  CA1=CA2.  
equals(element(N,A1,[H|T]), 
       element(N,A2,[H|T2])):-
  equals2(T,T2),
  canon(A1,CA1),
  canon(A2,CA2),
  CA1=CA2.
equals(text(X), text(X)).
equals(comment(X), comment(X)).
equals(pi(X), pi(X)).

% equals2::[Node]->[Node]->Boolean
equals2([], []).
equals2([H|T], [H|T2]):-
 equals(T,T2).
\end{verbatim}
\caption{Equality of documents in Prolog}
\label{fig:canonizedEquality}
\end{figure}

\subsubsection{Extended Operations}

In the following functions of other query languages and aggregate functions are designed.

\textbf{Implementing Joins}
Joins are essential operations in terms of the so-called \textit{Relational Algebra}.
The natural join ($\Join$) over two relations $R$ and $S$, can be defined as
$$t(z,x_{1},...x_{n},y_{1},...,y_{n}):-r(x_{1},...x_{n},z),s(y_{1},...y_{n},z).$$
(cf. sect.\ref{sect:modellingTransformations}), a relational selection $\sigma_{S}(R)$.
A $\Join$ can be defined over projection, selection and Cartesian product.
Other joins can be derived from $\Join$.
Joins are of high value if data is distributed, e.g. over multiple documents or even different computers.

\textbf{Aggregate Functions}

The aggregate function \texttt{count} determines in XPath the cardinality of a node-set.
\texttt{position} (see fig.\ref{fig:helperOperators}) determines the position of an element node within a hedge.
\texttt{last} and \texttt{sort} are different relatively intuitive XPath functions.
The operator \texttt{sort} and the numerical attribute \texttt{level} originate from XSLT and are implemented in Prolog, as shown in fig.\ref{fig:aggregationsoperators}.
\texttt{level} calculates some sequence \textit{$i_{0},...,i_{n-1}$}.
Each integer $i_{j}$ denotes the relative position for a given node within a hedge.
Although \texttt{level} allows different formattings in XSLT (\cite{ZVON}, example 29), it remains to the user whether to use or not and if which one to use.
It is even possible to mix formatting, e.g. Arabic with roman numerals.
However, this is of less interest to the purpose of this work.

\begin{figure}
\begin{tabular}{r c}
$last_{E}$: & \parbox{3.5cm}{\myrule{E=element(_,_,[...,C])}{C}}\\
$count_{E}$: & \parbox{4cm}{\myrule{E=element(_,_,[$C_{1},...,C_{n}$])}{n}}\\
$name_{E}$: & \parbox{3.5cm}{\myrule{E=element(Name,_,_)}{Name}}\\
$level_{X_{0},X_{n}}$: & \parbox{6.3cm}{
	\begin{center}
	$X_{0}$=element(_,_,[ ... , $C_{0,i_{0}}$ , ... ]),\\ $C_{0,i_{0}}=X_{1}$\\
	$X_{1}$=element(_,_,[ ... , $C_{1,i_{1}}$ , ... ]),\\ $C_{1,i_{1}}=X_{2}$\\
	$\vdots$\\
	$X_{n-1}$=element(_,_,[ ... , $C_{n-1,i_{n-1}}$ , ... ])\\
	$X_{n}$ = $C_{n-1,i_{n-1}}$\\
	\underline{$X_{n}$ = element(_,_,_)}\\
	$[i_{0},i_{1},...,i_{n-1}]$
	\end{center}
	}\\
\parbox{1.3cm}{$sortby-\\Name_{E}$}: & \parbox{6cm}{
	\begin{center}
	$\forall k: E_{k}$=element($N_{k},A_{k},C_{k}$)\\
	\underline{E=element(N,A,[$E_{1}$, ... ,$E_{n}$])}\\
	$\forall j: N_{i_{j}} \epsilon [N_{1},...,N_{n-1}] \wedge N_{i_{j}} \le N_{i_{j+1}}$\\
	$\forall k: E_{k}$=element($N_{i_{k}},A_{i_{k}},C_{i_{k}}$)\\
	element(N,A,[$E_{1}$, ... , $E_{n}$])
	\end{center}
	}
\end{tabular}
\caption{Aggregation functions as operators}
\label{fig:aggregationsoperators}
\end{figure}

\subsubsection{Helper Operators}

\textbf{Processing strings}

The boolean functions \texttt{upper_first}, \texttt{lower_first}, \texttt{contains} and \texttt{starts_with} are implemented as predicates according to the W3C-specification \cite{W3XSLTSpecification}.
The following four queries succeed.

\begin{verbatim}
?-upper_first('Cook','cook').
?-lower_first('Cook','cool').
?-contains('hallo','ll').
?-starts_with('hello\ world','h').
\end{verbatim}

The string-functions \texttt{string}, \texttt{substring}, \texttt{substring_after}, \texttt{substring_before}, \texttt{translate} and \texttt{normalize_space} follow.
The Prolog-operator \texttt{string} turns a number, an atom or a list into a string.
In contrast to XSLT, \texttt{normalize_space} does not apply to a string's inner but to the sides only.

\begin{verbatim}
?-X is string(1.3).
YES. X/'1.3'
?-X is substring('hallo',1,3).
YES. X/hal
?-X is substring_after('hello\ world',
   'hello ').
YES. X/'\ world'
?-X is substring_before('Hello\ world',
    '\ ').
YES. X/'Hello'
?-X is translate('goose','egos','EGOS').
YES. X/'GOOSE'
\end{verbatim}

In Prolog, the XSLT-operator \texttt{concate} is implemented by \texttt{cat}.
\texttt{cat} concatenates up to eight strings passed as arguments.
More elements may be passed as a list.
It is also used in order to convert numbers into strings as well as lists into strings.
For example, \texttt{X is cat('hello',' ','world','!')} returns \texttt{YES. X/'hello world!'}.

\textbf{Arithmetic Operations}

Arithmetic operators in Prolog differ much from XSLT's arithmetic operators.
In order to check whether a given element is a number, in Prolog, there are three predicates.
\texttt{isnumber/1} tests in general for a number (which is closest to XSLT's \texttt{number}).
\texttt{fnumber/1} tests if a given element is a floating-point number.
\texttt{inumber/1} tests if a given element is an integer.
The first three calls return \texttt{YES}:

\begin{verbatim}
?-isnumber(0).
?-fnumber(12.34).
?-inumber(1001).
\end{verbatim}

XSLT's base arithmetic operations \lit{+}, \lit{-}, \lit{*} and \lit{/} are defined in Prolog over text nodes and are expressed using the operators \texttt{plus, minus, mult} and \texttt{div}.
The query \texttt{Z is plus(X,Y)} calculates for the following example \texttt{Z} the value 104:

\begin{verbatim}
X=element(a,[],[text('100')]),
Y=element(b,[],[text('4')]).
\end{verbatim}

\textbf{Useful Helper Predicates}
Some operators in other template and query languages lack for no good reason in XSLT.
The operators proposed in this section are only due to practical needs.
For example, sometimes it is needed to sorting to distinguish words by lower and upper cases.
All words starting with a lower case shall appear first or last.
The predicates, \texttt{first_upper} and \texttt{first_lower}, do precisely this.
Sometimes it appears helpful in uppercase words – the predicate \texttt{upcase} does this.
Remarkably, \texttt{upcase} can be applied only in one direction: the input is any word coming as the first argument, the output is the word in upper-case coming second.
The inversion would not make sense, because there may be up to $2^{n}$ meaningful inputs here, where $n$ is the input word length. The exponential complexity searches for alternatives very inefficient here.

\begin{verbatim}
?-first_upper('cook','Cyber-space').
NO
?-first_lower('cook','Cook').
YES
?-upcase('HELLO','hello').
YES
\end{verbatim}

\textbf{Helper Operators over Nodes}

Often a user wants to know if a node contains some attribute identifier.
The operator \texttt{atts} can do this (see fig.\ref{fig:helperOperators}).

Another problem is with duplicates in the target document or a transformation, for instance, by accidentally accumulating the target element multiple times.
In XSLT, however, the determination of duplicates is not in balance with implementation efforts.
In order to check a document that contains duplicates, the predicate \texttt{distinct} is defined.

\texttt{nth} determines the position of a node.
In order to provide the user with XSLT's notation, the operator \texttt{position} is introduced.

\begin{figure}[h]
\begin{center}
\myrules
	{$atts_{E}$:}
	{\\ E=element(_,['$Att_{1}$="' a   "' ', ... , '$Att_{n}$="' b "' '],_)}
	{[$Att_{1}$, ... , $Att_{n}$]}\\
\end{center}
\ \\

\begin{center}
\myrules{$distinct_{E}$:}
	{\\ E=element(_,_,[$X_{1},...,X_{m}$])}
	{element(_,_,[$Y_{1},...,Y_{n}$]) \quad
	$\forall i,j: Y_{i},Y_{j} \epsilon \{X_{1},...,X_{m}\}$}\\
where $Y_{i} \not= Y_{j}$.
\end{center}

\ \\

\begin{center}
\myrules{$position_{L,X}$:}
	{L=[$C_{1}, ... , C_{x}, ... , C_{n}$]}
	{$C_{x}$}
\caption{Helper operators}
\label{fig:helperOperators}
\end{center}
\end{figure}

Whenever \texttt{xsl:value-of} references an element node, XSLT processes the pre-order document and concatenates all text nodes consecutively into the target document.
This circumstance is advantageous, especially when debugging without \texttt{xsl:message}.
For the sake of completeness, this is implemented by the predicate \texttt{printTree}.
The call to listing from fig.\ref{fig:subgoals1}b results in \texttt{Z/'hello world'}.

An unexpected difficulty occurs when the traversal order shall change from pre-order in XSLT to anything different from that.

The predicate \texttt{nodes} can perform the powerset.
The query from fig.\ref{fig:subgoals1}c succeeds.

\begin{figure}[h]
\begin{verbatim}
(a) X=element(\_,\_,
       [element(a,\_,[text(hello)]),
        text('\ world')]).
(b  printTree(X,Z).
(c) nodes(X,[X,element(a,\_[text(hello)]),
          text('\ world')]).
\end{verbatim}
 \caption{Example subgoal calls on nodes}
 \label{fig:subgoals1}
\end{figure}

\subsection{Functions}
\label{sect:aboutFunctions}

In XSLT there is, unfortunately, no easy way to define and call generalised functions.
Named templates currently simulate functions.
Templates are primarily transformation rules.
The mapping of a template x $\mapsto$ y can naturally be interpreted as a function.
However, the current realisation

$$\texttt{<xsl:template match="x">y</xsl:template>}$$

is a syntactic drawback.
Templates are suitable for transformations, but they are not good at all for functions in general.
That is easily demonstrated by fig.\ref{fig:gcdXslt} and fig.\ref{fig:gcdProlog} for the greatest common divisor.

\begin{figure}[h]
\begin{verbatim}
gcd(A,0,A).
gcd(A,B,C):-AB is A mod B, gcd(B,AB,C).

?-gcd(24,30,C).
\end{verbatim}
\caption{gcd in Prolog}
\label{fig:gcdProlog}
\end{figure}

\begin{figure}
\begin{verbatim}
<xsl:template match="/">
  <xsl:call-template name="gcd">
    <xsl:with-param name="a">
      <xsl:value-of select="number(24)"/>
    </xsl:with-param>
    <xsl:with-param name="b">
      <xsl:value-of select="number(30)"/>
    </xsl:with-param>
  </xsl:call-template>
</xsl:template>

<xsl:template name="gcd">
  <xsl:param name="a"/>
  <xsl:param name="b"/>
  <xsl:choose>
    <xsl:when test="$b=0">
      <xsl:value-of select="$a"/>
    </xsl:when>
    <xsl:otherwise>
      <xsl:call-template name="gcd">
        <xsl:with-param name="a">
          <xsl:value-of select="$b"/>
        </xsl:with-param>
        <xsl:with-param name="b">
          <xsl:value-of
select="$a mod $b"/>
            </xsl:with-param>
      </xsl:call-template>
    </xsl:otherwise>
  </xsl:choose>
</xsl:template>
\end{verbatim}
\caption{gcd in XSLT}
\label{fig:gcdXslt}
\end{figure}

\subsection{Context Environment}
\label{sect:contextEnv}

The term representation is chosen in sect.\ref{sect:modellingPrologDatastructure} to reach every node from every other element node above.
However, there is no possibility without additional information to infer the parent node if such exists.

A context environment may be proposed (cf. \cite{Kiselyov}) to resolve this problem.
The main idea behind this is to provide some set $\Gamma$ that collects all elements required during a derivation step.
Additional information is information that can not directly be inferred from input data.
The more significant issue behind this is type inference for given nodes.
So, some type is assumed to decide whether a given L-expression, for example, is compatible with some expression or not.
For example, in fig.\ref{fig:pre-post-order}, the result of string concatenation is put to the context environment \texttt{bindings} used by the following more profound instance (\cite{Kiselyov}, pp.17).

In practice, this means each deduction step needs to remind its caller with current $\Gamma$. Hence a link between caller and callee may need to be updated in the most general case.\\

For example, in Xalan/C, this problem is avoided by doubly-linked lists.
In Java, object references may be used, which are mimicking pointers.

The OBSERVER design pattern can implement the relation between model objects and dependent observing objects.
Prolog does not know about pointer arithmetics nor objects.
For this reason, Prolog is not appropriate for an immediate context environment in the way described.
However, integration in Java allows the implementation of Java and export contexts via a Java library within tuProlog.

In $\Gamma$, the node above and further neighbouring element nodes may follow (cf. fig.\ref{fig:xpathAxes}) are inserted.
However, while reasoning, $\Gamma$ may grow and shrink, and it needs to be passed to transformation rules because Prolog does not know a priori of global variables.
The user would have to pass every time a transformation rule is invoked $\Gamma$ and all affected nodes.
It is a severe disadvantage.
Apart from that, numerous unsolved questions related to the estimated failure rate and a high increase in complexity.

Context environments are an exciting alternative to side-effects in order to reference nodes.
However, they are essential when practically dropping the \texttt{up}-operator as in this work.
Implementation in Java may compensate for the disadvantage again because it may make a global state accessible via predicates.

\subsection{Typing}
\label{sect:typing}

Errors related to wrong typing are hard to catch, especially with no proper tool support at all. -- This refers to experiences from developers.
Unfortunately, tuProlog is untyped a priori.
It results in either a correct output document, an incorrect document or a questionable \texttt{NO} without any real explanation.
Meta-logical type-predicates were introduced checking for membership to a category, like lists, atoms, to improve operators in Prolog.
Range checks enriched these checks.
The transformation gets more complex but safer by this.
On the other side, due to its interpretation, Prolog is not as fast a priori as a highly-optimised C-application dealing, for instance, with matrix multiplication.
However, performance shall, in general, be considered very carefully.
There is a good optimisation for code generation available running in non-interpreting mode, and by all means, the problems addressed are not runtime critical here at all yet.

It is highly recommended to introduce error messages at appropriate positions.
If terms are correct, the helper predicate \texttt{checkSerializable/1} for XML-trees and \texttt{checkAttributes/1} for attributes are introduced to check before serialisation.
If the input is syntactically corrupt (cf. fig.\ref{fig:bnfprologdefinition}), an error is issued to standard output according to the error position.

Although it is not part of this work, a dynamic validation method would be beneficial here because each transformation step data constraints would be checked thoroughly.\\

As already mentioned, tuProlog has no type system.
Hence, transformations are strictly bound to nodes and lists and can hardly be abstracted beyond those data structures.
Notably, this is not necessarily a disadvantage.
However, perhaps a further parametrisation and grouping into type classes could be helpful, s.t. transformation rules become flexible.
By doing so, a parametrised predicate could be defined, for instance, only applicable to a specific type of class.

These problems could be resolved by using the predicate \texttt{count} accepting more categories as input.
From the standpoint of most flexible code, this still is not satisfactory because each case would need to be hard-coded explicitly.
Instead, it should abstract away syntax as much as possible in order to achieve maximum flexibility.
So, currently, a type is missing, which would be accepted by parametrised predicates if \textit{reified}.
That means a unified type construct would determine the information about the type needed.

The following ``\textit{Gedankenexperiment}'' should clarify several vital aspects here:
We initially assume a type constructor \lit{type} exists with arity 1.
A type composition is guarding the new type as a functor, s.t. if the old type were \textit{number}, then the new type would be a floating-point number \lit{float(number)}.
This way, type checking may effectively be done using pattern-matching over type classes.
Type polymorphism is achieved by extending the type constructor \lit{type} by one argument.
Herewith, the first argument denotes the base type, and the second argument denotes the type class using the functor-composition described.

The main disadvantage of type classes in Prolog was missing context environments (see sect.\ref{sect:contextEnv}), causing bloated type information.
It is desired to hide type information from the user.

\section{Implementation}
\label{sect:Implementation}

\subsection{Overview}

\subsubsection{Common}

Prolog as transformation language is implemented by predicates and functors (see sect.\ref{sect:TransformLibrary}).
Both are made available through a library.
This library can be selected by the user and be loaded using the tuProlog IDE window (see fig.\ref{fig:tuPrologIDE}) by selecting \textsf{Open Library Manager}.
This way, all Java-based predicates can be invoked from within the tuProlog IDE as if they were elements of the Prolog core.

The user-defined transformation script containing zero or more templates is loaded into the text area of the tuProlog IDE (fig.\ref{fig:tuPrologIDE}).
When all facts and rules are loaded successfully, implying they are syntactic correct and available to Prolog as theory, they can then be called.
A new theory can be (re-)defined from the IDE \textsf{Set Theory}.
As soon as the theory is successfully loaded, the transformation can be triggered by a Prolog query.
Here the query from the bottom text field (see fig.\ref{fig:tuPrologIDE}) is confirmed, and the result is emitted to the bottom output window.
If multiple solutions exist, then only the first is displayed first.
More solutions can be selected by pressing the lower buttons next to the output window.
Multiple solutions could also be an indicator for invalid transformations where only one solution is expected.

\subsubsection{Architecture}

\begin{figure}
\begin{center}
\includegraphics[height=8cm,angle=90]{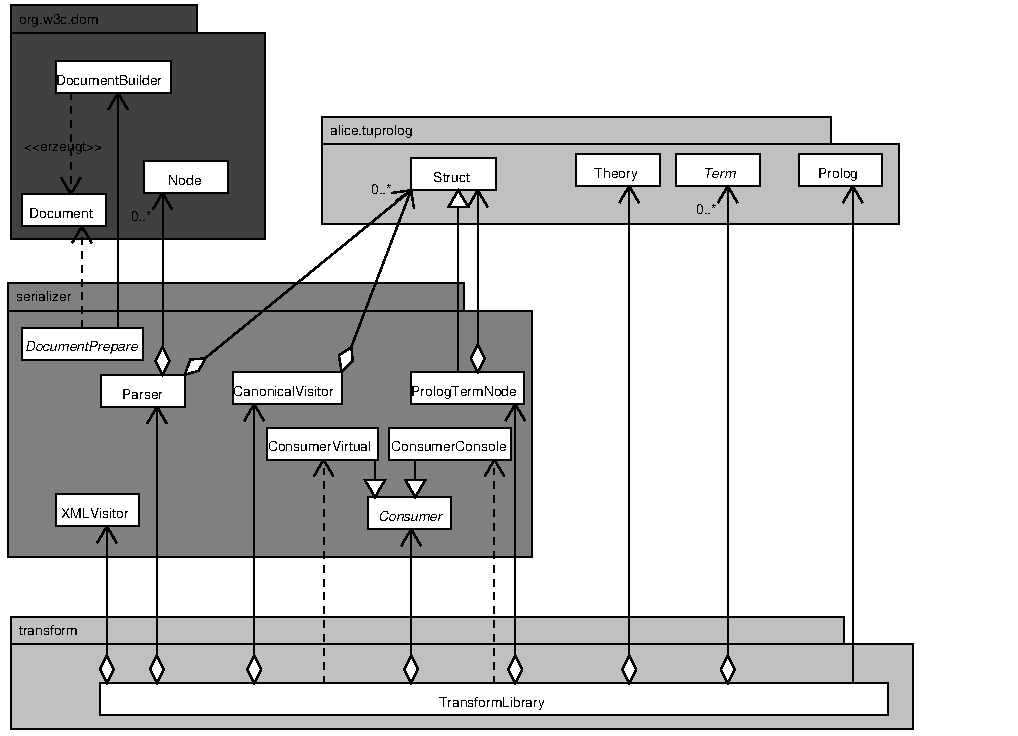}
\end{center}
\caption{Framework for Prolog XML-Transformations}
\label{fig:transformLibraryFramework}
\end{figure}

The Java-based part is shown in fig.\ref{fig:transformLibraryFramework}.
The layered architecture consists of the packages \texttt{transform}, \texttt{serialize},
\texttt{org.w3c.dom} and \texttt{alice.tuprolog}.
The packages with the highest abstraction are at the bottom.
Except for the white-box coupling in \texttt{PrologNodeTerm}, all packages are loosely coupled.
A strong inner cohesion would be present if a high number of classes and derivations were present. However, this is not the case here, which is good.

\begin{figure}
\begin{center}
\includegraphics[height=8cm,angle=90]{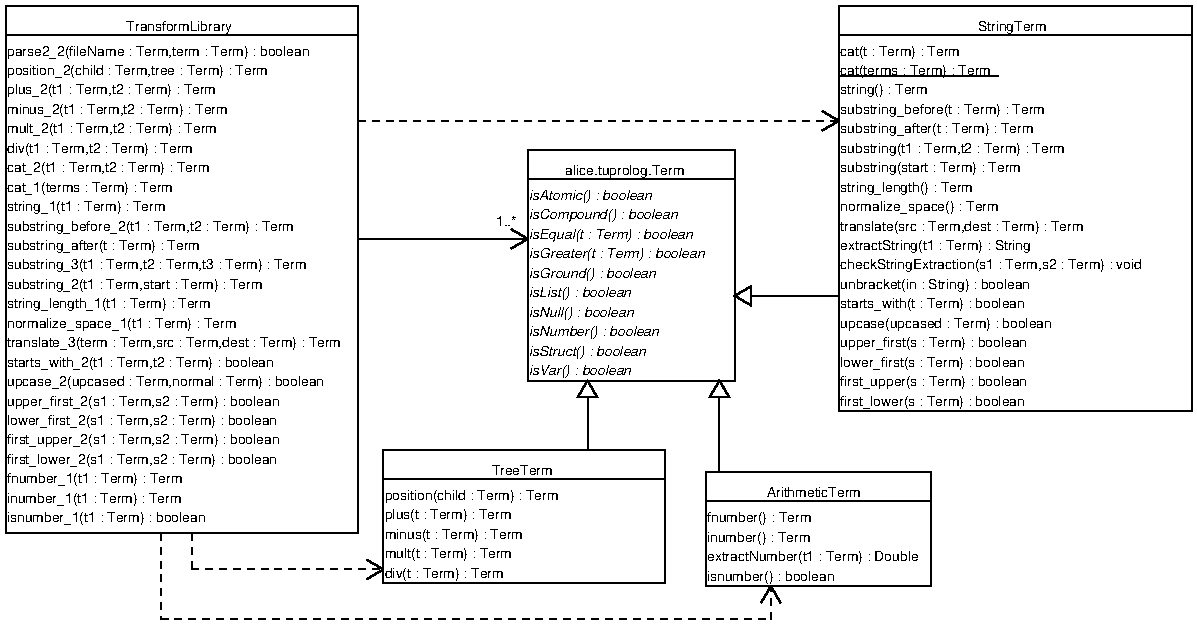}
\end{center}
\caption{Class \texttt{TransformLibrary}}
\label{fig:transformLibraryFacade}
\end{figure}

The package \texttt{transform} represents the top layer based on \texttt{serializer}
and \texttt{alice.tuprolog}.
The package \texttt{serializer} is located above \texttt{alice.tuprolog} and \texttt{org.w3c.dom}.
The layers are sorted in descending order.
This makes the framework simple w.r.t. complexity and extensibility.
\texttt{TransformLibrary} contains all transformation predicates discussed so far (see fig.\ref{fig:transformLibraryFacade}) and yet another bridging composition.

It delegates serialisation operations to \texttt{XMLVisitor}, read operations to \texttt{Parser} and canonisation operations to the concrete class \texttt{CanonicalVisitor}.
\texttt{Consumer} is for output.
The bridge to \texttt{PrologTermNode} is an interface towards visitable tuProlog-class instances.

The package \texttt{ser\-i\-al\-iz\-er} has six outgoing associations.
The abstract class \texttt{Do\-cu\-ment\-Pre\-pare} provides frequently used methods for \textit{Visitors}.
The class \texttt{Parser} parses incoming XML documents and generates corresponding \texttt{Struct} instances.
\texttt{CanonicalVisitor} sorts in a \texttt{Struct} attribute entries by attribute name.

The bridging compositions between class \texttt{Trans\-form\-Lib\-ra\-ry} in package \texttt{transform} and class \texttt{Term} in package \texttt{alice.tuprolog} and between \texttt{ser\-ial\-iz\-er.Parser} and class \texttt{org.\-w3c.\-dom.\-Node} is, in fact, extensions.
The user subclasses \texttt{Node} and \texttt{Term} without altering corresponding composited classes.
The arbitrary function may be implemented in any case.
The user defines which classes to instantiate.

Extensions to \texttt{Trans\-form\-Lib\-ra\-ry} can be achieved by adding further predicates or functors or by subclassing \texttt{Pro\-log\-Term\-Node} or \texttt{Con\-su\-mer}.
The error handling is simple.
When a Prolog query fails, tuProlog is notified, and at the end of the subgoal evaluation, \texttt{NO} is passed to the caller.
Unfortunately, syntactic errors of the transformation can only be noticed in Prolog by definitive tests or tracing.

\subsection{Integration into Java}

\subsubsection{TransformLibrary}
\label{sect:TransformLibrary}

The class \texttt{TransformLibrary} provides all predicates and functors in Prolog to transform XML documents (see fig.\ref{fig:transformLibraryFacade}).
Methods that return \texttt{boolean} are either predicates or methods.
Methods that return \texttt{Term} are functors.
Functors must be addressed by the operator \texttt{is}.
All Java operations to be exported must obey tuProlog's syntactic conventions (see fig.\ref{fig:declNewOperationsInJava}).
Methods are initially read using Java reflection and later assigned to concrete predicates and functors.

\begin{figure}[h]
\begin{grammar}
<OperationDecl> ::= <FunctorDecl>
	\alt <PredicateDecl>
	
<FunctorDecl> ::= boolean <Id> _ <Arity> ( <ParamDecls>  )

<PredicateDecl> ::= Term <Id> _ <Arity> ( <ParamDecls> )

<ParamDecls> ::= $\varepsilon$ 
	\alt <ParamDecl> <ParamDecl2>
	
<ParamDecls2> ::= $\varepsilon$ 
	\alt , <ParamDecl> <ParamDecls2>

<ParamDecl> ::= <Type> <Id>
\end{grammar}
\caption{Declaration for tuProlog-operators in Java}
\label{fig:declNewOperationsInJava}
\end{figure}

\texttt{Trans\-form\-Lib\-ra\-ry} represents a facade, which delegates queries further to 
\texttt{TreeTerm}, \texttt{ArithmeticTerm} and \texttt{StringTerm}.

The facade's associated classes share some methods of the class \texttt{TransformLibrary}, except a \texttt{Term}-instance is missing in each method's argument list.
Hence, the missing \texttt{Term}-instance is replaced by the associated class because it is a subclass of \texttt{Term}.
So, the concrete method becomes itself operation of the corresponding class.
The method \texttt{cat(terms: Term)} has no other relation to a term.
Hence it is implemented statically.

\subsubsection{PrologTermNode}

\begin{figure}
\begin{center}
\begin{tabular}{l}
\includegraphics[width=6cm]{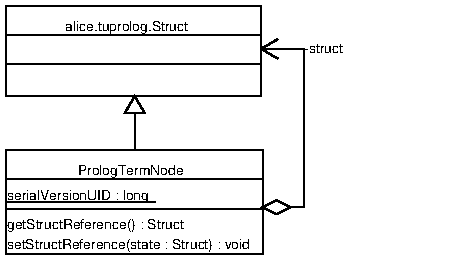}
\end{tabular}
\end{center}
\caption{PrologTermNode}
\label{fig:PrologTermNode}
\end{figure}

A given \texttt{Struct}-node is a \textit{Composite} recursively consisting of further instances of \texttt{Struct}, \texttt{Var} and \texttt{Number}.
The class \texttt{PrologTermNode} inherits all public and protected methods from \texttt{Struct} (see fig.\ref{fig:PrologTermNode}).
These methods of those two classes are visible to the user.
However, delegation allows only a partial insight.
\texttt{Struct} is immutable, and \texttt{PrologTermNode} is mutable.
So, it is a ``\textit{Template-Hook bridge}'' \cite{Pree}.

The traversal of \texttt{Struct} ought to be abstracted away as much as possible, s.t. each \texttt{Struct} accepts visitors, which implement own traversals (cf. sect.\ref{sect:SubsectionXMLVisitor}).
Hence, the shown \textit{Decorator} pattern is used from now on.
\texttt{Struct} acts as \textit{Component} and \texttt{PrologTermNode} acts as \textit{Decorator}.
\texttt{PrologTermNode} accesses a \texttt{Struct} instance by using the attribute \texttt{struct}.
The decoration is up to the user or \textit{Client}, who is instantiating the decorator (cf. \cite{Kerievsky}, p.169-192).

\subsubsection{XMLVisitor}
\label{sect:SubsectionXMLVisitor}

In order to traverse a \texttt{Struct}, a new method to be implemented would be most favourable.
As already mentioned, every Prolog representation of an XML document must be traversed at least in two different ways:

\begin{itemize}
 \item A Prolog tree is though entirely traversed on serialisation constructing a new \textit{DOM}.
 
 \item Attribute entries are sorted for canonisation.
 Afterwards, a new Prolog tree is returned.
\end{itemize}

Further traversals whose objective is a checking or outlining of certain properties are expected here.
It works in the opposite direction as on parsing.\\

\begin{figure}
\begin{center}
\begin{tabular}{l}
\includegraphics[width=7cm]{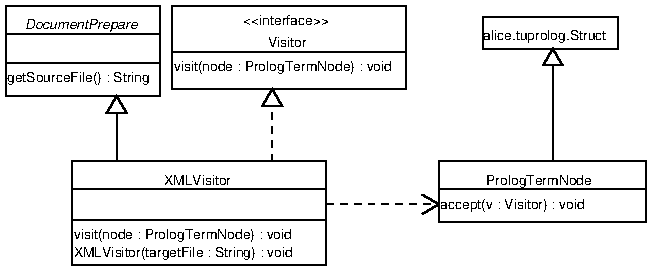}
\end{tabular}
\end{center}
\caption{Class \texttt{XMLVisitor}}
\label{fig:XMLVisitor}
\end{figure}

Hence, the traversal algorithm is separated from the data structure.
The \textit{Visitor}-pattern is applied (cf. \cite{Kerievsky}, p.345-363).
\texttt{alice.tuprolog.Struct} acts as \textit{Element}, 
\textit{ConcreteElement} is \texttt{PrologTermNode},
\textit{Visitor} implements the interface \texttt{Visitor}, and
\textit{ConcreteVisitor} is implemented by \texttt{XMLVisitor} (see fig.\ref{fig:XMLVisitor}).

Fig.\ref{fig:XMLVisitor} contains another pattern, namely the \textit{Class-Adapter} pattern.
\texttt{XMLVisitor} implements \textit{Adapter}, \texttt{Target} implements \textit{Visitor} and \texttt{DocumentPrepare} implements \textit{Adaptee}.
\texttt{DocumentPrepare} is for recurring DOM functions.

\subsubsection{Consumer}

\begin{figure}
\begin{center}
\begin{tabular}{l}
\includegraphics[width=8cm]{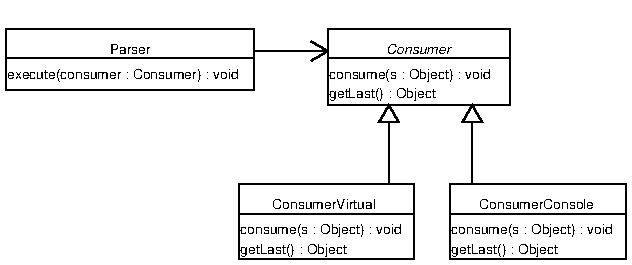}
\end{tabular}
\end{center}
\caption{Consumer}
\label{fig:Consumer}
\end{figure}

Fig.\ref{fig:Consumer} shows the abstract class \texttt{Consumer}, which defines methods \texttt{consume} and \texttt{getLast}.
The method "\texttt{consume}" issues messages to the system console or sends a message using a service, where \texttt{getLast} returns the last message consumed.
Depending on the application, the output may need to be buffered, e.g. in gauges.
The client holds one or more consumers, which are applied to one DOM instance of a document.

This model contains the \textit{Strategy}-pattern (cf. \cite{Kerievsky}, p.153-169).
\texttt{Parser} acts as \textit{Context}.
\texttt{Consumer} acts as \textit{Strategy} -- the concrete implementations \texttt{ConsumerVirtual} and \texttt{ConsumerConsole} act as \textit{ConcreteStrategy}.

\subsubsection{Parser}

The parser aims to recognise the language provided by the grammar from fig.\ref{fig:parserAttributedGrammar}.
First, the incoming XML document needs to be scanned for lexemes.
Scanning is done by the class \texttt{DocumentPrepare} by the method \texttt{prepareParser} by the call to \texttt{parse}.

\begin{verbatim}
factory = new DocumentBuilderFactoryImpl();
builder = factory.newDocumentBuilder();
this.sourceFile = sourceFile;
document = builder.parse(sourceFile);
\end{verbatim}

\textit{JAXP} also applies the \textit{AbstractFactory}-pattern.
An instance of \texttt{DocumentBuilderFactoryImpl} acts as a \textit{Concrete Factory}.
The generated factory contains instantiating methods, which implements the \textit{Builder}-pattern.
Some \textit{concrete builder} generates the scanned and parsed document given by an XML file name. 
The intermediate representation is a compound object hierarchy.

The \texttt{Parser}-class task is to translate a hierarchical object into a hierarchical Prolog term (cf. sect.\ref{sect:modellingPrologDatastructure}).
For the sake of simplicity, Prolog terms are directly translated into Prolog objects.
The target language objects of classes \texttt{Struct}, \texttt{Var}, \texttt{Int}, \texttt{Long} from the package \texttt{alice.tuprolog}.
A recursively descendant traversal is preferred for object traversal.
It is suggested to switch to a table-oriented parser in huge object hierarchies and performance issues \cite{Opeleva}.

\subsection{TuProlog Library}

All rules used within this work are listed in appendix \ref{appendix:PrologRules}.

Transformation requests are delegated to the tuProlog system.
Based on the ruleset and the incoming document, a true proposition is tried.
The staring predicate \texttt{parse2/2} is invertible.
\texttt{parse2(input,X)} binds the Prolog representation of an input XML document to variable symbol \texttt{X}.
\texttt{parse2(Output,Y)} serialises a given Prolog-term \texttt{Y} into the XML file named \texttt{Output}.

The comments provided show the typing as proposed in sect.\ref{sect:typing}.
Prolog-predicates always map onto a boolean value.
This means function \texttt{equal} has type \textit{Int$\rightarrow$Int$\rightarrow$Bool}, which can always be interpreted as predicate as long as its utmost-right type is a boolean.
The ordering of types corresponds to the ordering of declared arguments in the predicate head separated by commas.
Naturally, the super-type in Prolog is \texttt{Term}.\\

Parts of the system interpreting Prolog-transformations are written in Prolog.
That requires some explanation why at least.
First, all functors of the transformation system are written in Java.
Some predicates are implemented as \texttt{Library} in Java.

Others are in Prolog.
Prolog definitions are often significantly shorter than equivalent Java implementations (cf. fig.\ref{fig:PrologListingPosition}).
Let the XPath-operator \texttt{position/2} be given (see fig.\ref{fig:JavaListingPosition}), which has arity 3.
The predicate \texttt{nth/3} can be considered, even if parameters slightly divert in tuProlog from the ISO standard since a conform predicate was defined.
The Prolog-example is by far shorter, more straightforward to read and unadorned to understand.
Prolog allows fast prototyping \cite{SterlingPractice}.
So, these are simpler to use and less error-prone.
However, type checks as, for instance, in Java are initially not possible.
There is a possibility provided by tuProlog's library \texttt{JavaLibrary} to call external packages implemented in Java (see \cite{Denti01}, p.192f).
However, this approach is an additional indirection since Prolog-programs are realised as string that needs to be processed first.

\begin{itemize}
  \item Navigational operators can be composed, s.t. the rule definition is recursive (see fig.\ref{fig:recursiveNavigationOperator}

  \item The operator \texttt{sort} uses \texttt{quicksort} as subgoal and the predicate name \texttt{leAttributes} as an actual parameter.
They are used in tuProlog in order to simulate higher-order functions.

  \item In the last definition of the predicate \texttt{traverseElements}, the subgoal \texttt{compound(H)} checks $H$ is a compound list element.
For the sake of extensibility, only nodes are checked.
\end{itemize}

\begin{figure}[h]
\begin{verbatim}
position(Tree,Child,Pos):-
  Tree=element(_,_,C),
  nth(Pos,Child,C).
\end{verbatim}
\caption{Ternary predicate \texttt{position}}
\label{fig:PrologListingPosition}
\end{figure}

\begin{figure}
\begin{verbatim}
public Term position_2(Term child, 
               Term tree){
  Term result = null;
  Var C = new Var();
  if (tree.getTerm().unify(
       new Struct("element",
       new Var(),new Var(),C))){
    Iterator it = ((Struct)C.getTerm()).
                   listIterator();
    for (int i=0;it.hasNext();i++){
      if (((Struct)it.next()).
          match(child)){
        result = new Int(++i);
        break;
      }
    }
  }
  return result;
}
\end{verbatim}
\caption{Binary functor \texttt{position} in Java}
\label{fig:JavaListingPosition}
\end{figure}

\begin{figure}
\begin{verbatim}
transform(X / C,Y):-
  transform(X,X2),
  transform(X2 / C,Y).
\end{verbatim}
\caption{Chained navigator operator '\texttt{/}'}
\label{fig:recursiveNavigationOperator}
\end{figure}

\section{Comparison}
\label{sect:Comparison}

Before comparing languages, it needs to be checked whether the implementation behaves the same as XSLT.
So, the data model, parser, serialiser, traversing and operators need to be adequate.

The data model corresponds to the document structure.
The document is constructed successively (cf. sect.\ref{sect:elementNodes}-sect.\ref{sect:piNodes}).
Parser and serialiser base on the data model and implement the translation from a DOM structure to Prolog and vice versa.
The attributed grammars from fig.\ref{fig:parserAttributedGrammar} and fig.\ref{fig:bnfPrologNode} both implement both operations, whose correctness is assured by complete tests.
The traversal is defined intuitively by Prolog rules and is based on \cite{W3XSLTSpecification}, \cite{XPathSpec}.
Based on numerous tries and examples from test suites, such as \cite{ZVON}, a comparative examples base was assembled.
Functors and predicates simulate operators which denote path expressions and XSLT functions in Prolog.
After all, however, an entire cover is not possible (see sect.\ref{sect:baseOperations}).
The adaptions are due to missing features in Prolog and have been replaced by more pragmatic solutions instead.

Except for several cases on operators, the preconditions for comparison are fulfilled for a comparison.
The following questions shall be answered:
\begin{enumerate}
 \item Are all XSLT-stylesheet elements covered by Prolog?
 \item Which operators might be more expressible and easier to use?
 \item Can difference be quantified?
\end{enumerate}

\subsection{Criteria}

\begin{figure}
\begin{center}
\begin{tabular}{l}
\xymatrix{
 & & dynamic\\
 & objective \ar[ru] \ar[r] & static\\
qualitative \ar[ru] \ar[r] & subjective & \\
quantitative \ar[rd] \ar[r] & circumference \ar[r] \ar[rd] & measurable\\
 & \parbox{1.7cm}{$structural\\complexity$} & derivable
}
\end{tabular}
\end{center}
\caption{Comparison criteria}
\label{fig:comparisonCriteria}
\end{figure}
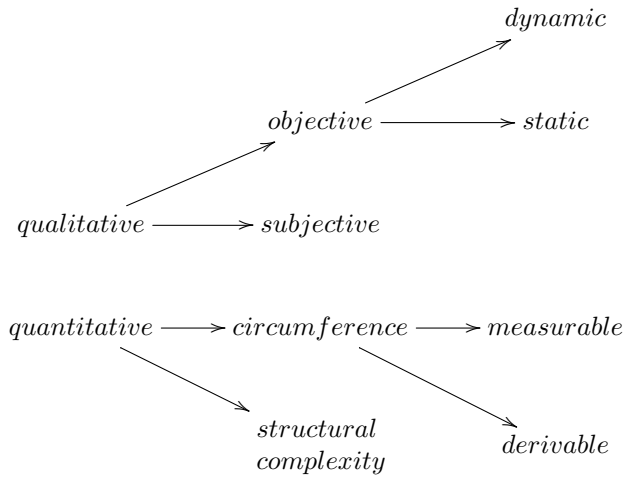

XSLT and Prolog are investigated on purpose under different criteria as XML-transformation language (see fig.\ref{fig:comparisonCriteria}).
The comparison is qualitative on the one side.
Analytical statements on program structure are being made. Those statements seem to be the most exact ones.
Qualitative statements can be distracted by dependencies between program parts.
Often single program parts are investigated only.
Combinatoric placements are often knowingly ignored.
However, these mistakes committed can be ignored in practice for the sake of overall qualitative judgements, which dominate.
On the other side, the program is measured.
So, by using software metrics, quantitative statements are ruled out.
However, these statements can heavily mislead if the considered amount of programs is too small or not representative.
This problem can be overcome by enriching more common test cases.

If the comparison is performed quantitatively, we mean the pure occurrence of certain symbols by the amount.
The measurement is then giving us a better understanding of the complexity of a given program.
Both terms allow us a deeper insight into the real nature of templates.
Other essential metrics on structures are often based upon statics, which becomes significant only after bigger case studies and generalisations and exact formulations.
Conventional quantitative measurements are, for example, the number of templates, redundancy and:
\begin{center}
\begin{tabular}{c c}
$\eta_{1}$ & total amount of operators\\
$N_{1}$ & cumulative sum of all operators counted\\
$\eta_{2}$ & total amount of operands\\
$N_{2}$ & cumulative sum of all operands counted
\end{tabular}
\end{center}

The ratio $N_{1} : N_{2}$ gives the level of functionality of an investigated language.
$N$ denotes the sum of all operators and operands, which is the empirically determined program length.
$N$ is more precise than the total amount of program lines \textit{LOC}.
Those metrics are better known as \textit{H\aa lstead-metrics} \cite{halstead77}.

The total program circumference $V=N ld(\eta)$ provides the number of bits needed to store the whole program minimally.
Here, mappings between operators and operands are assumed to be given.
The ratio $V:N$ is another criterion for redundancy apart from $LOC$ and file size.
The program length $N_{T}=\eta_{1}ld(\eta_{1})+\eta_{2}ld(\eta_{2})$ assumes ideally that each operator and operand is used exactly once only.
It would be interesting to know the deviation of experimental program length from the theoretic program length, which shall be done by $\Delta_{N}$.
The metric $\lambda=V^{*}L$ denotes the abstraction level of a language.
That shall always be critically questioned because statements are only made towards an ideal program $P^{*}$, which may be far from ideal in practice.
It is agreed upon that $P^{*}$ consists of two operators and two operands.
The estimated number of errors $B=V/300$ returns a good estimate in general.
It is further assumed the hypothetical programmer knows well XSLT and Prolog to the same amount and commits an error randomly after exactly 300 lines of code each.
In this model, each new error shall be independent of all previously committed errors.

Originally, H\aa lstead-metrics were defined on imperative programming languages \cite{halstead77}.
However, the metrics are obviously without any limitation applicable to Prolog and XSLT also, since both programming languages have operations and operands.
In XSLT, element node names can be encoded as operator.
Corresponding attribute values can be encoded as operands. 
Attribute names like \texttt{select} can be dropped since w.l.o.g. attribute entries can be assumed to be canonised.
In Prolog, it is agreed, predefined, user-specified operators and arguments are defined as operands.
Rule names are defined as arguments.

Both languages can be categorised.
An objective consideration allows lingual elements to judge in absolute terms.
It can be done statically or by running the program (dynamically).
A a given program listing is checked in a static comparison if selected data structures and operators behave the same as required and if a calculated term is valid.
Later both languages are investigated for variability and extensibility.
Variability denotes possibilities to integrate new functionality in existing lingual constructs and functions too.
Extensibility means design features allowing an arbitrary number of new functions to be integrated into existing.
Last, invertibility probes predicates to be ambiguous, s.t. a predicate may become overloaded with passed parameters becoming either input or output parameter depending on the factual data and the ordering (see in sect.\ref{intext:defInvertibility}).

In a dynamic comparison, the runtime behaviour of a program is investigated.
First, operations on reading and writing XML documents are checked.
This separation from the remaining operations seems reasonable since string processing functions are estimated slow.

A little consideration may reduce the significance of statements over lingual elements.
It is decided if and how much an element better fits than another.
Hence, the following usability criteria are considered.

\begin{center}
\begin{tabular}{l l}
1. & Readability/Plausibility\\
2. & Expressibility\\
3. & Lingual features\\
4. & Robustness/Stability\\
5. & Reusability
\end{tabular}
\end{center}

Their practical meaning sorts the criteria listed.
Readability reasons plausibility and hinders understanding.
Especially redundancy and a clear form are of importance here.
Expressibility investigates new features and ``\textit{syntactic sugar}'', both of which contribute to simpler expressions.
The hosting languages are probed.
The concepts concerned are unification, cuts, backtracking, pattern matching, inverse transformations, negation and typing.
Although many features already belong to the programming language Prolog, the impressions to be gained may be biased due to lacking comparability of selected programming features and non-corresponding concepts among compared languages.
In any case, the impression obtained may be biased.
Aspects of reliability and failure tolerance base on little experiences.
Reusability investigates if the language design is open enough, other libraries may easily be integrated and how good interoperability is.

\subsection{Commonalities}

Both languages, XSLT and Prolog, are descriptive and template-oriented.
In contrast to XSLT, Prolog can specify transformations without using templates.
Both transformation languages are Turing-computable (cf. \cite{Janssen}).
So, both are equally expressible and therefore are interchangeable.

An automatic transformation from Prolog to XSLT would not be unique due to helper predicates and cuts.
Given the gcd-example in fig.\ref{fig:gcdProlog} in Prolog, which shall be transformed into XSLT.
This transformation, however, can easily be performed without any real hinder in XSLT.

Nevertheless, the problem is about generalisations.
If, for instance, the given Prolog program contains cuts, the stylesheet would need to be transformed very specific to each template's intended algorithm, which may, in general, become extraordinary difficult in practice, and it depends on each problem to be resolved individually.
Non-terminating loops may not be entered in XSLT if only its corresponding fragment is cut off in Prolog (cf. sect.\ref{sect:baseConcepts}).

Computable also means non-terminating loops may be written in both Prolog and XSLT.
The gcd-example does not terminate if the base case is removed (cf. fig.\ref{fig:gcdXslt}).

On the one hand, XSLT is consequent.
On the other hand, XSLT heavily burdens practical usability.\\
On transformations, one or more input documents are scanned by template selection before being applied.
It is worth noting that increasing the number of documents processed within one sweep does not necessarily increase expressibility.
A transformation can be interpreted as a function having precisely one result.
Rule selection chooses one template rule to be applied next.
Priority avoids a non-deterministic rule selection.
In practice, templates with different signatures can be turned into other templates or helper predicates.

Both languages have in common they build up the target document bottom-up from smaller bits.
The rule application direction goes from left to right.
However, destructive operators may also have advantages in order to minimise the total amount of changes needed.
These non-monotonic operators are implemented by practical means in Prolog.
XSLT 1.0 lacks destructive operators.

\subsection{Language Features}

Namespaces are not considered, so ``\textit{extension namespaces}''
\cite{W3XSLTSpecification} are also not available.
Furthermore, \texttt{xsl:key}, \texttt{xsl:decimal-format}, and  \texttt{xsl:fallback} are dropped.
Output formatting using \texttt{xsl:output} is also not considered because an additional transformation step may express it.

\subsubsection{Path Expressions}

In XPath, the binary operators \lit{/}, \lit{//} access elements below the current element node (cf. sect.\ref{intext:navigationalOperators}).
The binary operator \lit{@} accesses attribute nodes.
XML is defined by recursive induction, and so are its path expressions.
The result of a path expression evaluation is either a node or a node-set.

In Prolog, the XPath-operator \lit{//} is substituted by \lit{\^}.
All accessory operators do not vary in arity nor result.
Since navigation operators are recursively defined (cf. appendix \ref{appendix:PrologRules}), corresponding predicates may also return several solutions.
Solutions may also require special treatment by the surrounding template and further helper predicates.
If operations are performed for multiple attributes, this may become more complicated after all in general than for ordinary element nodes.
For example, determining the total amount of attributes or attribute identifiers with a particular match is considerable.
As with element nodes, an application designer really would like to address attributes directly.
Therefore the \texttt{atts}-operator is introduced in Prolog, which is equivalent to XPath expression \texttt{./@*}.

\subsubsection{Node Tests}

Node tests map from a node onto a boolean value.
The boolean value denotes true for any node obeying a given predicate and denotes false otherwise.
The node-predicate is formulated as XPath-expression and may contain path expressions and membership functions.
XPath allows relations for equality and inequality (cf. sect.\ref{sect:expressions}).
This selection allows the XSLT-processor to skip selections from a supposed result set.

Prolog node tests are performed either immediately by providing an element node supposed to match with the given node or by tests possibly containing subtests.
So the node membership is implicitly determined by the functor.
Generally speaking, path expressions generate solutions, which then are restrained by node specifications and subtests (cf. fig.\ref{fig:generateNTest}).

In XSLT the membership to a certain node category is checked by the built-in functions \texttt{processing-instruction()}, \texttt{comment()}, \texttt{text()} and \texttt{element()}.
Testing in XSLT is rigid compared to Prolog because, for instance, only the current node w.r.t. the actual level in a document is allowed.
It implies node tests with deeper (grand-)children may only be formulated with much bigger effort --- if possible at all.
On the left of a transformation rule, a path expression is located denoting a node-set, which may be restricted further by node-predicates.
Once specified, path expressions may not be reused in further (sub-)paths within node-predicates.

\subsubsection{Axes}

Many XPath-axes could not be implemented due to a missing reference to the element node above in Prolog (cf. sect.\ref{sect:baseOperations}).

The axes \texttt{child} and \texttt{descendant} directly correspond to XPath-axes, except the concrete expression's syntax and semantics.
Since the transformation operators are designed simple, no background information is attached to any environment referring to the currently selected node (cf. sect.\ref{sect:contextEnv}).
That is why axis and used operators must both be passed to the predicate \texttt{transform}.

\subsubsection{Aggregates}

Aggregates (or aggregate functions) are functions over nodes and literals.
In XSLT, pre-defined XPath functions and templates for user-defined function implementation are available, which, however, are user-unfriendly (cf. sect.\ref{sect:aboutFunctions}).

For the reasons explained earlier, functors are provided explicitly in Prolog.
So, if in XSLT, an operator refers to the current node, then this node must explicitly be provided.
The aggregates \texttt{id()}, \texttt{last()}, \texttt{count()}, \texttt{name()} and \texttt{sum()} are implemented as operator and are available to the predicate \texttt{transform}.

The aggregate \texttt{level()} may be used for the \texttt{level} attribute inside an \texttt{xsl:number}-node.
It allows specifying one or more levels.

In contrast to XSLT, the Prolog implementation returns a positive integer.
The decision of which level to be used is up to the user.

Prolog has a profound number of basic built-in predicates and functors for lists and numbers, which improve the comfortable use of element and attribute nodes.
Besides, standard Prolog supports even more built-in predicates.
However, currently, tuProlog does not support all of them.
Examples of currently missing predicates include \texttt{nth}, \texttt{concat} and helper predicates \texttt{leAttributes}, \texttt{church},
\texttt{leStrings}, \texttt{checkSerializable}.
The predicate \texttt{nth} implements the list function with type $nth: x,y \mapsto z$ with some element node $x$ at position $y$ containing element $z$.
It is worth noting \texttt{nth} is fully invertible (cf. sect.\ref{sect:baseConcepts}).
The predicate \texttt{concat} concatenates all lists for a given set successively.
Please refer to sect.\ref{intext:XSLSort} and sect.\ref{sect:typing} for all remaining predicates.

Fig.\ref{fig:PrologExample1} demonstrates the use of aggregates.
The example provided is a complete and independent transformation program ready for execution, except input and output targets.
It returns the unary number for the number of occurring triples (red, green, blue) for a given child node-set.

\begin{figure}
\begin{verbatim}
template(element(_,_,C),[text(a)|As]):-
  append(_,[element(red,_,_),
            element(green,_,_),
            element(blue,_,_)|Post],C),
  traverse(element(_,_,Post),As).
\end{verbatim}
\caption{Prolog example for aggregation functions}
\label{fig:PrologExample1}
\end{figure}

\subsubsection{Expressions}
\label{sect:expressions}

XSLT has six kinds of expressions: \textit{node-expression}, \textit{template-name},
\textit{XPath-expression}, \textit{node-set-expression}, \textit{qname} and \textit{expression}
\cite{W3XSLTSpecification}.
In XSLT, \textit{node-expression} denotes a specific node, whereas \textit{node-set-expression} denotes in both languages the resulting node-set.
"\textit{node-expression}" is returned in Prolog either by a node specification or as a result of the \texttt{transform}-predicate.
An \textit{expression} is selected in matching attribute value (cf. sect.\ref{intext:XSLValueOf}, in sect.\ref{intext:XSLCopyOf}) in related XSL-tags.
In Prolog, this expression is covered by \texttt{transform}, aggregated sub-target as well as node specification.
\textit{qname} denotes a qualified name in XSLT.
It may be used for variables, parameters and templates (cf. sect.\ref{sect:varsAndParams}, sect.\ref{sect:aboutTemplates}).
Prolog symbols may be used everywhere, except in strings (which need to be concatenated separately).

Expressions in Prolog can do more than in XSLT, especially with lists and negated expressions.
Lists may be assigned to different categories.
Lists do not necessarily have to be node lists only.
The primary advantage is processing numbers instead of text nodes -- which in XSLT are the main default category.
The introduction of an explicit type system into Prolog would assist the user in development.
Negated expressions may be bound to certain operations only.
Negation may also affect predicates.
Atomic data types as text and numbers may now be used in the same way.
All expressions in this section need to be ground before serialisation.

\subsubsection{Constructors}

In XSLT, '$<$' and '$>$' denote the element constructor, containing attribute pairs separated by whitespaces.
Processing instructions are decorated with an additional question mark in order to distinguish them from element nodes.
Comment nodes contain an exclamation mark followed by two dashes.
Except for CDATA, all other symbols in between two element nodes is interpreted as text.

\textbf{xsl:element}

An element constructor may be \texttt{<a/>}.
Apart from the definition earlier, element constructors may also be explicitly defined using \texttt{xsl:element}.
This constructor assigns a name, attributes and child nodes to a specific element node.
Here, previously defined variables and parameters may be used (cf. sect.\ref{sect:varsAndParams}).
Attributes and child nodes that are not specified in further details are by default completed by empty lists.
Attributes and child nodes must be specified explicitly as such since ordering may change.
However, more flexibility does not bring any valuable addition since recurring long constructors do not benefit program readability.

Prolog provides only the essential bits here, namely attributes and child nodes.
Since its position within a children list is explicitly given, those do not need to be marked separately.
Symbols improve readability even further.
Instead of \texttt{xsl:value-of}, a symbol is replaced on runtime by its actual substitution (cf. sect.\ref{intext:XSLValueOf}).
This way, complicated documents may easily be composed.
As long as not all terms are ground, it may be used as a term-rewriting system or a document-generator.
However, the rewriting may only be successful if the term to be put instead is free of symbols already being replaced.
Finally, the derived term to be serialised shall be ground.

\textbf{xsl:attribute}\\
In both languages, attribute access is granted by the operator \lit{@}.
Herewith, for a given node and attribute name, the corresponding attribute value is sought.
If the correspondence can not be found, the path expression within is failing and so does the containing \texttt{transform}-clause.
From the user's perspective, attribute pairs do not have to obey an ordering.

In Prolog, attribute entries are implemented as lists (cf. sect.\ref{sect:attsList}).
That is why the predicate \texttt{canon} needs to be applied before a document comparison.
Attributes as lists allow a comfortable way to add, delete, reorder, alter and localise.
In combination with pattern-matching, neighbouring attributes, and in general, patterns over attributes, may be detected.
Due to a missing ordering, \texttt{append} may be applied to attribute lists without any prior sorting.

In both fig.\ref{fig:XSLTExample1} and fig.\ref{fig:PrologExample2}, the same element node is meant.
In the XSLT-example \texttt{xsl:element}, the element node can be built without \texttt{xsl:element}.
However, the attribute \texttt{person} shall be replaceable by a variable.
The example in XSLT contains 320 symbols.
The Prolog example has 144 symbols.
The Prolog example is almost entirely free of symbols not really needed for a correct representation.
If attributes are used at several locations, then the difference becomes even more significant.
To relax the limitation of XSLT version 1.0, so-called ``\textit{attribute-sets}'' were introduced, allowing binding attributes to variables similar to element nodes (cf. sect.\ref{intext:xslVar}).
However, the problem with \texttt{xsl:attribute-set} may be there already co-exists plenty of specific identifiers, and it is sometimes quite hard not to lose control over vast namespaces and identifier usages as it initially may look.
The evaluation happens simultaneously, as in Prolog when the binding is used (cf. \cite{W3XSLTSpecification}).

\begin{figure}
\begin{verbatim}
<xsl:template match="/">
  <xsl:element name="person">
    <xsl:attribute name="name">Rene
    </xsl:attribute
    <xsl:attribute name="profession">
     student</xsl:attribute>
    <xsl:element name="address">
      <xsl:attribute name="city">
        Dresden</xsl:attribute>
      <xsl:attribute name="country">
        Germany</xsl:attribute>
    </xsl:element>
  </xsl:element>
</xsl:template>
\end{verbatim}
\caption{XSLT example for element constructor}
\label{fig:XSLTExample1}
\end{figure}

\begin{figure}
\begin{verbatim}
template(_,element(person,A1,[A])):-
 A1=['name="Rene"','profession="student"'],
 A2=['city="Dresden"','country="Germany"'],
 A=element(address,A2,[]).
\end{verbatim}
\caption{Prolog example for element constructor}
\label{fig:PrologExample2}
\end{figure}

\textbf{xsl:text, xsl:processing-instruction, xsl:comment}\\
In addition to element constructors, constructors are also defined for text, pi and comment nodes.
All three new constructors are of a \textit{kind} \mbox{$\nabla \rightarrow \nabla$}.
The first argument denotes a string, which is represented by an atom in Prolog.
The second argument denotes the node to be constructed.

\begin{center}
\begin{tabular}{l l}
XSLT-tag	&	Prolog constructor\\
\hline
xsl:text	&	text\\
xsl:processing-instruction	&	pi\\
xsl:comment	& comment
\end{tabular}
\end{center}

\subsubsection{Variables and Parameters}
\label{sect:varsAndParams}

\textbf{xsl:variable}
\label{intext:xslVar}

A variable assignment is invalid as soon as a variable appears on both sides of an assignment.
It may be avoided in Prolog by a preceding ``occurs-check'' \cite{Sterling}.
This test, however, is deactivated by default in tuProlog for the sake of runtime performance.
XSLT would issue an error message at a non-declared variable.
In both XSLT and Prolog, a variable denotes element nodes.

Variables are declared in XSLT by
$$\texttt{<xsl:variable name=qname/>}$$

In XSLT, variables can be bound only once.
The following assignments are just ignored.
Also, partial assignment is not foreseen.
These could otherwise be used as subterms.
In XPath expressions, variables are addressed by name with a preceding dollar sign.

In Prolog, symbols are variables similar to those in XSLT.
An assignment or \textit{term unification} is possible more than once only if old and new terms match each other.
In case both terms do not unify, then the general clause fails.

\textbf{xsl:param, xsl:with-params}

As already mentioned, templates can be considered as a function or a named template with an arbitrary number of parameters.
In both languages, parameters are defined formally and used inside a template body.
Parameters are bound in the same way they are bound to a template call.

Function definitions are not appropriate in XSLT (cf. sect.\ref{sect:aboutFunctions}) since those are heavily restricted.
Despite the fact, templates are parametrised with \texttt{xsl:param}.
\texttt{xsl:call-template} calls templates with \texttt{xsl:with-param}.
Remarkably, parameters are only used in named templates.
XSLT also allows not all formal parameters are bound to actual ones.
So, unbound actual parameters are treated by the XSLT-processor as unused and are initialised with some default value which may differ \cite{W3XSLTSpecification}.
From the standpoint of other programming languages, this is not satisfactory because most popular imperative programming languages consider a different procedure signature as being some different function.
Another drawback could be the different way of handling parameters, variables, attribute identifiers and attribute sets.
Each of the placeholders mentioned is valid only in certain tags and is addressed sometimes with, but sometimes without dollar signs.
Attribute sets have an environment scope, which differs significantly from parameters and variables \cite{W3XSLTSpecification}.
It looks counter-intuitive at first glance.
In many other programming languages, like Prolog, there is just no such distinction between scopes.
The distinction should be given up for a simpler syntax and semantics.

If a symbol is only referenced in Prolog in a clause, this corresponds to the modular programming languages' actual parameters.
If a parameter is referenced as an incoming and outgoing symbol, then this corresponds to referential parameters.
Parameters are declared in a clause's head and represent a subterm.
These can be described further inside the rule's body.

It is recommended to distinguish clearly between template and function.
Only functions should have parameters.
Function calls should only be valid if the parameter list entirely matches because otherwise, overloading functions do not increase expressibility.

\textbf{xsl:copy, xsl:copy-of}
\label{intext:XSLCopyOf}

Copies may be created in two different ways.
First, if only the considered element node itself with all associated attributes of that particular element node is of interest, then \texttt{copy} shall be used.
Second, if an exact deep copy of an element node is needed, then \texttt{copy-of} in XSLT or \texttt{copy_of} in Prolog shall be used.

An example for attribute replacement:
 The colour of tables shall be changed from red to blue for a given document.
Other attributes shall not appear in the newer tables.
The XSLT example from fig.\ref{fig:XSLTExample2} selects by using \texttt{match} red tables and builds up a blue table in which all child nodes are copied.
The Prolog-example in fig.\ref{fig:PrologExample3} does the same but is more readable.
It is easy to see now that documents differ only in their attributes.

\begin{figure}
\begin{verbatim}
<xsl:template
   match="//table[@hcol="FF0000"]">
 <table hcol="0000FF">
   <xsl:for-each select="./child::*">
     <xsl:value-of select="."/>
   </xsl:for-each>
 </table>
</xsl:template>
\end{verbatim}
\caption{Attribute substitution in XSLT}
\label{fig:XSLTExample2}
\end{figure}

\begin{figure}
\begin{verbatim}
template(element(table,A,C),
         element(table,A2,C)):-
  append(X,['hcol="FF0000"'|Y],A),
  append(X,['hcol="0000FF"'],A2).
\end{verbatim}
\caption{Attribute substitution in Prolog}
\label{fig:PrologExample3}
\end{figure}

\subsubsection{Assignments}

Introduced assignments in this section are typical idioms in modern imperative programming languages.
However, XSLT, a functional programming language, shall avoid those idioms and provide instead perhaps more functional concepts (cf. \cite{Kiselyov}).
If side effects are not foreseen in XSLT in the first instance, then bonding towards languages with side effects would not make much sense.
So, users familiar with imperative programming languages would otherwise wonder why there are no genuine assignments available, for instance.
That is the reason why multiple XSLT assignments are not implemented yet in Prolog.

\textbf{xsl:choose, xsl:when, xsl:otherwise}
\label{intext:xslChoose}

\texttt{xsl:choose} (see fig.\ref{fig:XSLTExample3}) attempts to mimic \texttt{if-then-else} constructs as introduced initially to \cite{Wirth}.

\switchPascal
\begin{grammar}
<if-statement> ::= if <expression> then <statement>
	\alt if <expression> then <statement> else <statement>
\end{grammar}

A condition is thoroughly evaluated before a call in XSLT, as done in Pascal.
That is why an \texttt{if-then-else} may, in general, not be fully substituted by a \texttt{case} construct.

Further \texttt{xsl:when}-cases can be evaluated in \texttt{else}'s by adding further conditions.
If a branch condition is selected in XSLT, this implies that all other branches are ignored.
Prolog tests other \texttt{when}-branches, iff branches are given by clauses.
That is why in Prolog, a cut needs to be added explicitly (see fig.\ref{fig:PrologExample4}).

\begin{figure}
\begin{verbatim}
<xsl:choose>
  <xsl:when test="condition1">
    ...
  </xsl:when>
  <xsl:when test="condition2">
    ...
  </xsl:when>
  <xsl:otherwise>...</xsl:otherwise>
</xsl:choose>
\end{verbatim}
\caption{Constructor \texttt{choose} in XSLT}
\label{fig:XSLTExample3}
\end{figure}

\begin{figure}
\begin{verbatim}
choose(pattern1,I,O):-... , !.
choose(pattern2,I,O):-... , !.
choose(_,I,O):-... .
\end{verbatim}
\caption{Constructor \texttt{choose} in Prolog}
\label{fig:PrologExample4}
\end{figure}

\textbf{xsl:for-each, xsl:if}

The tag \texttt{for-each} is similar to a counting loop.
A counting loop in terms of XSLT is a loop with a bound number of iterations.
The number of iterations is fixed before entering the loop.
The step width may vary.
In contrast to XSLT, the iteration is often associated with a count-up variable in imperative programming languages.
Counting loops can be mimicked by primitive recursion.
However, the opposite does not work in general.
If the number of iterations is unknown before entering the loop, other loop variants should be considered instead.

\texttt{xsl:for-each} denotes a counting loop, where the counter may be requested by \texttt{position()}.
In nested counting loops, nodes must be bound to variables.
Conditions with no alternatives may often be rewritten, s.t. no \texttt{xsl:if} is needed anymore, for instance, by moving node tests into new templates or helper predicates.

\textbf{xsl:value-of}
\label{intext:XSLValueOf}

In XSLT, variables must be referenced only within XPath expressions.
Direct use in XSLT is not appropriate.
XSLT allows numbers, text and trees.
The XSLT-processor does not distinguish in \texttt{xsl:value-of} between text and trees.
On trees, text nodes are always issued concatenated.

Prolog distinguishes between categories.
Numbers may not be integrated into a document but only as a string embedded to a text node.
\texttt{printTree} processes element nodes the way described, where \texttt{printChildren} does the same for child node lists.
The introduction would cause Prologue programs to bloat compared to XSLT.
However, Prolog allows a direct and straightforward notation for operators.

\textbf{xsl:sort}
\label{intext:XSLSort}

Sorting relates primarily to attributes.
The ordering has little significance because results may be inverted.
The comparison criterion is in both languages specified by \texttt{upper-first} and \texttt{lower-first}.
In addition to that, Prolog provides an ASCII-wise sorting for both.

\texttt{sortbyName} does sorting by element node name.
For the sake of extensibility, \texttt{sort} is generic.
It is functional, which by default uses \texttt{leAttributes}.
If a new node type is introduced, then just a new \texttt{transform} rule must be defined, referring to the new comparison relation over the new node type.

In \texttt{xsl:sort}, this is not possible, so the sorting needs to be implemented newly as a named template.
According to the mechanism introduced by \cite{Novatchev}, a corresponding functional may be defined.\\

As an example, a sorted list of all names shall be created.
It is worth noting that sorting follows \texttt{xsl:sort}.
In fig.\ref{fig:XSLTExample4}, \texttt{xsl:sort} may not be moved in the \texttt{th}-node.
The Prolog example in fig.\ref{fig:prologExampleSort} does not need templates.
The helper predicate \texttt{getTRs} guards each element of the sorted list \texttt{[H|T]} twice, \texttt{X} contains the input document, and \texttt{Res} contains the output document.

\begin{figure}[h]
\begin{verbatim}
<xsl:template match="/">
  <table>
    <xsl:for-each select="//name">
      <xsl:sort order="ascending"
                select="."/>
        <tr>
          <th>
            <xsl:value-of select="."/>
          </th>
        </tr>
    </xsl:for-each>
  </table>
</xsl:template>
\end{verbatim}
\caption{Sorting in XSLT}
\label{fig:XSLTExample4}
\end{figure}

\begin{figure}
\begin{verbatim}
getTRs([],[]).
getTRs([H|T],[H2|T2]):-
  H2=element(tr,[],
       [element(th,[],[text(H)])]),
  getTRs(T,T2).
	
findall(Y,transform(X^name#1,Y),Ys),
quicksort(Ys,leStrings,S),
getTRs(S,TRs),
Res=element(table,[],TRs).
\end{verbatim}
\caption{Sorting in Prolog}
\label{fig:prologExampleSort}
\end{figure}

\textbf{xsl:message}

This feature corresponds to the Prolog predicate \texttt{write/1}.

\textbf{xsl:number}

In order to alter the representation of numbers, programming languages are aware of different formats.
\texttt{xsl:number} does this job in XSLT, and it may also provide a node with its position in a hedge, for instance, for a table of content.
\texttt{xsl:number} recursively determines for a given element node all predecessors, including the upper node.
In the "\lit{single}"-mode, the parent node is determined, where \lit{multiple} determines all predecessors.
This tag can also be used as a floating-point cast.
In Prolog, casts are performed by the built-in predicates \texttt{integer} and \texttt{float}.
Numbers are transformed into text by the operator \texttt{string} before adding it to the result set.
The \texttt{level}-operator determines in a tree all predecessors for a given sub-node.
The relative positions are added as integers to the result set.
The relative path for a parent node is the first element in a result set.
Different from XSLT, formatting is done by the user by need.

\textbf{distinct}

The removal of duplicates from a list is a frequent operation.
In SQL, the operator \texttt{distinct} was already introduced.
This operator seeks all elements having no duplicates and sorts the result set afterwards with represents of the duplicates set.

There is no \texttt{distinct} in XSLT 1.0.
However, this operator is missing.
Prolog knows a simple workaround here by using \texttt{append}, for instance.

\textbf{flatten, nodes}

\texttt{flatten} and \texttt{nodes} are used to obtain a linear ordering of document nodes for any given tree.
The predicate \texttt{flatten} traverses the document in preordering and appends a shallow copy of that node into the result set for each visited node.
\texttt{xsl:copy} is applied to each element node.
All other nodes are copied as are.
The predicate \texttt{nodes} copies every node precisely as found.

Although both operators should be available in XSLT, they can easily be implemented by the user.

\subsubsection{Templates}
\label{sect:aboutTemplates}

This section compares templates in both languages based on the rule-based transformation approach (see sect.\ref{sect:rulebasedTransformApproach}).

XSLT has a meta-node as its top document node.
This node allows specifying a fully independent template of any incoming concrete document at the beginning of every transformation.
This meta-node allows a unique definition of path expressions.
Otherwise, templates may also be expressed without meta-node.
Even so, they are still allowed to allow processing instructions and comments at the top level of a document.
W.l.o.g. this can always be true without the need to be at the top by moving those nodes below.
The data model in Prolog (see sect.\ref{sect:modellingPrologDatastructure}) does not allow a meta-node because template-predicates and functors by convention traverse nodes uniquely.

\textbf{xsl:template}

Stylesheets contain templates.
A \texttt{stylesheet} feature is not required in Prolog.
XSLT cannot drop the guarding tag because of the required XML well-formedness.
Multiple documents can only be managed heterogeneously in XSLT.
The first document is used implicitly in templates.

Further documents may be referred to by the function \texttt{document(sname)}.
The XML document \texttt{sname} denotes the stylesheet located absolutely or relatively to the current path.

Prolog does not require meta-nodes.
Documents can be located homogeneously because the predicate \texttt{parse2} allows arbitrary symbols as the second parameter.

In XSLT, templates can be executed in different modes.
It corresponds to different procedures with the same list of formal parameters.
Modes are introduced to achieve higher flexibility.
The concrete implementation is selected on a call by the switch \texttt{mode}.
It corresponds precisely with different procedure calls and brings no additional flexibility, which named templates cannot simulate.
The only exception is a shorter notation for the default template.
For example, the second template

\begin{verbatim}
<xsl:template name="a"
      mode="mode2" priority="4" />
<xsl:template name="a"
      mode="mode1" priority="5" />
\end{verbatim}

can be rewritten as
$$\texttt{<xsl:template name="amode1"/>}$$
The first line from the example would analogously be transformed into a template named \texttt{amode2}.
The attribute \texttt{priority} explicitly provides priorities.
In Prolog, new predicates would be introduced.
If arguments are passed as parameters, then we would get \texttt{amode1(A_{1},...,A_{n},Res)}.
The arguments from $A_{1}$ until $A_{n}$ do not have to be nodes, in contrast to XSLT.
An argument $A_{j}$ can be used as an ingoing and outgoing parameter.

Alternatively, the \texttt{template}-predicate can be extended by additional parameters.
This results in an ignore during traversal, so templates must be called explicitly.
For the sake of improved readability, this should be avoided.
Explicit priority rules are not needed in Prolog since appearance provides the priority.

\textbf{xsl:apply-templates}

An implicit traversal applies templates to deeper element nodes.
In Prolog, node traversal proceeds in order by \texttt{traverse} and \texttt{traverseElements} for node sets.
A traversal would be advantageous if it could be specified which node shall be selected next.
This behaviour reifies the \textit{Iterator}-design pattern since the document acts as \textit{Aggregate}, traversed by the \textit{Iterator}.
This iterator should provide some function \texttt{next} over a child node list.
The following listing shows two possible implementations of \texttt{next}:

\begin{verbatim}
next(L,A,N):-append(_,[A,N|_],L).
next(L,A,N):-append(_,[N,A|_],L).
X=element(v1,_,[element(v2,_,C),X2]),
X2=element(v5,_,[element(v6,_,_ )|_]),
append(_,[element(v3,_,_ ),
  element(v4,_,_ )|_],C).
\end{verbatim}

The first \texttt{next} implements the pre-order traversal, where the second does post-order.

For example, document \texttt{X} shall be checked, whether \texttt{v2} is indeed the immediate successor of \texttt{v1}, and the entire node \texttt{v3} shall be determined.
The corresponding query would be

\begin{verbatim}
?- X=element(_,_,[element(_,_,[V3|_]),
                  element(v5,_,_)]).
\end{verbatim}

Document \texttt{X} is unifiable and binds \texttt{element(v3,_,_)} to variable \texttt{V3}.
The Prolog query illustrates that implicit calls can lead to bloated programs (see fig.\ref{fig:XSLTExample6}).

\begin{figure}[h]
\begin{verbatim}
<xsl:template match="/">
  <xsl:if test="//v1//v2">
    <xsl:apply-templates select="//v3" />
  </xsl:if>
</xsl:template>

<xsl:template match="v3">
  <xsl:value-of select="." />
</xsl:template>
\end{verbatim}
\caption{Implicit template call}
\label{fig:XSLTExample6}
\end{figure}

\textbf{xsl:call-template}

The explicit traversal has procedural calls.
XSLT offers the possibility to call templates with or without parameters side-effect-free.
The result of the call is a node-set, which is substituted for the call.
The tag \texttt{xsl:call-template} is transformed in Prolog to

\begin{verbatim}
template(_,_):-
  ..., 
  qname(X1,...,Xn,Result).
\end{verbatim}

Herewith \textit{qname} denotes the template name. It is implemented as a predicate.
The template call is easier in Prolog because the predicate only is needed with corresponding parameters.
In contrast to XSLT, Prolog may have side effects.
That does not need to be a disadvantage.
However, on more complex transformations, this may hinder the user indeed due to its complexity.
Hence, symbols in predicates need to obey common \textit{conventions in Prolog}.
W.l.o.g. it should be agreed upon the first parameter(s) are \lit{+} and the last parameter only is \lit{-}.
It is worth noting multiple \lit{-} may always be summarised as one comprehensive list of functor with a flexible amount of arguments within.

The previous section raised the question of when would it be appropriate to substitute big and hard to handle templates with helper predicates.
These are the most important cases in a nutshell:

\begin{enumerate}
 \item A new aggregate function can easily be defined.
 \item The function considered has non-monotone characteristics.
 \item A template has a few branches only.
 \item A XPath expression would be too bloated.
\end{enumerate}

Whenever different operations are supposed to be applied to differing node sets in templates, those shall also be moved to helper predicates instead.
It improves the program's logical structure and increases readability and reuse --- especially when it comes to listing functions.
If a template contains many branches, then Prolog's template would have plenty of templates.
An elegant way to avoid this is to factorise branches.
Branches would be hoisted from the template into a new helper predicate.
Cancellation conditions of a \texttt{for-each} loop are difficult to formulate because the loop counter shall not be used explicitly.
Thus predicates become hard to read and error-prone.
In these cases, iterations over a node-set shall be replaced by helper predicates.

\subsubsection{Non-monotone Operations}

In Prolog, non-monotone operators were introduced for insertion and deletion.
The tag \texttt{xsl:copy} represents a non-monotone operator.
The operator \texttt{insertBefore} is defined according to fig.\ref{fig:XSLTExample7}.

A matching predicate looks like this:

\begin{verbatim}
template(element(N,_,C),
         [element(N,_,[text(a)|C])]).
\end{verbatim}

This example demonstrates how simple insertion in Prolog can be.
The same XSLT program becomes by far more complicated, especially if the element to be inserted is supposed to be placed second or third within the target document.
Due to the predicate \texttt{append}, this can be done carefree in Prolog.

\begin{figure}
\begin{verbatim}
<xsl:template match="/">
  <xsl:element name="name()">
    <xsl:text>a</xsl:text>
      <xsl:for-each select="/child::*">
        <xsl:value-of select="."/>
      </xsl:for-each>
  </xsl:element>
</xsl:template>
\end{verbatim}
\caption{Insertion of a text node at the first position}
\label{fig:XSLTExample7}
\end{figure}

\subsection{Non-biased Static Criteria}

\subsubsection{Variability}

Templates can achieve variability in both languages.
Templates contain user-defined elements specifying the source and target document.
There is, however, the constraint that source elements denote only nodes from the source documents, where target elements exclusively denote result sets.

XSLT excludes the variability of operators.
The user is only free in XSLT to define functions as named templates.
These differ from integrated XPath functions in practice, so this does not represent natural variability.\\

In contrast to that, Prolog allows user-defined transformations and even to (re-)define (built-in) operators.
By skipping meta-logical predicates, generalisation may be achieved since predicates may permit even more categories.
Input data or objects are no longer checked for membership to a particular category but are being processed as long as there is no conflict with term incompatibility.
For instance, the operator \texttt{sort} is parametrised by the predicate \texttt{leAttributes}.
By overloading \texttt{leAttributes}, the user can adapt orderings over specific data structures.

The Java classes mentioned next allow variability.
For instance, subclassing of \texttt{PrologTermNode} classes can be built up, which alter the tuProlog-class \texttt{Struct} by the \textit{Decorator}-pattern.
An implementation of the interface \texttt{Visitor} may refine visitor behaviour even more.
The classes \texttt{ConsumerVirtual} and \texttt{ConsumerConsole} implement the concrete behaviour of methods \texttt{consume} and \texttt{getLast}.

Further subclasses would have even further refinement in consequence, which is as expected --- the composition bridges from fig.\ref{fig:transformLibraryFramework} permit other behaviour to be defined by further subclassing.

In Prolog, a non-ground assignment to a variable symbol may be used as a  template for fragments of a document \cite{Wallace} (cf. sect.\ref{sect:baseConcepts}).
That allows instantiating values from a chosen domain into the template without specifying the target document's overall structure in a bottom-up direction.

\subsubsection{Extensibility}
\label{sect:extensibility}

Since XSLT and XPath both do not allow any new operators, they are often considered closed \cite{Kiselyov}.

In Prolog, new operators may be defined without limitation.
Transformation operators are uniquely defined without any convention by the \texttt{transform}-predicate.
The same goes for arbitrary user-defined functions.
All this is possible without the need of redesigning or compiling the transformation language.
Attention is needed when overloading existing functions.
Overloading whilst execution has in consequence new alternatives may appear which require special treatment.

\subsubsection{Invertibility}

Invertibility is defined in sect.\ref{intext:defInvertibility}.
It shall be researched in which cases input and output may be swapped without changing the transformation rules.

In general, an inversion requires in XSLT a complete rewriting of transformation rules.
Templates need to be newly defined, which often looks completely different then.
As soon as two text nodes are concatenated, the origin inputs are no more reconstructable.
Same counts for the origin inputs to the list operator \texttt{cat}.
Without additional information, it is not even possible to determine the original number of arguments.

However, identity is invertible.
The mapping $\{x \mapsto y, y \mapsto x\}$ is also invertible if \textit{x} does not occur free in \textit{y}.

In Prolog, template-free transformations can be specified, which are better for invertibility since numerous predicates are invertible by default.
The example in fig.\ref{fig:PrologExample7} shows predicate \texttt{nth}, which for some list \texttt{L} determines the \texttt{N}-th list element \texttt{E}.
Predicate \texttt{church} represents in the first argument Church's representation of some given natural number \texttt{X} which comes as the second argument.
\texttt{nth} is fully invertible since mappings exist, s.t.

$\{ N \mapsto (L,E), L \mapsto (N,E), $ \mbox{$E \mapsto (N,L), $}$ (N,L) \mapsto E,
(N,E) \mapsto L, (L,E) \mapsto N, (N,L,E) \mapsto boolean\}$.

The mapping $ \mapsto (N,L,E)$ is no valid mapping since \texttt{church} is defined only for ground terms in the first argument.
The last mapping $(N,L,E) \mapsto boolean$ can also be read as $(N,L,E) \mapsto boolean$.
The boolean value can be obtained by interpreting the function as Prolog relation.
If the transformation applies predicate \texttt{nth}, then the transformation is invertible.
If besides the result, for instance, some list $L$, at least one more argument is passed to the target document, e.g. $E$.
When transforming back due to the invertibility of \texttt{nth}, the missing bit is determined.
A representation as the predicate is preferred over a functor representation in Java because "\texttt{is}" is unidirectional and allows only one valid parameter binding, where invertible predicates may be used bidirectional (referring to the convention made earlier using \lit{+} and \lit{-}).

Besides, arithmetic function and cuts may also cause restrictions in Prolog to invertibility.
If a transformation determines a solution in a finite amount of time, and all required arguments are passed to the target document, and each deduction step is invertible, then the considered transformation is invertible too.
If this is the case, then, e.g. the \texttt{church}-predicate may be used for enforcing invertibility.
If the last matches with the invertible predicate's signature, then the predicate continues the desired parameter.
The general problem with cuts is that not the correct node may be chosen.
In that case, the start will not be reached by inverse deduction, so the reasoning fails.
Other inverse derivation will be searched.
If a cut appears whilst inverse deduction, it is very likely the correct derivation path was chosen.

\begin{figure}
\begin{center}
\includegraphics[width=6cm]{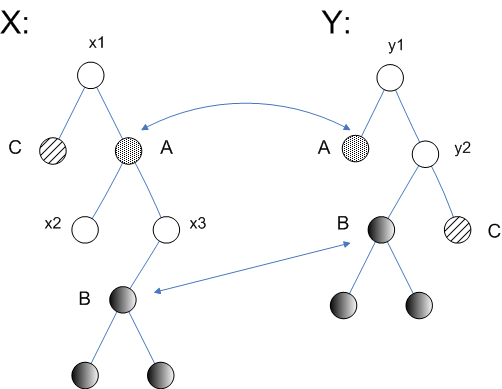}
\end{center}
\caption{Isomorphic transformation}
\label{fig:isomorphicTransformation}
\end{figure}

The problem is the ordering of neighbouring element nodes in the original document.
The original document structure may be reproduced, but not all element nodes would be known because while transformation and inverse transformation, only a few nodes would be considered (cf. \cite{Vion-Dury03}). 
In order to avoid this problem, each node of the original should appear in the node specification.
This way, gaps in the reconstructable documents may be excluded.
Fig.\ref{fig:isomorphicTransformation} shows an example where the original document is thoroughly reproduced by generalised unification.
The nodes $y_{1},y_{2}$ are inserted during transformation.
$x_{1},x_{2},x_{3}$ may be reproduced from both nodes as long as some \textit{x} correlates.
That assumes if it was specified in the transformation rule.

\begin{these}
A transformation is ``\textit{practically invertible}'', iff
\begin{enumerate}
 \item all used base functions are invertible, and
 \item each node from the input document is uniquely corresponding to a node in the output document, and
 \item each node from the output document is uniquely corresponding to a node in the input document
\end{enumerate}
\end{these}

Test cases can effectively check invertibility.
Points 2 and 3 can effectively be checked by renaming.
Transformations using the \lit{\^}-operator are in general not invertible.
Invertibility holds if \lit{\^} is congruent with the identity mapping.

\begin{figure}
\begin{verbatim}
nth0(s(zero),[X|_],X).
nth0(s(M),[_|L],X):-nth0(M,L,X).

nth(N,L,E):-
  var(N),
  nth0(N1,L,E),
  church(N1,N).
nth(N,L,E):-
  church(N1,N),
  nth0(N1,L,E).
\end{verbatim}
\caption{The predicate \texttt{nth}}
\label{fig:PrologExample7}
\end{figure}

\subsection{Biased Criteria}

The examples in this section are taken from the XSLT tutorial \cite{ZVON} and may be modified accordingly.
The examples deal with typical transformations, and the overall coverage of W3C's specification on XSLT is nearly representative \cite{W3XSLTSpecification}.

\subsubsection{Readability}

As still to be shown quantitatively, Prolog is easier to read than XSLT due to a much higher abstraction level.
One reason lies in the small gap between document language and transformation language.
The document language is XML -- a markup language.
The transformation language is either XSLT or Prolog.

The transformation language is a meta-language w.r.t. XML.
It means the abstraction level should be higher.
The language may even be descriptive.
XSLT in contrast is a XML-dialect.
XPath is embedded into XSLT.
XSLT has a gap in granularity, namely between abstraction level and syntax.
XPath is a fixed part of XSLT, but it is not a markup language.

The problem becomes obvious when a transformation is described.
The components of a transformation must be specified within tags.
Parts of a document may be put into a template without any explanation.
That implies nodes are returned as a result of a template application.
Besides, there are closing tags, which also increase the redundancy in a document.
The Prolog representation of transformation rules on the other side is closer to a more formal, mathematical notation.

Transformation rules are turned into helper predicates, s.t. premises consist of subgoals and allow one deduction step at a time.

Functions are an essential part of transformations.
In Prolog, arbitrary functions can be interpreted as predicates with $n$ inputs and one output.
Binary operators particularly achieve higher readability.
Functions can be defined shorter in Prolog than in XSLT.
Hence, they are closer to a mathematical notation.

The simple relative expressions in Prolog allow concise notations at a time.
For example, the concatenation of two lists:

\begin{verbatim}
concat([],L,L).
concat([H|T],Y,[H|T2]):-concat(T,Y,T2).
\end{verbatim}

There are numerous other examples in favour of Prolog.
More detailed statistics on transformations can be found later.

\subsubsection{Expressibility}

The insertion and deletion of attribute entries in XSLT are bloated because the surrounding nodes' specification seems awkward.
In Prolog, one can use either "\textit{do not care}" or "\textit{do not know}"-symbols to specify fragments of the documents or to specify that there are arbitrary nodes around, which makes specification flexible in comparison to XSLT.\\

In Prolog, the operators \texttt{atts} and \texttt{distinct} are introduced, which are immensely complex to introduce in XSLT.
The operator \texttt{atts} seems promising in cases when statements about the number of attributes filtered shall be made. 

\textit{Syntactic sugar} stands for shortened axes that exist in both languages.
The shortened operator for \texttt{child::*} is \lit{/}, and \lit{@} is the shortened operator for \texttt{attribute::*}.
Pattern-matching leads to a shortened and formally exact case distinction (see in sect.\ref{intext:xslChoose}).
So, nodes may be specified more comfortable without navigation operators nor additional statements.
Neighbouring relationships and node instances can so be specified.
Another remark of second-order Prolog-predicates is the accumulation of the result set of generative subgoals.

Moreover, solutions are recursively accumulated.
The predicates \texttt{findall} and \texttt{setof} search for (distinct) solutions.\\

Despite its power, expressions are limited.
So, Prolog variables can not be assigned different values, even after several successful unification.
Hence no decimal counters may be implemented.
Hence, from a functional perspective, Prolog is restricted towards functions for a reason just explained.
The representation ``\textit{Functions as predicates}'' is getting hard to read, especially when functions get composed.
Because of the opposite representation ``\textit{Predicates as functions}'', predicates may clash with names from other predicates.

The functional paradigm is violated by Prolog, especially when it comes to the output mode of variables.
Functions that alter incoming parameters violate referential transparency because a different instance may alter callee variables.

\subsubsection{Language Features}

In this section, for each Prolog feature a partial characterisation is given, which differs significantly from XSLT.

\textbf{Unification}

The input document \texttt{X} is unified with

\begin{verbatim}
element(top,[],[
  element(a,['id="1"'],[]),
  element(b,['name="rene"'],[text(hello)])
])
\end{verbatim}

The XSLT-program is in fig.\ref{fig:unificationInXSLT}, and the corresponding Prolog-program is in fig.\ref{fig:unificationInProlog}.

Correct would be the result

\begin{verbatim}
 Res/element(c,['id="1"'],[text(hello)])
\end{verbatim}

The disadvantage of XSLT over Prolog is that XSLT makes additional assumptions.
First, it is assumed \texttt{id} is an attribute within an element node (line 4).
In Prolog, it could also be an attribute with the name \texttt{id1}, s.t. unification does not fail.
Second, element node \texttt{b} may contain one text node.
If no assumption is true, then the text of other nodes will be issued.
The alternative would be counting all child nodes whom all need to be text nodes simultaneously.

Moreover, XSLT has the disadvantage to be relatively complex since it requires navigation operators instead of intermediate results.
From \texttt{Res} (line 4) and the rules from lines 1-3, the original structure can be reproduced either fully or partially.
Copying attributes (line 2) is more complex in XSLT: First, the element node must be copied.
Second, all attributes over \lit{@} need to be copied manually.
The problem occurs again that attribute identifiers may be unknown.
In Prolog determined intermediate results may be reused multiple times.
In contrast to XSLT, variables do not have to be separately defined when used.

\begin{figure}
\begin{verbatim}
<xsl:template match="/">
  <xsl:element name="c">
    <xsl:attribute name="id">
      <xsl:value-of select="//a/@id" />
    </xsl:attribute>
    <xsl:value-of select="//b" />
  </xsl:element>
</xsl:template>
\end{verbatim}
\caption{Unification in XSLT}
\label{fig:unificationInXSLT}
\end{figure}

\begin{figure}
\begin{verbatim}
X=element(_,_,[A,B|_]),
A=element(_,Att,_),
B=element(,_,_,[T|_]),
Res=element(c,Att,[T]).
\end{verbatim}
\caption{Unification in Prolog}
\label{fig:unificationInProlog}
\end{figure}

\textbf{Backtracking}

There is no backtracking in XSLT.
Multiple solutions are determined and managed by the callee in a result list.
If a transformation generates an unforeseen list of elements, the corresponding XPath expression is often not correct.
Often it is too coarse.
Further conditions restrict the solution set, so incorrect or irrelevant solutions are excluded.

Comparing generated documents may quickly lead the user to the incorrect location.
If a mistake is deep within a call-stack of Prolog rules, then in the worst case debugging may still be very complicated, and intermediate documents and solutions may be overwritten on serialisation.
In order to exclude multiple solutions as early as possible cuts, shall be used.

Hence, the potential source of errors shall be well commented on and traced with appropriate debug output (cf. \cite{Wallace}, \cite{Vion-Dury02}).

\textbf{Joins}

In distributed applications, often document fragments from different sources need to be placed together.
It is assumed, documents on different computers may be joined naturally.
Joins are defined over relations.
Applying joins on XML documents (also on terms) is not always wanted since subordinate attribute entries, amount, and types may be hard to check.
Rows in a table denote facts.
Facts can be read by predicates using variables.
In terms, rows are accessed by an index.
A subtree is interpreted as a row of that relation to map a term expression onto a relation.
So, attributes correspond to tree attributes.
Possible child nodes are ignored.
Moreover, a tree is interpreted as a relation.
The element node name turns into the relation's name.
Attributes are either ignored or can be used as relation description in distributed applications.
The child nodes correspond to the lines of relation.
The only condition is that each child node has the same attribute names and child nodes are not recursive.

In tuProlog, this can effectively be provided by the class \texttt{DCGLibrary}.
The example from fig.\ref{fig:twoXMLdocsInProlog} shows two XML documents as Prolog terms.
DCGs turn \texttt{X} into the relation \texttt{x} (see fig.\ref{fig:relationsAsPrologFacts}, \texttt{Y} accordingly), and the natural join over \texttt{x} and \texttt{y} can now be defined as

\begin{verbatim}
natural_join(Id,Name,FirstName):-
  x(Id,Name),
  y(Id,FirstName).
\end{verbatim}

Further joins can be defined analogously.
Outer and further joins can be defined in XSLT using named templates.
Tuples and relations are not part of the data models of XSLT (cf. \cite{W3XSLTSpecification}).

\begin{figure}[h]
\begin{verbatim}
X=element(x,_,[
  element(_,['id="123"',
    'name="hallo"'],[]),
  element(_,['id="4"','name="welt"'],[]),
  element(_,['id="789"','name="!"'],[])]),
Y=element(y,_,[
  element(_,['id="789"',
    'name="hello"'],[]),
  element(_,['id="5"','name="world"'],[]),
  element(_,['id="123"','name="?"'],[])]),
\end{verbatim}
\caption{Two XML documents in Prolog}
\label{fig:twoXMLdocsInProlog}
\end{figure}

\begin{figure}
\begin{verbatim}
x(123,hallo).
x(4,welt).
x(789,'!').
y(789,hello).
y(5,world).
y(123,'?').
\end{verbatim}
\caption{Relations \texttt{x}, \texttt{y} as Prolog-facts}
\label{fig:relationsAsPrologFacts}
\end{figure}

\subsubsection{Stability}

G\"{o}del's ``\textit{Entscheidungsproblem}'' implies that a program analysing another program will terminate or not is undecidable in general.
In practice, a helper predicate may not terminate for reasons, e.g. including a missing base case or wrong ordering of rules.
All considered documents are finite and are traversed top-down.
Mutual recursion of ascending and descending navigation operators are excluded in our considerations on documents, except general recursion but with general bounds.
Hidden cycles in predicates may be guarded by \texttt{write/1}.
If non-termination is still the case, it may any time be stopped using ``\textit{stop}'' from tuProlog's IDE.
XSLT automatically cancels any recursion after a certain depth is reached for stability reasons.

\subsubsection{Reuse}

Java-classes and Prolog-predicates allow a variation of existing functions (see in sect.\ref{intext:xslVar}) and an extension of new functions (see sect.\ref{sect:extensibility}).

The implementation of the transformation language in Prolog is multi-paradigmal (cf. \cite{Denti05}).
The Java-class \texttt{TransformLibrary} offers all methods and Prolog-predicates for transformation.
The Prolog-part bases on tuProlog libraries, which are entirely written in Java.
Hence, the language synthesised is platform-independent.
Prolog-functors implemented in Java allow rich operations initially not available to GNU Prolog, such as database and network access.
In addition to that, tuProlog-rules offer possibilities to implement \texttt{JavaLibrary}-interfaces in Prolog.

\subsection{Metrology}

All evaluations and measures discussed in this section are placed in the appendix.\\

Mainly the examples from \cite{ZVON} were used.
Since all examples cover nearly entirely the whole XSLT-specification \cite{W3XSLTSpecification}, they count as a great inspirational source for simple and complex transformation.
All examples were taken, except for the examples 11, 16, 41, 57-62 and 66-71.
Because those were either modified slightly, replaced by redundant other examples, or removed because of lack of direct relation to the specification.
Each XSLT example was implemented in Prolog with and without templates.
In some examples, even a third alternative was provided where appropriate with documentation.
The generic predicate \texttt{go} was introduced to provide each explicit example with a unique starting point.

The \textit{Prolog Measurement Tool} \cite{PMT} was used to simplify the metrology slightly to determine metrics for Prolog-programs. 
All other data was only manually obtained. Data obtained from the PMT was all thoroughly be checked manually.
Hand-written XSLT scripts partially analysed XSLT examples but most often also manually counted.\\

\textbf{Measurable Metrics:}\\

\begin{figure}
\begin{center}
\begin{tabular}{l}
 \includegraphics[width=8cm]{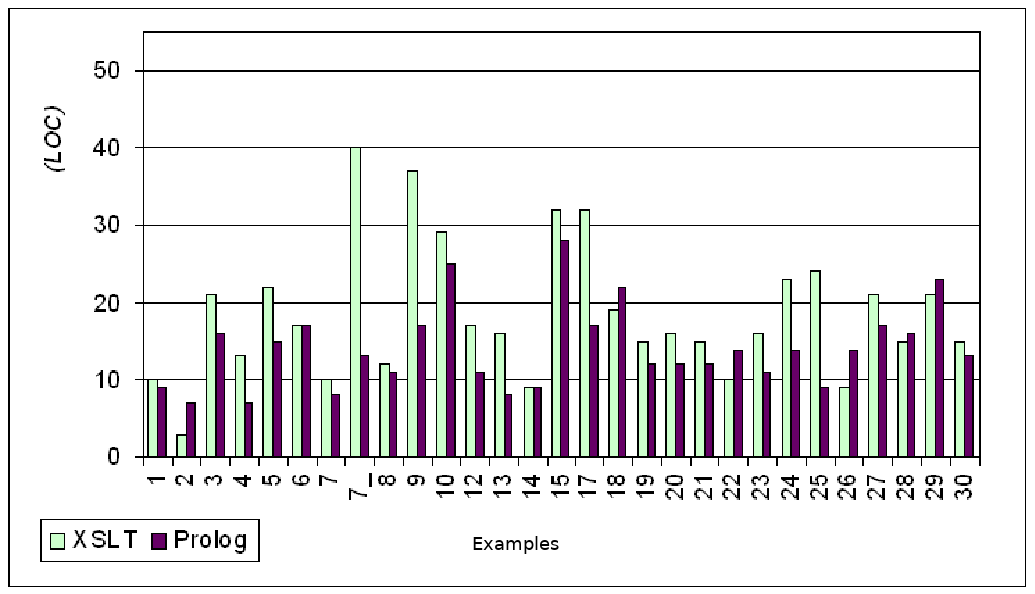}
\end{tabular}
\end{center}
\caption{LOC for examples 1-7,7\_2,8-10,12-30}
\label{fig:metricsLOC1}
\end{figure}

\begin{figure}
\begin{center}
\begin{tabular}{l}
 \includegraphics[width=8cm]{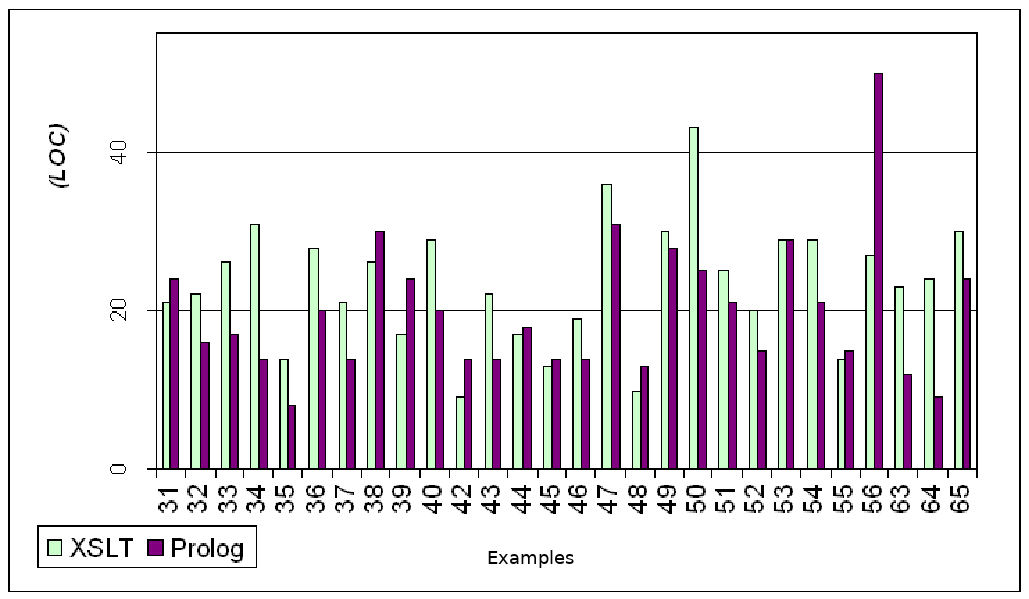}
\end{tabular}
\end{center}
\caption{LOC for examples 31-40,42-56,63-65}
\label{fig:metricsLOC2}
\end{figure}

Fig.\ref{fig:metricsLOC1} shows the first 29 examples by lines of codes ($LOC$).
The four examples 3,5,8,10 clearly show the shortened representation.
Nearly in all examples, Prolog is significantly shorter.
Exceptions are sporadic and have in those cases almost no effect at all.
The examples from fig.\ref{fig:metricsLOC2} represent solutions to much more complex tasks.
In these cases, only Prolog and XSLT are getting closer.
Example 56 is a statistical anomaly, which is significantly worse than XSLT, but this is because of the lack of an ``\textit{up}''-operator as described earlier.
The lack is why the whole problem requires a relatively complex implementation compared to XSLT’s implementation using the parental operator.
This single problem is an exception -- because this kind of problems occurs very seldom in practice.
Another issue is that string operators in Prolog are more complex because Prolog distinguishes explicitly on nodes between numbers and strings, where XSLT automatically converts everything to a string.
--- This may be considered an advantage and disadvantage. 
It is worth noting operations over strings and number conversions are better avoided if not needed.
So, examples 38 and 39 can be represented slightly worse because explicit conversion would be required.
Examples 29 and 31 are lengthier because number conversions can only be evaluated by using the predicate \texttt{transform}.
Examples 24, 25, 32-34 and 64 illustrate variables and parameters are easier to use in Prolog than in XSLT, which is a significant improvement.\\

The program length in bytes enforces the metric $LOC$ (cf. appendix \ref{appendix:metricsTables}).
Prolog programs are often significantly shorter due to their tag-free notation but may lead to more program lines due to sophisticated node constructions.
Only the combination of both LOC and bytes leads to a profound statement about redundancy.

$\eta$ illustrates the language circumference is more balanced in Prolog.
So, it may be more comfortable for the user if a language has many extraordinary operands and relatively few operators, indicating poor expressibility.
XSLT has vast differences between $\eta_{1}$ and $\eta_{2}$ compared to Prolog (cf. examples 6-12).
That means XSLT quickly tends to a monotone program style.

The first ten examples were investigated separately.
XSLT examples from Appendix \ref{appendix:metricsTables} only consider immediate XSLT-constructs.
XSLT-examples marked with ``\textit{XSLT2}'' are the same as those marked with ``\textit{XSLT}'', except the metrics obtained refer now to both XSLT and XPath expressions.
In ``\textit{XSLT2}'', the less-equal and greater-equals signs, as well as the division sign, are counted as operators.
An element node's name is counted twice for a tag, where "\textit{XSLT}" counts only once.
This distinction is required because XSLT-programs are evaluated only on XSLT-level, where XPath-expressions are interpreted as operands.
XPath-expressions must be measured differently due to a differing syntax.

\begin{figure}[h]	]
\begin{center}
\begin{tabular}{l}
\includegraphics[width=8cm]{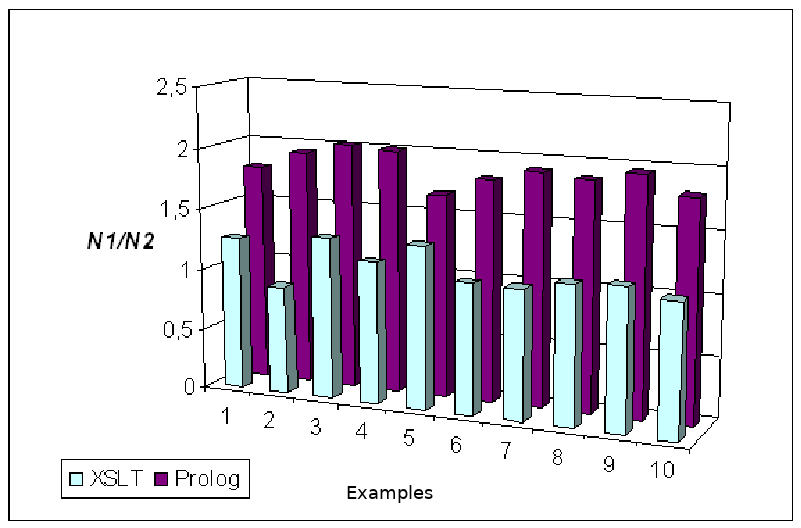}
\end{tabular}
\end{center}
\caption{Functionality as ratio $N_{1}$ : $N_{2}$}
\label{fig:metricsN1N2}
\end{figure}

Fig.\ref{fig:metricsN1N2} shows the ratio of $N_{1}$ : $N_{2}$ for the first ten examples.
On average, Prolog is approximately 30\% more functional than XSLT.
Apart from that, examples 2 and 22 are significant.
In Prolog, an equivalent representation is 2x or even 3x shorter than in XSLT because stylesheet definitions may be saved and shorter node constructors.
Path expressions and ``\textit{If-then}''-statements are on average 50\% shorter than in XSLT (cf. examples 6, 7, 9, 18, 26).
However, string operations seem to be a bit more than twice as flexible in XSLT than in Prolog because additional conversions are not needed (cf. example 52) due to multiple conversions in strings, which otherwise are required in Prolog \texttt{translate}.\\

\textbf{Derivable Metrics:}

\begin{figure}[h]
\begin{center}
\begin{tabular}{l}
\includegraphics[width=8cm]{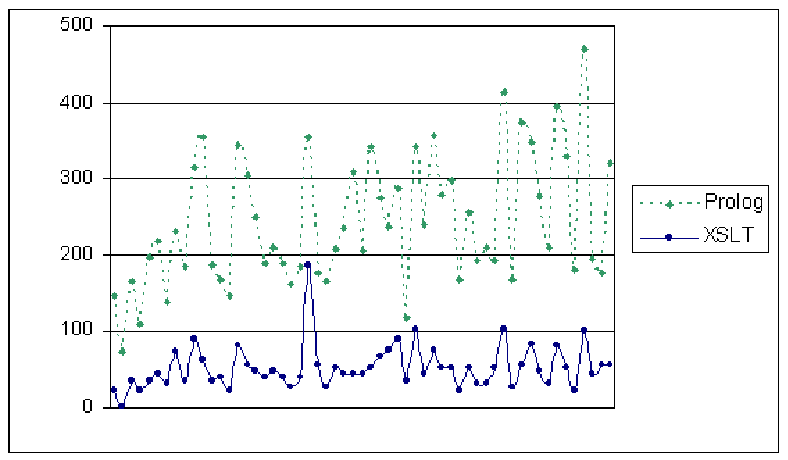}
\end{tabular}
\end{center}
\caption{Theoretical program length $N_{T}$}
\label{fig:metricsNT}
\end{figure}

In fig.\ref{fig:metricsNT} is the comparison between the theoretical program length of Prolog and XSLT.
It can be noticed that Prolog programs seem to be by far ``longer'' than XSLT programs.
However, this effect is a consequence of the encapsulated XPath within XSLT.
$\eta_{1}$ and $\eta_{2}$ determine the program length.
These measures can strongly deviate by only XSLT operators and operands.
A consideration of XPath path expressions and operands belonging to them leads to balancing the difference between both languages, as seen in the first examples of ``\textit{XSLT2}'' of appendix \ref{appendix:metricsTables}.

The positive length deviation in percent  $\Delta_{N}$ is defined by
$$\Delta_{N}=\frac{100 \cdot (max(N_{T},N)-min(N_{T},N))}{max(N_{T},N)}$$

\begin{figure}
\begin{center}
\begin{tabular}{l}
\includegraphics[width=8cm]{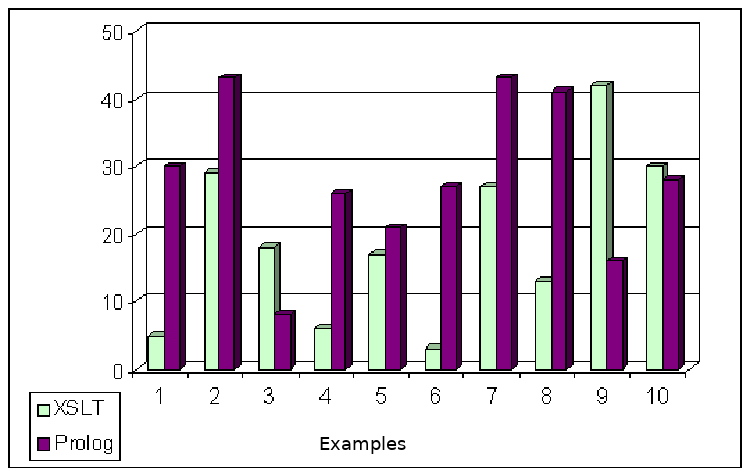}
\end{tabular}
\end{center}
\caption{Deviation $\Delta_{N}$}
\label{fig:metricsDeviation}
\end{figure}

\begin{figure}
\begin{center}
\begin{tabular}{l}
\includegraphics[width=8cm]{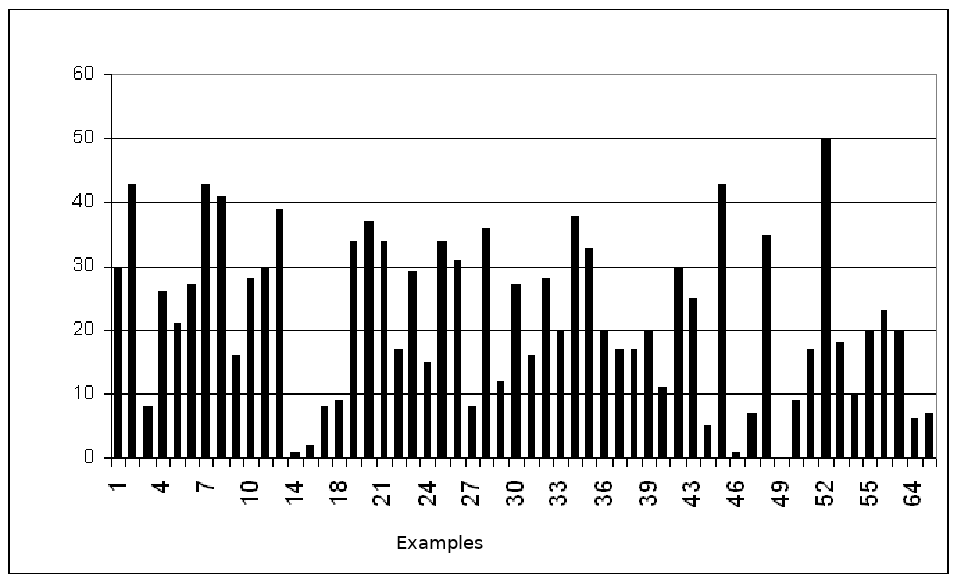}
\end{tabular}
\end{center}
\caption{Sorted length deviation in Prolog}
\label{fig:metricsSortedDeviation}
\end{figure}

The deviation $\Delta_{N}$ is substantial (see fig.\ref{fig:metricsDeviation}).
That is why it is harder to estimate Prolog, whether a complex task requires a complex program or an easy program.
In fig.\ref{fig:metricsSortedDeviation}, all examples are sorted by ascending deviation.
On average, the deviation is approximately 22\%.
There is one statistical exception at \texttt{translate}.

The examples in ascending order have the same increase in deviations.
Hence, the set is normal-distributed.
Differences of more than 60\% can be detected in examples 19, 20, 28, 30, 31, 33, 34, 39 and 56.
They can be explained by the short, implicit representation of conventional statements in Prolog.

Moreover, in Prolog, \texttt{value-of} has to be enforced by additional conversions of tree nodes and attributes.\\

For the reasons mentioned, the abstraction level $\lambda$ is representable only for the first ten examples.
Here only examples 7 and 7\_2 are appropriate enough in order to make generalisations on transformations.
The remaining examples illustrate only one aspect and are too short in order to make generalisations.
The expected number of errors $B$ supersedes in Prolog programs the first ten XSLT-programs by about 12.7\%, except for some stability restrictions due to a small number of operands.
There is a more considerable deviation in favour of Prolog in example 7\_2.
Apart from this, the number of errors expected in Prolog supersedes the number for XSLT programs.
The user is expected to commit errors more often in Prolog than in XSLT for an equivalent length of the program listing.
In other words: programs are shorter with the same expected number of errors.
From a correction perspective, XSLT is worse because programs are much longer.
On the other side, prologue programs may even be several times shorter but are harder to maintain because of variables' input and output behaviour.

\subsection{Comparison of Prolog with XQuery and SXSLT}

XQuery is an XML-based query language \cite{XQuerySpec}.
XQuery cannot only be used for queries but also transformations.
XQuery queries are in curly brackets in XML tags.
A query consists of a \lit{FOR}-loop, a variable declaration \lit{LET}, an output ordering \lit{ORDER BY}, a conditional part \lit{WHERE} and a result document \lit{RETURN}.
The result of an XQuery-query is only an element node, not a node-set as in XSLT.
XQuery knows of templates -- \lit{FOR} returns the matching pattern and \lit{RETURN} the node-set.
XQuery uses XPath as a self-sustained language but is no markup language.
So it is easier to read at first glance, mainly because it offers numerous aggregate functions and processes sets and joins.
XQuery is free of side effects, the same as XSLT.
It does not offer node unification.
Multiple solutions are listed linearly, and backtracking is not allowed.\\

The language SXSLT \cite{Kiselyov} is a functional template-oriented transformation language written in Scheme and is free of side effects.
SXSLT mimics Scheme.
It extends XSLT by higher-order functions and other Scheme's language core features.
The use of functions makes it extraordinary expressible and flexible.
Unfortunately, both SXSLT and tuProlog do not have a type inference system.
It makes usability more complicated than it actually could be.
Scheme partially offers pattern matching in specifications.
However, this matching is not as expressible, s.t. variables could be unified with ground terms.
Scheme's prefix term notation makes SXSLT very user-unfriendly or very obscure compared to other popular transformation languages.
For example, statements and function call, even in small functions, get heavily overloaded with brackets.

\section{Summary}
\label{sect:Summary}

This work compares a Prolog-based transformation language with XSLT.
First, the Prolog-transformation language was designed and implemented.
It consists of XML read- and write operations, transformation operators and a pre-defined traversal order.
Second, the Prolog-based transformation language is analysed towards previously defined quality measures and compared with XSLT.

The transformation language is partially written in Prolog and Java.
Thus the transformation language obtained is platform-independent, and extensions may be organised within modules.
Due to its openness, Prolog is restricted to transformations and fits tasks beyond this domain.\\

This work's goal was initially to investigate if and which attempts exist in logical and functional languages for XML transformations.
The essential lingual features of Prolog should be considered for transformations.

Next, an appropriate data structure should be researched in Prolog for parsing and serialisation of XML documents.
Essential transformation operators should be implemented in Java and Prolog and thoroughly be tested afterwards.
Deviations and new operations of XSLT were discussed.

From comparisons with other transformation languages and query languages, new operators were derived.
According to numerous selected examples, each XSLT program was turned into an equivalent (or multiple equivalents) Prolog program(s) and probed.
Already introduced operators were refined and extended by new operators and predicates.

Afterwards, comparison criteria were defined.
Biased and non-biased criteria played an important role.
Quantitative metrics were chosen for logical and functional programs.
The invertibility of predicates was researched.

In conclusion, both languages were compared with each other, and evaluations were made.\\

Not surprising, the comparison showed different results.
Prolog's short notations are remarkable.
Besides the XSLT-model, the data model in Prolog allows tuples, atoms and lists in a comprehensive form.
Many statements, as known from imperative languages, are in a shortened form.
Symbols allow convenient reuse without definitions and further agreements, as well as a flexible use within programs.
The node constructors are easy to read and contribute to a better abstraction of ``\textit{before}'' and ``\textit{after}'' the transformation.
Aggregate functions are compact. Its return value needs to be projected from relations.
Unification allows multiple applications over term operations, one of which is invertibility.
Backtracking allows a flexible search.
Pattern-matching can be applied at any position in a Prolog program.
The openness of Prolog is of advantage.

Implementations without templates are now preferred, especially with many branches and invertible functions.

XSLT programs are only shorter when many implicit string operations need to be performed due to built-in functions.
Functions were simulated in both XSLT and Prolog.
Formatting functions and settings can easier be changed in XSLT.
Unparametrised constructors can often be represented shorter.
In XPath, path expressions often are a bit shorter and concise.
However, this sight initial advantage is quickly lost with XML tags and inconvenient placeholder declarations again.
Templates are preferred if the given document structure is partially or unknown and only a few elements are processed.
Some of the weaknesses of XSLT are strengthened in version 2.0.
For instance, user-defined functions may be defined.\\

Prolog does not fully substitute XSLT as a transformation language for XML documents, mainly due to less popular tool support.
This problem could be relaxed if powerful tools and plugins are more popularised.
Then even a rise in complex XML schema transformation, content management and text retrieval seems promising.

\newpage
\section*{Glossary}

\myglsentry{Arity}
\myglsdesc{Arity is the number of parameters a predicate or a functor has ($\nearrow$ functor)}

\myglsentry{Backtracking}
\myglsdesc{Recursive search, which continues until a solution is found or a contradiction. If a contradiction occurs, then backtracking jumps back to the last valid branching and tries all alternatives always from the closest last recent.}

\myglsentry{Cut}
\myglsdesc{A Prolog primitive cutting off alternative solutions that may occur during a goal is processed.}

\myglsentry{Decorator}
\myglsdesc{Is a design-pattern decorating existing sub-class instances (see \cite{Kerievsky}).}

\myglsentry{DOM}
\myglsdesc{Short-form for \underline{D}ocument \underline{O}bject \underline{M}odel. Represents an XML document as data structure. The JAXP-library offers classes for manipulating DOMs.}

\myglsentry{IDE}
\myglsdesc{\underline{I}ntegrated \underline{D}evelopment \underline{E}nvironment is a development environment that contains tools like editor, compiler and debugger.}

\myglsentry{JAXP}
\myglsdesc{stands for \underline{J}ava \underline{A}PI for \underline{X}ML \underline{P}rocessing; is a Java library for processing XML documents ($\nearrow$ DOM).}

\myglsentry{Higher-order functions (functional)}
\myglsdesc{Synonym for a function that accepts functions as input and output. Functionals allow abstraction of program logic, so that recursion may be resolved. Thus higher flexibility is achieved in contrast to conventional functions as in Pascal or C. Typical list functionals are left folding, mapping and filtering.}

\myglsentry{Functor}
\myglsdesc{Prolog-operator used for the evaluation of arithmetic expressions.
User-defined functors must be embedded into a predicate. The definition of a functor can be in Prolog or as Java-method with a return value of type \texttt{Term} within the tuProlog-library. The Prolog definition op(\textit{Precedence}, \textit{Arity}, \textit{Name}) is noted as fact ($\nearrow$ precedence, arity).}

\myglsentry{Ground and non-ground terms}
\myglsdesc{Ground terms are terms that contain no free occurrences of variables.}

\myglsentry{H\aa lstead-Metric}
\myglsdesc{A $\nearrow$ metric for getting the circumference and complexity of a program.
To this metric belong the total amount of operators $\eta_{1}$, and operands $\eta_{2}$, and the sum of all used operators $N_{1}$, the sum of all used operands $N_{2}$, program length $N$, theoretical program length $N_T$, the experimental program circumference $V$ and further measures about the circumference of a program, but not about the program structure. \textit{LOC} is not a H\aa lstead metric because it also considers comments and whitespaces ($\nearrow$ LOC).}

\myglsentry{Introspection}
\myglsdesc{Mechanism reflecting on a program as input. Herewith the input is first checked for syntax and second semantically. If successful, the intermediate result may be manipulated. In Java, introspection is implemented by the Reflection-API.}

\myglsentry{Kind}
\myglsdesc{A type or a type parameter in case of a composed type. It may be used in order to express the type of an element constructor. An example would be $\nabla \rightarrow \nabla \rightarrow \nabla$.}

\myglsentry{Lazy evaluation}
\myglsdesc{Evaluation meta-schema which calculates branches of a calculation when they are needed.}

\myglsentry{Lexem}
\myglsdesc{synonym for a token or least lingual unit, it is the result of the lexical analysis.}

\myglsentry{LOC}
\myglsdesc{stands for \underline{L}ines \underline{O}f \underline{C}ode; is a program metric ($\nearrow$ metric).}

\myglsentry{Markup}
\myglsdesc{is a text fragment guarded by tags, separators with a special syntax}

\myglsentry{Meta-information}
\myglsdesc{the information which not directly relates to an XML document but which are still used by applications interpreting XML.}

\myglsentry{Meta-logical type predicate}
\myglsdesc{predicates capable of determining the membership to a particular category for a given Prolog-term. Examples are \texttt{var}, \texttt{list}, \texttt{atom}.}

\myglsentry{Metric}
\myglsdesc{A program measure that allows to characterise a program quantitatively ($\nearrow$ H\aa lstead-metric). Metrics can be by circumference or by structure.}

\myglsentry{Monad}
\myglsdesc{Programming feature  which encapsulates the view to a data type. In contrast to arbitrary objects, monads are at least half-groups. Monads allow assignments and error handling without entirely rewriting a function.}

\myglsentry{Monotone and non-monotone predicates}
\myglsdesc{Non-monotone predicates or functors ($\nearrow$ functor) alter a given data structure. Monotone predicates or functions keep data structures mostly as they are with only minor local changes.}

\myglsentry{Non-strict function}
\myglsdesc{A non-strict function terminates, even if one of its arguments would not terminate if evaluated before executing that function ($\nearrow$ lazy evaluation).}

\myglsentry{Parse operation}
\myglsdesc{The process which reads an XML document in Prolog, s.t. a corresponding term is successfully bound to a symbol.}

\myglsentry{Pattern Matching}
\myglsdesc{A programming feature in Prolog allows terms to be used anywhere in a rule and with symbols and subterms ($\nearrow$ unification) instead of having explicit ``\textit{if-then}''-checks.}

\myglsentry{PI-nodes}
\myglsdesc{Processing-Instruction nodes are part of an XML document the same as element nodes. They may be used in different ways depending on their domain, e.g. for graphic output or view settings.}

\myglsentry{Precedence}
\myglsdesc{Precedence or inverted priority defines in which ordering functors are evaluated ($\nearrow$ functor). The smaller precedence of a functor is, the higher gets its priority.}

\myglsentry{Prolog-conventions}
\myglsdesc{Symbols and variables have in Prolog different characteristics. \lit{+} denotes a term that may only be used as input, where \lit{-} for output. \lit{?} denotes a term that may be used for both input and output. \lit{@} insists a term must be provided, which excludes anonymous symbols.}

\myglsentry{Referential transparency}
\myglsdesc{Referential transparency is obeyed, when from an outer scope, no changes may be applied to the inner scope of some function.}

\myglsentry{Reification}
\myglsdesc{In general, reification stands for objectifying an activity that most often can be described by some verb.
Problem reification to a varying refactoring (cf. \cite{Kerievsky}) complexity in Java, depending on the problem granularity.
A method may be hoisted into a \texttt{Strategy}-class.
}

\myglsentry{Relational Algebra}
\myglsdesc{Mathematical formalism where operations are defined over relations or tables. Base operations are union, minus, Cartesian product, projection, selection and renaming (see sect.\ref{sect:modellingTransformations}). }

\myglsentry{Sequential search}
\myglsdesc{Forward sequential search leads to a derivation chain. A backward sequential derivation searches for trivial solutions matching with the immediate last subgoal.}

\myglsentry{Serialisation operation}
\myglsdesc{Inverse to the $\nearrow$ parse operation. The Prolog term is serialised into an XML output document.}

\myglsentry{Stylesheet}
\myglsdesc{XML file containing all templates needed for one transformation step.}

\myglsentry{Template}
\myglsdesc{Transformation rule specifying original and target nodes. On $\nearrow$ traversal, the actual node is $\nearrow$ pattern-matched against the original node in one of the templates.}

\myglsentry{Transformation language}
\myglsdesc{Language, which often is descriptive, transforming an incoming XML document into an outgoing XML document by using templates as mapping ($\nearrow$ XSLT, Template).}

\myglsentry{Traversal}
\myglsdesc{Here, an iteration of a tree-structured data structure/model.}

\myglsentry{Invertible predicates}
\myglsdesc{are predicates that may be called with parameters used as input or output or both. If predicates are then still defined, they are overloaded.
Mainly they are invertible if input terms may be swapped with output terms. If any combination is defined, then the considered predicate is fully invertible. Invertible predicates are of use in inverse mappings ($\nearrow$ invertibility).}

\myglsentry{Invertibility}
\myglsdesc{
The property of invertibility here refers to transformations. If given a target document the original document may be reconstructed with the same transformation $\nearrow$ stylesheet, then the considered transformation is invertible. Often contributed aggregate functions being used also need to be invertible ($\nearrow$ invertible predicates).}

\myglsentry{UML}
\myglsdesc{\underline{U}nified \underline{M}odelling \underline{Language}; is a graphical/textual modelling language, often used with recent object-oriented background, like complex software systems.}

\myglsentry{Unification}
\myglsdesc{Is a term-preserving term operation, often when comparing two terms.
 If two terms may be generalised, namely unassigned symbols are substituted, s.t. the two terms become equal. Then both terms are called unifiable. Within a program using $\nearrow$ pattern-matching substitutions may be performed locally, but with local change effects.}

\myglsentry{Unsafe clause}
\myglsdesc{Clauses that may not terminate are called unsafe. Left recursion and calls to clauses with invalid input are one cause for lack of safety.}

\myglsentry{Visitor}
\myglsdesc{The design pattern \textit{Visitor} $\nearrow$ reifies operations on a data structure with the two objectives: variability and transparency (see \cite{Kerievsky}).}

\myglsentry{Full invertibility}
\myglsdesc{($\nearrow$ invertible predicates).}

\myglsentry{W3C}
\myglsdesc{The \underline{W}orld \underline{W}ide \underline{W}eb \underline{C}onsortium standardised XML, XPath, XSLT and many other XML technologies, which are standard now.}

\myglsentry{Embedded language}
\myglsdesc{An embedded language is a language embedding domain-specific languages. For example, XSLT is an embedded language, the language where it is embedded is XML. The part of Prolog processing XML described in this work is an embedded language.}

\myglsentry{XML}
\myglsdesc{E\underline{X}tensible \underline{M}arkup \underline{L}anguage is a $\nearrow$ markup language for representing semi-structured data.}

\myglsentry{XPath}
\myglsdesc{Navigation language for addressing elements within an XML document. XPath is not $\nearrow$ embedded into XML.}

\myglsentry{XSLT}
\myglsdesc{E\underline{X}ensible \underline{S}tylesheet \underline{L}anguage for \underline{T}ransformation is a template-centric transformation language $\nearrow$ embedded into XML ($\nearrow$ Template, transformation language, XML).}

\section*{APPENDIX A: Metrics Evaluation}
\label{appendix:metricsTables}

\ \newpage

\begin{tabular}{|l l| l l | l l l l }
\hline
\# &  Example &  LOC & Bytes & $\eta_{1}$ & $\eta_{2}$ & $N_{1}$ & $N_{2}$\\
\hline
\hline
1	&	Prolog	&	13	&	326 & 14	&	20	&	62	&	36\\
	&	Prolog2	&	9	&	233 & 14	&	17	&	50	&	28\\
	&	XSLT	&	10	&	290 & 5	        &	5	&	6	&	5\\
	&	XSLT2	&	10	&	290 & 5	        &	12	&	29	&	23\\
\hline
2	&	Prolog	&	7	&	125 & 12	&	10	&	29	&	15\\
	&	Prolog2	&	7	&	106 & 11	&	10	&	24	&	13\\
	&	XSLT	&	3	&	110 & 1	&	2	&	1	&	2\\
	&	XSLT2	&	3	&	110 & 4	&	6	&	8	&	9\\
\hline
3	&	Prolog	&	16	&	351 & 11	&	21	&	81	&	40\\
	&	XSLT	&	21	&	522 & 6	&	7	&	12	&	9\\
	&	XSLT2	&	21	&	522 & 5	&	18	&	57	&	49\\
\hline
4	&	Prolog	&	10	&	277 & 13	&	17	&	58	&	29\\
	&	Prolog2	&	8	&	245 & 11	&	16	&	48	&	25\\
	&	Prolog3	&	7	&	129 & 12	&	12	&	28	&	16\\
	&	XSLT	&	13	&	331 & 5	&	5	&	7	&	6\\
	&	XSLT2	&	13	&	331 & 5	&	15	&	36	&	31\\
\hline
5	&	Prolog	&	15	&	405 & 15	&	23	&	80	&	48\\
	&	XSLT	&	22	&	522 & 6	&	7	&	12	&	9\\
	&	XSLT2	&	22	&	522 & 5	&	18	&	57	&	48\\
\hline
6	&	Prolog	&	17	&	486 & 16	&	24	&	82	&	45\\
	&	XSLT	&	17	&	543 & 6	&	9	&	11	&	12\\
	&	XSLT2	&	17	&	543 & 6	&	23	&	60	&	56\\
\hline
7	&	Prolog	&	8	&	192 & 14	&	14	&	40	&	21\\
	&	XSLT	&	10	&	323 & 5	&	7	&	6	&	7\\
	&	XSLT2	&	10	&	323 & 6	&	18	&	34	&	32\\
\hline
$7\_{2}$	&	Prolog	&	13 & 559	&	14	&	23	&	116	&	57\\
	&	XSLT	&	40	&	1205 & 6	&	15	&	28	&	24\\
	&	XSLT2	&	40	&	1205 & 6	&	27	&	137	&	122\\
\hline
8	&	Prolog	&	11	&	285 & 16	&	20	&	58	&	31\\
	&	XSLT	&	12	&	394 & 6	&	7	&	8	&	7\\
	&	XSLT2	&	12	&	394 & 6	&	18	&	41	&	38\\
\hline
9	&	Prolog	&	17	&	926 & 15	&	33	&	178	&	91\\
	&	XSLT	&	37	&	1176 & 6	&	18	&	26	&	27\\
	&	XSLT2	&	37	&	1176 & 7	&	22	&	138	&	119\\
\hline
10	&	Prolog	&	27	&	781 & 24	&	37	&	135	&	75\\
	&	Prolog2	&	25	&	680 & 22	&	37	&	134	&	75\\
	&	XSLT	&	29	&	967 & 6	&	13	&	20	&	25\\
	&	XSLT2	&	29	&	967 & 7	&	21	&	111	&	102\\
\hline
12	&	Prolog	&	12	&	352 & 16	&	24	&	79	&	43\\
	&	Prolog2	&	11	&	304 & 13	&	23	&	69	&	36\\
	&	XSLT	&	17	&	439 & 6	&	7	&	10	&	7\\
\hline
13	&	Prolog	&	8	&	217 & 14	&	18	&	52	&	27\\
	&	XSLT	&	16	&	437 & 7	&	7	&	10	&	9\\
\hline
14	&	Prolog	&	9	&	223 & 14	&	17	&	52	&	27\\
	&	Prolog2	&	22	&	516 & 19	&	22	&	122	&	58\\
	&	Prolog3	&	16	&	359 & 17	&	22	&	81	&	43\\
	&	XSLT	&	9	&	266 & 5	&	5	&	5	&	5\\
\hline
15	&	Prolog	&	29	&	918 & 23	&	32	&	189	&	85\\
	&	Prolog2	&	28	&	903 & 21	&	37	&	187	&	92\\
	&	XSLT	&	32	&	969 & 10	&	13	&	20	&	15\\
\hline
17	&	Prolog	&	17	&	664 & 18	&	34	&	139	&	67\\
	&	Prolog2	&	21	&	790 & 21	&	36	&	170	&	87\\
	&	XSLT	&	32	&	917 & 9	&	9	&	20	&	11\\
\hline
18	&	Prolog	&	22	&	627 & 21	&	24	&	129	&	56\\
	&	XSLT	&	19	&	657 & 6	&	10	&	12	&	13\\
\hline
\end{tabular}

\begin{tabular}{l l| l l | l l l |}
\hline
\# &  Example & $\frac{N_{1}}{N_{2}}$ & $N_{T}$& $\Delta_{N}$ & $\lambda$ & B\\
\hline
\hline
1	&	Prolog	&	1.7	&	139.7	&	30	&	0.1	&	1.7	\\
	&	Prolog2	&		&	122.8	&	37	&	0.2	&	1.3	\\
	&	XSLT	&	1.2	&	23.2	&	54	&	1.8	&	0.1	\\
	&	XSLT2	&	1.3	&	54.6	&	5	&	0.3	&	0.7	\\
\hline
2	&	Prolog	&	1.9	&	76.2	&	43	&	0.3	&	0.7	\\
	&	Prolog2	&	1.8	&	71.3	&	49	&	0.4	&	0.5	\\
	&	XSLT	&	0.5	&	2.0	&	33	&	13.5	&	0.0	\\
	&	XSLT2	&	0.9	&	23.5	&	29	&	1.1	&	0.2	\\
\hline
3	&	Prolog	&	2.0	&	130.3	&	8	&	0.1	&	2.0	\\
	&	XSLT	&	1.3	&	35.2	&	42	&	0.8	&	0.3	\\
	&	XSLT2	&	1.2	&	86.7	&	18	&	0.1	&	1.6	\\
\hline
4	&	Prolog	&	2.0	&	117.6	&	26	&	0.2	&	1.4	\\
	&	Prolog2	&	1.9	&	102.1	&	29	&	0.2	&	1.2	\\
	&	Prolog3	&	1.8	&	86.0	&	49	&	0.3	&	0.7	\\
	&	XSLT	&	1.2	&	23.2	&	46	&	1.5	&	0.1	\\
	&	XSLT2	&	1.2	&	70.2	&	6	&	0.2	&	1.0	\\
\hline
5	&	Prolog	&	1.7	&	162.6	&	21	&	0.1	&	2.2	\\
	&	XSLT	&	1.3	&	35.2	&	42	&	0.8	&	0.3	\\
	&	XSLT2	&	1.2	&	86.7	&	17	&	0.1	&	1.6	\\
\hline
6	&	Prolog	&	1.8	&	174.0	&	27	&	0.1	&	2.3	\\
	&	XSLT	&	0.9	&	44.0	&	49	&	0.7	&	0.3	\\
	&	XSLT2	&	1.1	&	119.6	&	3	&	0.1	&	1.9	\\
\hline
7	&	Prolog	&	1.9	&	106.6	&	43	&	0.2	&	1.0	\\
	&	XSLT	&	0.9	&	31.3	&	59	&	1.4	&	0.2	\\
	&	XSLT2	&	1.1	&	90.6	&	27	&	0.2	&	1.0	\\
\hline
$7\_{2}$	&	Prolog	&	2.0	&	157.3	&	9	&	0.1	&	3.0	\\
	&	XSLT	&	1.2	&	74.1	&	31	&	0.3	&	0.8	\\
	&	XSLT2	&	1.1	&	143.9	&	44	&	0.0	&	4.4	\\
\hline
8	&	Prolog	&	1.9	&	150.4	&	41	&	0.1	&	1.5	\\
	&	XSLT	&	1.1	&	35.2	&	58	&	1.2	&	0.2	\\
	&	XSLT2	&	1.1	&	90.6	&	13	&	0.2	&	1.2	\\
\hline
9	&	Prolog	&	2.0	&	225.1	&	16	&	0.0	&	5.0	\\
	&	XSLT	&	1.0	&	90.6	&	42	&	0.3	&	0.8	\\
	&	XSLT2	&	1.2	&	117.8	&	54	&	0.1	&	4.2	\\
\hline
10	&	Prolog	&	1.8	&	302.8	&	31	&	0.1	&	4.2	\\
	&	Prolog2	&	1.8	&	290.9	&	28	&	0.1	&	4.1	\\
	&	XSLT	&	0.8	&	63.6	&	30	&	0.3	&	0.6	\\
	&	XSLT2	&	1.1	&	111.9	&	47	&	0.1	&	3.4	\\
\hline
12	&	Prolog	&	1.8	&	174.0	&	30	&	0.1	&	2.2	\\
	&	Prolog2	&	1.9	&	152.1	&	31	&	0.1	&	1.8	\\
	&	XSLT	&	1.4	&	35.2	&	53	&	1.0	&	0.2	\\
\hline
13	&	Prolog	&	1.9	&	128.4	&	39	&	0.2	&	1.3	\\
	&	XSLT	&	1.1	&	39.3	&	53	&	0.9	&	0.2	\\
\hline
14	&	Prolog	&	1.9	&	122.8	&	36	&	0.2	&	1.3	\\
	&	Prolog2	&	2.1	&	178.8	&	1	&	0.1	&	3.2	\\
	&	Prolog3	&	1.9	&	167.6	&	26	&	0.1	&	2.2	\\
	&	XSLT	&	1.0	&	23.2	&	58	&	1.9	&	0.1	\\
\hline
15	&	Prolog	&	2.2	&	264.0	&	3	&	0.0	&	5.3	\\
	&	Prolog2	&	2.0	&	285.0	&	2	&	0.0	&	5.4	\\
	&	XSLT	&	1.3	&	81.3	&	57	&	0.4	&	0.5	\\
\hline
17	&	Prolog	&	2.1	&	248.0	&	17	&	0.1	&	3.9	\\
	&	Prolog2	&	2.0	&	278.4	&	8	&	0.0	&	5.0	\\
	&	XSLT	&	1.8	&	57.1	&	47	&	0.5	&	0.4	\\
\hline
18	&	Prolog	&	2.3	&	202.3	&	9	&	0.1	&	3.4	\\
	&	XSLT	&	0.9	&	48.7	&	49	&	0.6	&	0.3	\\
\hline
\end{tabular}

\newpage

\begin{tabular}{|l l| l l | l l l l}
\hline
19	&	Prolog	&	12	&	270 & 17	&	19	&	66	&	33\\
	&	XSLT	&	15	&	459 & 8	&	6	&	8	&	7\\
\hline
20	&	Prolog	&	12	&	286 & 16	&	22	&	67	&	36\\
	&	XSLT	&	16	&	513 & 9	&	7	&	9	&	8\\
\hline
21	&	Prolog	&	15	&	355 & 19	&	23	&	80	&	43\\
	&	Prolog2	&	12	&	283 & 17	&	19	&	66	&	32\\
	&	XSLT	&	15	&	470 & 8	&	6	&	8	&	7\\
\hline
22	&	Prolog	&	14	&	338 & 18	&	15	&	74	&	37\\
	&	XSLT	&	10	&	323 & 5	&	6	&	5	&	6\\
\hline
23	&	Prolog	&	11	&	289 & 15	&	20	&	62	&	34\\
	&	Prolog2	&	14	&	411 & 18	&	24	&	85	&	47\\
	&	XSLT	&	16	&	507 & 8	&	6	&	9	&	7\\
\hline
24	&	Prolog	&	14	&	426 & 16	&	23	&	95	&	49\\
	&	Prolog2	&	14	&	445 & 16	&	26	&	98	&	52\\
	&	XSLT	&	23	&	563 & 8	&	7	&	14	&	8\\
\hline
25	&	Prolog	&	9	&	276 & 11	&	19	&	49	&	30\\
	&	XSLT	&	24	&	826 & 7	&	11	&	15	&	16\\
\hline
26	&	Prolog	&	17	&	387 & 20	&	23	&	79	&	44\\
	&	Prolog2	&	14	&	300 & 16	&	18	&	61	&	35\\
	&	Prolog3	&	14	&	326 & 18	&	22	&	65	&	37\\
	&	XSLT	&	9	&	293 & 5	&	6	&	5	&	6\\
\hline
27	&	Prolog	&	17	&	463 & 16	&	21	&	94	&	50\\
	&	Prolog2	&	22	&	617 & 21	&	28	&	122	&	64\\
	&	XSLT	&	21	&	699 & 9	&	8	&	12	&	9\\
\hline
28	&	Prolog	&	16	&	387 & 19	&	24	&	79	&	44\\
	&	Prolog2	&	18	&	459 & 21	&	28	&	92	&	50\\
	&	XSLT	&	15	&	486 & 7	&	8	&	8	&	8\\
\hline
29	&	Prolog	&	23	&	774 & 21	&	34	&	156	&	78\\
	&	XSLT	&	21	&	611 & 9	&	6	&	12	&	6\\
\hline
30	&	Prolog	&	19	&	457 & 22	&	25	&	105	&	53\\
	&	Prolog2	&	13	&	286 & 18	&	20	&	68	&	37\\
	&	XSLT	&	15	&	471 & 8	&	7	&	8	&	7\\
\hline
31	&	Prolog	&	24	&	722 & 23	&	36	&	159	&	85\\
	&	XSLT	&	21	&	644 & 9	&	8	&	12	&	8\\
\hline
32	&	Prolog	&	23	&	607 & 23	&	31	&	123	&	63\\
	&	Prolog2	&	16	&	399 & 19	&	27	&	86	&	48\\
	&	XSLT	&	22	&	783 & 9	&	11	&	14	&	12\\
\hline
33	&	Prolog	&	24	&	632 & 22	&	31	&	126	&	68\\
	&	Prolog2	&	17	&	423 & 19	&	19	&	89	&	40\\
	&	XSLT	&	26	&	1112 & 12	&	10	&	16	&	12\\
\hline
34	&	Prolog	&	14	&	368 & 20	&	24	&	78	&	45\\
	&	XSLT	&	31	&	1043 & 14	&	11	&	18	&	14\\
\hline
35	&	Prolog	&	11	&	249 & 14	&	17	&	52	&	30\\
	&	Prolog2	&	8	&	153 & 10	&	13	&	33	&	22\\
	&	XSLT	&	14	&	425 & 4	&	9	&	9	&	10\\
\hline
36	&	Prolog	&	20	&	506 & 21	&	30	&	121	&	70\\
	&	XSLT	&	28	&	981 & 10	&	17	&	20	&	21\\
\hline
37	&	Prolog	&	23	&	633 & 23	&	30	&	141	&	67\\
	&	Prolog2	&	14	&	427 & 20	&	24	&	92	&	47\\
	&	XSLT	&	21	&	621 & 8	&	7	&	12	&	7\\
\hline
38	&	Prolog	&	30	&	855 & 18	&	39	&	136	&	99\\
	&	XSLT	&	26	&	908 & 5	&	16	&	20	&	18\\
\hline
39	&	Prolog	&	24	&	528 & 22	&	27	&	118	&	64\\
	&	XSLT	&	17	&	585 & 8	&	9	&	10	&	9\\
\hline
40	&	Prolog	&	20	&	723 & 21	&	31	&	147	&	71\\
	&	XSLT	&	29	&	916 & 8	&	9	&	18	&	9\\
\hline
\end{tabular}

\begin{tabular}{l l| l l | l l l |}
\hline
19	&	Prolog	&	2.0	&	150.2	&	34	&	0.1	&	1.7	\\
	&	XSLT	&	1.1	&	39.5	&	63	&	1.1	&	0.2	\\
\hline
20	&	Prolog	&	1.9	&	162.1	&	37	&	0.1	&	1.8	\\
	&	XSLT	&	1.1	&	48.2	&	65	&	0.9	&	0.2	\\
\hline
21	&	Prolog	&	1.9	&	184.8	&	34	&	0.1	&	2.2	\\
	&	Prolog2	&	2.1	&	150.2	&	35	&	0.1	&	1.7	\\
	&	XSLT	&	1.1	&	39.5	&	63	&	1.1	&	0.2	\\
\hline
22	&	Prolog	&	2.0	&	133.7	&	17	&	0.1	&	1.9	\\
	&	XSLT	&	0.8	&	27.1	&	61	&	1.7	&	0.1	\\
\hline
23	&	Prolog	&	1.8	&	145.0	&	34	&	0.1	&	1.6	\\
	&	Prolog2	&	1.8	&	185.1	&	29	&	0.1	&	2.4	\\
	&	XSLT	&	1.3	&	39.5	&	60	&	1.1	&	0.2	\\
\hline
24	&	Prolog	&	1.9	&	168.0	&	15	&	0.1	&	2.5	\\
	&	Prolog2	&	1.9	&	186.2	&	20	&	0.1	&	2.7	\\
	&	XSLT	&	1.8	&	43.7	&	50	&	0.7	&	0.3	\\
\hline
25	&	Prolog	&	1.6	&	118.8	&	34	&	0.2	&	1.3	\\
	&	XSLT	&	0.9	&	57.7	&	47	&	0.5	&	0.4	\\
\hline
26	&	Prolog	&	1.8	&	190.5	&	36	&	0.1	&	2.2	\\
	&	Prolog2	&	1.7	&	139.1	&	31	&	0.1	&	1.6	\\
	&	Prolog3	&	1.8	&	173.2	&	41	&	0.1	&	1.8	\\
	&	XSLT	&	0.8	&	27.1	&	61	&	1.7	&	0.1	\\
\hline
27	&	Prolog	&	1.9	&	156.2	&	8	&	0.1	&	2.5	\\
	&	Prolog2	&	1.9	&	226.8	&	18	&	0.1	&	3.5	\\
	&	XSLT	&	1.3	&	52.5	&	60	&	0.7	&	0.3	\\
\hline
28	&	Prolog	&	1.8	&	190.7	&	36	&	0.1	&	2.2	\\
	&	Prolog2	&	1.8	&	226.8	&	37	&	0.1	&	2.7	\\
	&	XSLT	&	1.0	&	43.7	&	64	&	1.0	&	0.2	\\
\hline
29	&	Prolog	&	2.0	&	265.2	&	12	&	0.0	&	4.5	\\
	&	XSLT	&	2.0	&	44.0	&	60	&	0.9	&	0.2	\\
\hline
30	&	Prolog	&	2.0	&	214.2	&	27	&	0.1	&	2.9	\\
	&	Prolog2	&	1.8	&	161.5	&	35	&	0.1	&	1.8	\\
	&	XSLT	&	1.1	&	43.7	&	66	&	1.1	&	0.2	\\
\hline
31	&	Prolog	&	1.9	&	290.2	&	16	&	0.0	&	4.8	\\
	&	XSLT	&	1.5	&	52.5	&	62	&	0.8	&	0.3	\\
\hline
32	&	Prolog	&	2.0	&	257.6	&	28	&	0.1	&	3.6	\\
	&	Prolog2	&	1.8	&	209.1	&	36	&	0.1	&	2.5	\\
	&	XSLT	&	1.2	&	66.6	&	61	&	0.6	&	0.4	\\
\hline
33	&	Prolog	&	1.9	&	251.7	&	23	&	0.1	&	3.7	\\
	&	Prolog2	&	2.2	&	161.4	&	20	&	0.1	&	2.3	\\
	&	XSLT	&	1.3	&	76.2	&	64	&	0.5	&	0.4	\\
\hline
34	&	Prolog	&	1.7	&	196.5	&	38	&	0.1	&	2.2	\\
	&	XSLT	&	1.3	&	91.4	&	65	&	0.4	&	0.5	\\
\hline
35	&	Prolog	&	1.7	&	122.8	&	33	&	0.2	&	1.4	\\
	&	Prolog2	&	1.5	&	81.3	&	33	&	0.3	&	0.8	\\
	&	XSLT	&	0.9	&	36.5	&	49	&	0.9	&	0.2	\\
\hline
36	&	Prolog	&	1.7	&	239.4	&	20	&	0.1	&	3.6	\\
	&	XSLT	&	1.0	&	102.7	&	60	&	0.3	&	0.7	\\
\hline
37	&	Prolog	&	2.1	&	251.2	&	17	&	0.1	&	4.0	\\
	&	Prolog2	&	2.0	&	196.5	&	29	&	0.1	&	2.5	\\
	&	XSLT	&	1.7	&	43.7	&	57	&	0.9	&	0.2	\\
\hline
38	&	Prolog	&	1.4	&	281.2	&	17	&	0.0	&	4.6	\\
	&	XSLT	&	1.1	&	75.6	&	50	&	0.4	&	0.6	\\
\hline
39	&	Prolog	&	1.8	&	226.5	&	20	&	0.1	&	3.4	\\
	&	XSLT	&	1.1	&	52.5	&	64	&	0.8	&	0.3	\\
\hline
40	&	Prolog	&	2.1	&	245.8	&	11	&	0.1	&	4.1	\\
	&	XSLT	&	2.0	&	52.5	&	49	&	0.6	&	0.4	\\
\hline
\end{tabular}

\newpage

\begin{tabular}{|l l| l l | l l l l}
\hline
42	&	Prolog	&	14	&	347 & 17	&	18	&	66	&	35\\
	&	Prolog2	&	14	&	343 & 19	&	18	&	67	&	34\\
	&	XSLT	&	9	&	299 & 5	&	5	&	5	&	5\\
\hline
43	&	Prolog	&	14	&	428 & 18	&	27	&	103	&	51\\
	&	XSLT	&	22	&	678 & 9	&	8	&	13	&	8\\
\hline
44	&	Prolog	&	18	&	482 & 17	&	21	&	103	&	51\\
	&	Prolog2	&	18	&	472 & 19	&	23	&	102	&	53\\
	&	XSLT	&	17	&	460 & 5	&	7	&	9	&	8\\
\hline
45	&	Prolog	&	14	&	336 & 18	&	23	&	66	&	37\\
	&	XSLT	&	13	&	387 & 5	&	7	&	7	&	7\\
\hline
46	&	Prolog	&	14	&	445 & 14	&	20	&	99	&	40\\
	&	Prolog2	&	21	&	532 & 19	&	26	&	106	&	58\\
	&	XSLT	&	19	&	638 & 7	&	10	&	11	&	10\\
\hline
47	&	Prolog	&	31	&	903 & 21	&	41	&	186	&	103\\
	&	XSLT	&	36	&	1188 & 6	&	20	&	28	&	28\\
\hline
48	&	Prolog	&	13	&	324 & 13	&	21	&	52	&	39\\
	&	XSLT	&	10	&	300 & 5	&	6	&	5	&	6\\
\hline
49	&	Prolog	&	28	&	1078 & 23	&	40	&	206	&	110\\
	&	XSLT	&	30	&	941 & 8	&	10	&	18	&	10\\
\hline
50	&	Prolog	&	25	&	1071 & 16	&	38	&	146	&	95\\
	&	XSLT	&	43	&	1403 & 6	&	17	&	31	&	23\\
\hline
51	&	Prolog	&	21	&	745 & 18	&	31	&	127	&	64\\
	&	XSLT	&	25	&	812 & 8	&	8	&	14	&	8\\
\hline
52	&	Prolog	&	15	&	463 & 11	&	29	&	34	&	55\\
	&	XSLT	&	20	&	569 & 4	&	8	&	12	&	8\\
\hline
53	&	Prolog	&	29	&	812 & 19	&	43	&	164	&	93\\
	&	XSLT	&	29	&	864 & 9	&	14	&	19	&	17\\
\hline
54	&	Prolog	&	24	&	939 & 19	&	41	&	175	&	96\\
	&	Prolog2	&	21	&	885 & 17	&	39	&	161	&	88\\
	&	XSLT	&	29	&	784 & 6	&	11	&	18	&	11\\
\hline
55	&	Prolog	&	15	&	438 & 14	&	23	&	84	&	43\\
	&	XSLT	&	14	&	366 & 4	&	6	&	8	&	6\\
\hline
56	&	Prolog	&	50	&	1797 & 28	&	43	&	302	&	174\\
	&	XSLT	&	27	&	871 & 12	&	15	&	17	&	16\\
\hline
63	&	Prolog	&	14	&	428 & 16	&	23	&	91	&	45\\
	&	Prolog2	&	12	&	376 & 14	&	22	&	69	&	43\\
	&	XSLT	&	23	&	563 & 8	&	7	&	14	&	8\\
\hline
64	&	Prolog	&	12	&	341 & 15	&	22	&	67	&	38\\
	&	Prolog2	&	15	&	384 & 11	&	19	&	69	&	43\\
	&	Prolog3	&	9	&	276 & 12	&	19	&	44	&	28\\
	&	XSLT	&	24	&	826 & 7	&	11	&	15	&	16\\
\hline
65	&	Prolog	&	25	&	878 & 22	&	34	&	173	&	81\\
	&	Prolog2	&	24	&	810 & 21	&	34	&	165	&	81\\
	&	XSLT	&	30	&	818 & 8	&	10	&	18	&	10\\
\hline
\hline
\end{tabular}

\newpage

\begin{tabular}{l l| l l | l l l|}
\hline
42	&	Prolog	&	1.9	&	144.5	&	30	&	0.1	&	1.7	\\
	&	Prolog2	&	2.0	&	155.8	&	35	&	0.1	&	1.8	\\
	&	XSLT	&	1.0	&	23.2	&	58	&	1.9	&	0.1	\\
\hline
43	&	Prolog	&	2.0	&	203.4	&	25	&	0.1	&	2.8	\\
	&	XSLT	&	1.6	&	52.5	&	60	&	0.7	&	0.3	\\
\hline
44	&	Prolog	&	2.0	&	161.7	&	5	&	0.1	&	2.7	\\
	&	Prolog2	&	1.9	&	184.8	&	16	&	0.1	&	2.8	\\
	&	XSLT	&	1.1	&	31.3	&	47	&	1.1	&	0.2	\\
\hline
45	&	Prolog	&	1.8	&	179.1	&	43	&	0.1	&	1.8	\\
	&	XSLT	&	1.0	&	31.3	&	56	&	1.3	&	0.2	\\
\hline
46	&	Prolog	&	2.5	&	139.7	&	1	&	0.1	&	2.4	\\
	&	Prolog2	&	1.8	&	202.9	&	19	&	0.1	&	3.0	\\
	&	XSLT	&	1.1	&	52.9	&	60	&	0.7	&	0.3	\\
\hline
47	&	Prolog	&	1.8	&	311.9	&	7	&	0.0	&	5.7	\\
	&	XSLT	&	1.0	&	101.9	&	45	&	0.2	&	0.9	\\
\hline
48	&	Prolog	&	1.3	&	140.3	&	35	&	0.1	&	1.5	\\
	&	XSLT	&	0.8	&	27.1	&	61	&	1.7	&	0.1	\\
\hline
49	&	Prolog	&	1.9	&	316.9	&	0	&	0.0	&	6.3	\\
	&	XSLT	&	1.8	&	57.2	&	52	&	0.5	&	0.4	\\
\hline
50	&	Prolog	&	1.5	&	263.4	&	9	&	0.0	&	4.6	\\
	&	XSLT	&	1.3	&	85.0	&	36	&	0.3	&	0.8	\\
\hline
51	&	Prolog	&	2.0	&	228.6	&	17	&	0.1	&	3.6	\\
	&	XSLT	&	1.8	&	48.0	&	54	&	0.7	&	0.3	\\
\hline
52	&	Prolog	&	0.6	&	178.9	&	50	&	0.1	&	1.6	\\
	&	XSLT	&	1.5	&	32.0	&	38	&	0.9	&	0.2	\\
\hline
53	&	Prolog	&	1.8	&	314.0	&	18	&	0.0	&	5.1	\\
	&	XSLT	&	1.1	&	81.8	&	56	&	0.4	&	0.5	\\
\hline
54	&	Prolog	&	1.8	&	300.4	&	10	&	0.0	&	5.3	\\
	&	Prolog2	&	1.8	&	275.6	&	10	&	0.0	&	4.8	\\
	&	XSLT	&	1.6	&	53.6	&	46	&	0.5	&	0.4	\\
\hline
55	&	Prolog	&	2.0	&	157.3	&	20	&	0.1	&	2.2	\\
	&	XSLT	&	1.3	&	23.5	&	42	&	1.4	&	0.2	\\
\hline
56	&	Prolog	&	1.7	&	367.9	&	23	&	0.0	&	9.8	\\
	&	XSLT	&	1.1	&	101.6	&	68	&	0.4	&	0.5	\\
\hline
63	&	Prolog	&	2.0	&	168.0	&	20	&	0.1	&	2.4	\\
	&	Prolog2	&	1.6	&	151.4	&	26	&	0.1	&	1.9	\\
	&	XSLT	&	1.8	&	43.7	&	50	&	0.7	&	0.3	\\
\hline
64	&	Prolog	&	1.8	&	156.7	&	33	&	0.1	&	1.8	\\
	&	Prolog2	&	1.6	&	118.8	&	6	&	0.1	&	1.8	\\
	&	Prolog3	&	1.6	&	123.7	&	42	&	0.2	&	1.2	\\
	&	XSLT	&	0.9	&	57.7	&	47	&	0.5	&	0.4	\\
\hline
65	&	Prolog	&	2.1	&	271.1	&	7	&	0.0	&	4.9	\\
	&	Prolog2	&	2.0	&	265.2	&	8	&	0.0	&	4.7	\\
	&	XSLT	&	1.8	&	57.2	&	52	&	0.5	&	0.4	\\
\hline
\hline
\end{tabular}

Source: \texttt{https://rhaber123.\-github\-.io/web-page/}

\newpage

\section*{APPENDIX B: Prolog Rules Set}
\label{appendix:PrologRules}
\subsection{Internal Rules}

traverse/2,traverseElements/2

\begin{verbatim}
% traverse::Node -> [Node]
traverse(pi(_),[]):-!.
traverse(comment(_),[]):-!.
traverse(X,Res):-template(X,Res), !.
traverse(element(_,_,L),Res):-
  traverseElements(L,Res).

% traverseElements::[Node] -> [Node]
traverseElements([],[]).
traverseElements([H|T],Res):-
  not(list(H)), compound(H),
  traverse(H,Res1),
  traverseElements(T,Res2),
  append(Res1,Res2,Res).
\end{verbatim}

\subsection{Transformation Operators}

\lit{/}/2, \lit{\textasciicircum}/2, \lit{@}/2, \lit{?}/2, id/3, \lit{\#}/2, 
c/2, atts/1, sort/2, sortbyName/1, child/1, descendant/1, copy/1, copy_of/1,
level/2, last/1, count/1, name/1, distinct/1

\begin{verbatim}
:-op(100,yfx,'/').
% /:: ElementNode -> Name -> ElementNode
transform(E1 / Child,element(Child,A,C)):-
  E1=element(Name,AttList,Children),
  append(_,[(element(Child,A,C))|_],
         Children).
transform(X / Child,Y):-transform(X,X2),
  transform(X2 / Child,Y).

:-op(100,yfx,'^').
% ^:: ElementNode -> Name -> ElementNode
transform(_ ^ Name,_):-
  (var(Name);list(Name)), !, fail.
transform(element(Name,A,C) ^ Name,
          element(Name,A,C)).
transform(element(N,_,[H|_]) ^ Name,X):-
  transform(H ^ Name,X).
transform(element(N,A,[_|T]) ^ Name,X):-
  N\=Name,
  transform(element(N,A,T) ^ Name,X).
transform(X ^ Name,Y):-
  transform(X,X2),
  transform(X2 ^ Name,Y).
\end{verbatim}

\begin{verbatim}
:-op(100,yfx,'@').
% @:: ElementNode -> AttName -> AttName
transform(element(_,AttList,_) @ Att,X):-
  append(_,[A|_],AttList),
  atom_codes(Att,AttCodes),
  atom_codes(A,ACodes),
  append(Pre,[61,34|X2],ACodes),
  append(X3,[34],X2),
  Pre=AttCodes, !, atom_codes(X,X3).
transform(X @ Att, Y):-
  transform(X,X2),
  transform(X2 @ Att, Y).

:-op(100,fy,atts).
% atts:: ElementNode -> [AttribName]
transform(atts element(_,L,_),Y):-
  not(list(L)), !, fail.
transform(atts element(_,L,_),_):-
  findall(X,selectattribute(X,L),[]),
  !, fail.
transform(atts element(_,L,_),Y):-
  findall(X,selectattribute(X,L),Y).
transform(atts E,Y):-
  transform(E,E2),
  transform(atts E2,Y).

:-op(100,yfx,'?').
% ?:: ElementNode -> String
transform(X ? Att1):-
  atom(Att1), transform(atts X,X2),
  member(Att1,X2).

:-op(100,yfx,id).
% id:: ElementNode -> AttValue -> AttName
transform(X id S,Attrib):-
  X=element(_,AL,_), 
  transform(atts X,AttribNames),
  member(Attrib,AttribNames),
  transform(X @ Attrib,S).
transform(X id S,Id):-
  transform(X,X2),
  transform(X2 id S,Id).	

:-op(100,yfx,'#').
% #:: ElementNode -> Integer -> String
transform(element(_,_,L) # N,Y):-
  integer(N), N>=1,
  findall(X,member(text(X),L),Z),
  nth(N,Z,Y).
transform(X # N,Y):-
  transform(X,X2), transform(X2 # N,Y).

% ?:: ElementNode -> Integer -> String
transform(element(_,_,L) ? N,Y):-
  integer(N), N>=1,
  findall(X,member(pi(X),L),Z),
  nth(N,Z,Y).
transform(X ? N,Y):-
  transform(X,X2),
  transform(X2 ? N,Y).

:-op(100,yfx,'c').
% c:: ElementNode -> Integer -> String
transform(element(_,_,L) c N,Y):-
  integer(N), N>=1, 
  findall(X,member(comment(X),L),Z),
  nth(N,Z,Y).
transform(X c N,Y):-
  transform(X,X2),
  transform(X2 c N,Y).

:-op(100,yfx,sort).
% sort:: ElementNode -> AttName
          -> ElementNode
transform(element(N,A,L)
          sort AttName,
          element(N,A,Y)):-
  extendStructure(L2,AttName,L),
  quicksort(L2,leAttributes,L3),
  extendStructure(L3,AttName,Y).
	
:-op(100,fy,sortbyName).
% sortbyName:: ElementNode -> ElementNode
transform(sortbyName element(N,A,L),
          element(N,A,Y)):-
  quicksort(L,le,Y).

:-op(100,fy,child).
% child:: ElementNode -> ElementNode
transform(child element(_,_,C),Y):-
  member(Y,C).
transform(child X,Y):-
  transform(X,X2),
  transform(child X2,Y).

:-op(100,fy,descendant).
% descendant:: ElementNode -> ElementNode
transform(descendant X,Y):-
  transform(child X,Y).
transform(descendant X,Y):-
  transform(child X,Y2),
  transform(descendant Y2,Y).	

:-op(100,fy,copy).
% copy:: ElementNode -> ElementNode
transform(copy element(N,_,_),
          element(N,[],[])).
transform(copy text(T),text(T)).
transform(copy comment(C),
          comment(C)).
transform(copy pi(P),pi(P)).
transform(copy X,Y):-
  transform(X,X2),
  transform(copy X2,Y).

:-op(100,fy,copy_of).
% copy_of:: ElementNode -> ElementNode
transform(copy_of X,X):-
  X=element(_,_,_);
  X=text(_);
  X=comment(_); X=pi(_).
transform(copy_of X,Y):-transform(X,Y).

:-op(100,yfx,level).
% level:: ElementNode -> ElementNode 
          -> [Integer]
transform(Tree level Node,Y):-
  level1(Tree,Node,Y).
transform(Tree level Node,Y):-
  transform(Tree,Tree2),
  transform(Node,Node2),
  level1(Tree2,Node2,Y).
	
:-op(100,fy,last).
% last:: ElementNode -> ElementNode
transform(last element(_,_,C),Y):-
  last(C,Y).
transform(last X,Y):-
  transform(X,X2),
  transform(last X2,Y).

:-op(100,fy,count).
% count:: ElementNode -> Integer
transform(count element(_,_,C),Len):-
  length(C,Len).
transform(count X,Y):-
  transform(X,X2),
  transform(count X2,Y).

:-op(100,fy,name).
% name:: ElementNode -> String
transform(name element(Name,_,_),_):-
  (var(Name);list(Name)), !, fail.
transform(name element(Name,_,_),Name).
transform(name X,Y):-
  transform(X,X2),
  transform(name X2,Y).

:-op(100,fy,distinct).
% distinct:: ElementNode -> ElementNode
transform(distinct element(N,A,L),
          element(N,A,Z)):-
  reverse(L,L2),
  removeDuplicates(L2,L3),
  reverse(L3,Z).
\end{verbatim}

\subsection{Non-Monotone Predicates}

removeElement/3, remove/3, removeAttribute/3, insertBefore/4, insertAfter/4

\begin{verbatim}
% removeElement:: ElementNode -> 
   ElementNode -> ElementNode
removeElement(element(N,As,L),
              Name,element(N,As,L2)):-
  delete(element(Name,_,_),L,L2).

% remove:: ElementNode -> String 
           -> ElementNode
remove(element(N,As,L),Node,
       element(N,As,L2)):-
  delete(Node,L,L2).

% removeAttribute:: ElementNode 
    -> AttribName -> ElementNode
removeAttribute(E,Att,element(N,As2,L)):-
  E=element(N,As,L),
  transform(E @ Att,Val),
  atom_codes(Att,AttCodes),
  atom_codes(Val,ValCodes),
  append(AttCodes,[61,34|ValCodes],Res2),
  append(Res2,[34],Res),
  atom_codes(Selected,Res),
  delete(Selected,As,As2).
\end{verbatim}

\begin{verbatim}
% insertBefore:: ElementNode -> Node
    -> Node -> ElementNode
insertBefore(_,_,RecentNode,_):-
  (var(RecentNode);
   list(RecentNode);
   number(RecentNode);
   atom(RecentNode)),
  !, fail.
insertBefore(_,NewNode,_,_):-
  (var(NewNode);
   list(NewNode);
   number(NewNode);
   atom(NewNode)),
  !, fail.
insertBefore(E1,NewNode,RecentNode,
             element(N,A,List2)):-
  E1=element(N,A,List),
  compound(RecentNode), 
  !, compound(NewNode),
  append(Pre,[RecentNode|Post],List),
  append(Pre,[NewNode,RecentNode|Post],
         List2).

% insertBefore:: ElementNode -> Node
   -> Integer -> ElementNode
insertBefore(E1,NewNode,Position,
             element(N,A,List2)):-
  E1=element(N,A,List),
  integer(Position),
  !, Position>=1,
  nth(Position,List,X),
  append(Pre,[X|Post],List),
  append(Pre,[NewNode,X|Post],List2).
	
% insertAfter:: ElementNode -> Node
      -> Node -> ElementNode
insertAfter(_,_,RecentNode,_):-
  (var(RecentNode); list(RecentNode);
   number(RecentNode); atom(RecentNode)),
  !, fail.
insertAfter(_,NewNode,_,_):-
  (var(NewNode); list(NewNode);
   number(NewNode); atom(NewNode)),
  !, fail.
insertAfter(E1,NewNode,RecentNode,
    element(N,A,List2)):-
  E1=element(N,A,List),
  compound(RecentNode), !,
  append(Pre,[RecentNode|Post],List),
  append(Pre,[RecentNode,NewNode|Post],
         List2).

% insertAfter:: ElementNode -> Node 
    -> Integer -> ElementNode
insertAfter(E1,NewNode,Position,
            element(N,A,List2)):-
  E1=element(N,A,List), integer(Position),
  !, 
  Position>=1, nth(Position,List,X),
  append(Pre,[X|Post],List),
  append(Pre,[X,NewNode|Post],List2).
\end{verbatim}

\subsection{Protected Helper Predicates}

level1/3, level0/4, levels0/5, nth0/3, selectattribute/2, removeDuplicates/2,
lexicalle/2, le/2, ge/2, concat0/3, extendStructure/3, checkSerializable/1,
checkSerializables/1, checkAttributes/1

\begin{verbatim}
% level1:: ElementNode -> Node -> [Integer]
level1(Tree,Node,Result):-
  level0(Tree,Node,[],Result).

% level0:: ElementNode -> Node 
    -> [Integer] -> [Integer]
level0(element(_,_,Children),Y,Res0,Res):-
  nth(N,Children,Y), Res=[N|Res0].
level0(element(N,A,[H|T]),Y,Res0,Res):-
  level0(H,Y,Res0,Res1),
  Res=[1|Res1];
  levels0([H|T],T,Y,Res0,Res).

% levels0:: [Node] -> [Node] -> Node
   -> [Integer] -> [Integer]
levels0(L,[H|T],Y,Res0,Res):-
  level0(H,Y,Res0,Res1),
  nth(N,L,H), Res=[N|Res1];
  levels0(L,T,Y,Res0,Res).

% nth0:: ChurchTerm -> [Term] -> Term
nth0(s(zero),[X|_],X).
nth0(s(M),[_|L],X):-nth0(M,L,X).

% selectattribute:: AttName -> AttEntry
selectattribute(_,L):-
  (var(L);number(L);
   atom(L), not(list(L))),
  !, fail.
selectattribute(X,List):-
  member(Y,List),
  atom_codes(Y,YCodes2),
  append(X2,[61,34|YCodes],YCodes2),
  append(_,[34],YCodes), atom_codes(X,X2).
			 
% removeDuplicates:: [Term] -> [Term]
removeDuplicates(L1,L2):-not(list(L1)),
  !, fail.
removeDuplicates([],[]).
removeDuplicates([H|T],T2):-
  member(H,T),
  removeDuplicates(T,T2).
removeDuplicates([H|T],[H|T2]):-
  not(member(H,T)),
  removeDuplicates(T,T2).

% lexicalle:: [ANumerical] -> [ANumerical]
lexicalle([],[]).
lexicalle([],[H2|_]).
lexicalle([H|_],[]):-fail.
lexicalle([H|_],[H2|_]):-
  nonvar(H), nonvar(H2), H>H2, fail.
lexicalle([H|T],[H2|T2]):-
  nonvar(H), nonvar(H2), H=H2,
  lexicalle(T,T2), !.
lexicalle([H|T],[H2|T2]):-
  var(H), var(H2), !, fail.
lexicalle([H|_],[H2|_]):-
  nonvar(H), nonvar(H2), H<H2, !.
lexicalle([H|T],[H2|T2]):-
  H=H2, lexicalle(T,T2), !.

% le:: ElementNode -> ElementNode
le(element(N,_,_),element(N2,_,_)):-
  atom(N), not(list(N)),
  atom(N2), not(list(N2)),
  atom_codes(N,NCodes),
  atom_codes(N2,N2Codes),
  lexicalle(NCodes,N2Codes).

% ge:: ElementNode -> ElementNode
ge(X,Y):-le(Y,X).

% concat0:: [[Term]] -> [Term] ->  [Term]
concat0([],X,X).
concat0([H|T],X,Y):-list(H),
	append(X,H,X2), concat0(T,X2,Y).

% exendStructure:: [Node] -> Term
   -> [ElementNode]
extendStructure([],_,[]).
extendStructure(L,_,L2):-
  not(ground(L)),
  not(ground(L2)), !, fail.
extendStructure([E1|T2],Extension,[E2|T]):-
  E1=element(N,A,C,Extension),
  extendStructure(T2,Extension,T),
  E2=element(N,A,C).

% checkSerializable:: Node	
checkSerializable(pi(_)):-!.
checkSerializable(comment(_)):-!.
checkSerializable(text(_)):-!.
checkSerializable(element(N,A,C)):-
  not(list(N)), atom(N),
  checkAttributes(A),
  checkSerializables(C), !.
checkSerializable(X):-
  write('\nError: '), write(X), 
  write(' was not expected here!'), fail.

% checkSerializables:: [Node]	
checkSerializables([]).
checkSerializables([H|T]):-
  checkSerializable(H),
checkSerializables(T).

% checkAttributes:: [AttribEntry]
checkAttributes([]):-!.
checkAttributes([H|T]):-
  atom_codes(H,HCodes), 
  append(_,[61,34|HCodes1],HCodes),
  append(_,[34],HCodes1),
  checkAttributes(T), !.
checkAttributes(X):-
  write('\nError in remaining attributes
           list: '),
  write(X), fail.
\end{verbatim}

\subsection{Helper Predicates}

sum/2, last/2, nth/3, concat/2, church/2, leAttributes/2, leStrings/2,
checkSerializable0/1, concat/3, printTree/2, printChildren/2,
flatten/2, flattenList/2, nodes/2, nodesList/2

\begin{verbatim}
% sum:: [Number] -> Number
sum([],0).
sum([H|T],X):-sum(T,X2), X is X2+H.

% last:: [Term] -> Term
last([_|T],L):-last(T,L).
last([H],H).

% nth:: Integer -> [Term] -> Term
nth(N,L,E):-
  var(N), nth0(N1,L,E), church(N1,N).
nth(N,L,E):-
  church(N1,N), nth0(N1,L,E).

% concat:: [[Term]] -> [Term]
concat(L,X):-concat0(L,[],X).

% church:: ChurchTerm -> Integer
church(zero,0):-!.
church(s(X),N):-
  var(N),
  church(X,N1),
  N is N1+1.
church(s(X),N):-
  not(var(N)),
  N1 is N-1,
  church(X,N1).

% leAttributes:: ElementNode -> ElementNode
leAttributes(element(N,AL1,_,Att1),
             element(N2,AL2,_,Att1)):-
  transform(element(_,AL1,_) @ Att1,A1),
  transform(element(_,AL2,_) @ Att1,A2),
  atom_codes(A1,E1Codes),
  atom_codes(A2,E2Codes),
  lexicalle(E1Codes,E2Codes).

% leString:: String -> String
leStrings(S1,S2):-
  atom(S1),
  not(list(S1)), 
  atom(S2),
  not(list(S2)),
  atom_codes(S1,S1Codes),
  atom_codes(S2,S2Codes),
  lexicalle(S1Codes,S2Codes).

% checkSerializable0:: ElementNode
checkSerializable0(element(N,A,C)):-
  checkSerializable(element(N,A,C)), !.
checkSerializable0(X):-
  write('\nError: element()-constructor was 
           expected, but '),
  write(X),
  write(' was found!'),
  fail.

% concat:: String -> String -> String
concat(E1,E2,A1):-var(A1),
  A1 is cat(E1,E2).
concat(E1,E2,A1):-var(E1),
  atom_codes(E2,E2Codes),
  atom_codes(A1,A1Codes),
  append(E1Codes,E2Codes,A1Codes),
  atom_codes(E1,E1Codes).
concat(E1,E2,A1):-var(E2), 
  atom_codes(E1,E1Codes),
  atom_codes(A1,A1Codes),
  append(E1Codes,E2Codes,A1Codes),
  atom_codes(E2,E2Codes).

% printTree:: Node -> String
printTree(text(T),T):-
  !, atom(T), not(list(T)).
printTree(comment(_),'').
printTree(pi(_),'').
printTree(element(_,_,Children),Res):-
  printChildren(Children,Res), !.

% printChildren:: [Node] -> String
printChildren([],'').
printChildren([H|T],Res):-
  printTree(H,Res1),
  printChildren(T,Res2),
  Res is cat(Res1,Res2).

% flatten:: Node -> [Node]
flatten(X,_):-
  (var(X);list(X);number(X)),
  !, fail.
flatten(element(N,A,L),
        [element(N,A,[])|T2]):-
  !, flattenList(L,T2).
flatten(X,[X]):-
  X=text(_);X=pi(_);X=comment(_).

% flattenList:: [Node] -> [Node]
flattenList([],[]).
flattenList([H|T],L):-
  flatten(H,L1),
  !, flattenList(T,L2), append(L1,L2,L).

% nodes:: Node -> [Node]
nodes(X,_):-
  (var(X);list(X);number(X)),
  !, fail.
nodes(element(N,A,L),
      [element(N,A,L)|T2]):-
  !, nodesList(L,T2), !.
nodes(X,[X]):-
  X=text(_);
  X=pi(_);
  X=comment(_).

% nodesList:: [Node] -> [Node]
nodesList([],[]).
nodesList([H|T],L):-
  nodes(H,L1),
  nodesList(T,L2),
  append(L1,L2,L).
\end{verbatim}



\begin{thebibliography}{10}
  \providecommand{\url}[1]{#1}
  
\bibitem{Ata:2004}
Frank Atanassow and Johan Jeuring. \textit{Customizing an XML-Haskell Data Binding with Type Isomorphism inference in Generic Haskell}. Sci. Comput. Program., 65(2):72–107, 2007.

  \bibitem{Baumeister} Joachim Baumeister, Frank Puppe, Dietmar Seipel. \textit{Refactoring Methods for Knowledge Bases}. Universit\"at W\"urzburg 2004.\\
\textit{http://ki.informatik.uni-wuerzburg.de/papers/baumeister/2004/\\Refactoring-EKAW04.pdf}

  \bibitem{Belli} Fevzi Belli, Oliver Jack. \textit{Implementation-Based Analysis and Testing of Prolog Programs}. ACM SIGSOFT 1993.
				  
  \bibitem{Bruno} Emmanuel Bruno, Jacques Le Maitre, Elisabeth Murisasco. \textit{Extending XQuery with Transformation Operators}. Proceedings of the 2003 ACM symposium on Document engineering. ACM Press 2003.

  \bibitem{Christensen04} Aske Simon Christensen, Christian Kirkegaard, Anders M\o ller. \textit{A Runtime System for XML Transformations in Java}. International XML Database Symposium (XSym). Springer 2004.
				
  \bibitem{XSLTMark} Data Power XSLTMark 2001.\\
				 \textit{http://www.datapower.com/xmldev/xsltmark.html}
				 
  \bibitem{Denti01} Enrico Denti, Andrea Omicini, Alessandro Ricci. \textit{tuProlog: A Light-weight Prolog for Internet Applications and Infrastructures -- Practical Aspects of Declarative Languages}. 3rd International Symposium (PADL'01). Springer 2001.

  \bibitem{Denti05} Enrico Denti, Andrea Omicini, Alessandro Ricci. \textit{Multi-paradigm Java-Prolog integration in tuProlog}. Elsevier 2005.

  \bibitem{Grosof} Benjamin N. Grosof, Yannis Labrou, Hoi Y. Chan. \textit{A Declarative Approach to Business Rules in Contracts: Courteous Logic Programs in XML}. Proceedings of the 1st ACM conference on Electronic commerce. ACM Press 1999.

  \bibitem{halstead77} Maurice H. H\r{a}lstead. \textit{Elements of Software Science}. New York, USA: Elsevier Science, 1977. ISBN 0444002057.

  \bibitem{Heumesser} Bernd Heumesser, Andreas Ludwig, Dietmar Seipel. \textit{Web Services based on Prolog and XML}. Proc. 15th Intl. Conference on Applications of Declarative Programming and Knowledge Management (pp. 369--378). INAP 2004.\\
	\textit{http://www-db.informatik.uni-tuebingen.de/forschung/papers/\\inap2004.pdf}
				 
  \bibitem{Hosoya} Haruo Hosoya, Benjamin C. Pierce. \textit{Xduce: A statically typed XML processing language}. ACM Transactions on Internet Technology. ACM 2003.

  \bibitem{Janssen} Wim Janssen, Alexandr Korlyukov, Jan Van den Bussche. \textit{On the tree-transformation power of XSLT}. arXiv.org 2006.\\
	\textit{http://arxiv.org/pdf/cs.PL/0603028}
				  
  \bibitem{Kerievsky} Joshua Kerievsky. \textit{Refactoring to Patterns}. Addison-Wesley, 2005.
				  
  \bibitem{Kirkegaard} Christian Kirkegaard, Anders Moeller, Michael I. Schwartzbach. \textit{Static Analysis of XML Transformations in Java}. IEEE Transactions on Volume 30, Issue 3. IEEE 2004.
				  
  \bibitem{Kiselyov} Oleg Kiselyov, Shriram Krishnamurthi. \textit{SXSLT: Manipulation Language for XML}. Proc. Fifth Intl. Sym. Practical Aspects of Declarative Languages (PADL2003).\\
	 \textit{http://www.cs.brown.edu/\textasciitilde sk/Publications/Papers/Published/kk-sxslt}

  \bibitem{Leslie} Donald M. Leslie. \textit{Transforming Documentation from the XML Doctypes used for the Apache Website to DITA: a Case Study}. Proceedings of the 19th annual international conference on Computer documentation, ACM Press 2001.

  \bibitem{May} Wolfgang May. \textit{XPath-Logic and PathLog: A Logic-Based Approach for Declarative XML Data Manipulation}. Technical Report No. 149, Institut f\"ur Informatik, Freiburg, Universit\"at Freiburg 2001.\\
	\textit{ftp://ftp.informatik.uni-freiburg.de/documents/reports/report149/\\report00149.ps.gz}
				 
  \bibitem{Meijer} Erik Meijer, Mark Shields. \textit{XM$\lambda$: A Functional Language for Constructing and Manipulating XML Documents}. CiteSeer 2000\\
	\textit{http://citeseer.ist.psu.edu/meijer00xmlambda.html}
				
  \bibitem{Moertel} Tom Moertel. \textit{XSLT, Perl, Haskell \& a Word on Language Design}.
	Kuro5hin.org from 15.06.2002.\\
	\textit{http://www.kuro5hin.org/print/2002/1/15/1562/95011}
				
  \bibitem{Novatchev} Dimitre Novatchev. \textit{The Functional Programming Language XSLT -- A proof through examples}, November 2001.\\
	\textit{http://www.topxml.com/xsl/articles/fp}
  				 
  \bibitem{Opeleva} Elvira Opaleva. \textit{Yazyki programmirovaniya i metody translyacii}. BHV Peterburg, Russia 2005.

  \bibitem{Pree} Wolfgang Pree. \textit{Framework Development and Reuse Support}. Prentice Hall 1995.
				
  \bibitem{PrologWiki} Prolog (Programmiersprache). \textit{Beitrag im deutschsprachigen Wikipedia -- download from 15.05.2006}.\\
	\textit{http://de.wikipedia.org/wiki/Prolog_(Programmiersprache)}
			
  \bibitem{PMT} \textit{Prolog Measurement Tool}. Universit\"at Magdeburg, 2006\\
	  \textit{ftp://irb.cs.uni-magdeburg.de/pub/local/papers/smlab/tools/pmt/\\pmt_deu.zip}

  \bibitem{Renar} Allen Renear, David Dubin, C. M. Sperberg-McQueen, Claus Huitfeldt.
	\textit{Towards a Semantics for XML Markup}. Proceedings of the 2002 ACM symposium on Document Engineering, ACM Press 2002.

  \bibitem{Seipel02} Dietmar Seipel. \textit{Processing XML Documents in PROLOG}. Proc. 17th Workshop on Logic Programming, WLP 2002.\\
	\textit{http://www-info1.informatik.uni-wuerzburg.de/database/papers/\\wlp\_2002.ps.gz}

  \bibitem{Seipel05} Dietmar Seipel, Klaus Pr\"ator. \textit{XML Transformations Based on Logic Programming}. Ulmer Informatik-Berichte 2005.\\
	 \textit{http://www-info1.informatik.uni-wuerzburg.de/database/papers/\\wlp\_2005\_seipel\_praetor.pdf}
				 
  \bibitem{Sterling} Leon Sterling, Ehud Shapiro. \textit{The Art of Prolog}. MIT Press 1994
				
  \bibitem{SterlingPractice} Leon Sterling. \textit{The Practice of Prolog}. MIT Press 1990

  \bibitem{Vion-Dury02} Jean-Yves Vion-Dury, Veronika Lux, Emmanuel Pietriga. \textit{Experimenting with the Circus Language for XML Modelling and Transformation}. Proceedings of the ACM on Symbolic Document Engineering, Xerox 2002.\\
	 \textit{http://www.xrce.xerox.com/Publications/Attachments/2002-032/\\CircusDocEng02.pdf}
				 
  \bibitem{Vion-Dury03} Jean-Yves Vion-Dury.
         \textit{XPath on Left and Right Sides of Rules: Toward Compact XML Tree Rewriting through Node Patterns}. Proceedings of the 2003 ACM symposium on Document Engineering, ACM Press 2003.

  \bibitem{Wadler} Phil Wadler. \textit{A Formal Semantics of Patterns in XSLT,	Markup Technologies}. CiteSeer 1999.\\
	\textit{http://citeseer.ist.psu.edu/204315.html}
				
  \bibitem{Wallace} Malcolm Wallace, Colin Runciman. \textit{Haskell and XML: Generic Combinators or Type-Based Translation?} ACM SIGPLAN ICFP 1999. ACM Press 1999
				
  \bibitem{Wirth} Niklaus Wirth.  \textit{The Programming Language Pascal}. Acta Informatica (1), 1971.
				
  \bibitem{Xact} \textit{Xact XML-Transformation Framework for Java}.\\
	 \textit{http://www.brics.dk/Xact}
  
  \bibitem{XalanJ} \textit{Xalan/J. XML Project}.\\
	\textit{http://www.xml.apache.org/xalanj}
				 				 
  \bibitem{XMLSpy} XMLSpy.\\
	\textit{http://www.xmlspy.com}
	
  \bibitem{XPathSpec} XPath 1.0 Specification, 1999.\\
	\textit{http://www.w3c.org/TR/xpath}
				 
  \bibitem{XQuerySpec} XQuery 1.0 Specification, 2006.\\
	\textit{http://www.w3c.org/TR/xquery}
	
  \bibitem{W3XSLTSpecification} XSLT 1.0 Specification, 1999.\\
	\textit{http://www.w3.org/TR/xslt}								 

  \bibitem{ZVON} ZVON.org -- Online XSLT tutorial. \textit{http://www.zvon.org}\\

\end{thebibliography}
\end{document}